%% file: main.tex
\input{src/preamble}
\begin{document}
\frontmatter
\input{src/front}
\newpage
\input{src/dedication}
\clearemptydoublepage
\input{src/abstracts}
\input{src/acknowledgements}
\input{src/nomenclature}
\clearemptydoublepage
\tableofcontents
\mainmatter
\input{src/chapters/intro}
\input{src/chapters/gj3998d}
\clearemptydoublepage
\input{src/chapters/gj526}
\clearemptydoublepage
\input{src/chapters/barnard_letter}
\clearemptydoublepage
\input{src/chapters/conclusions}
\appendix
\clearemptydoublepage
\input{src/chapters/other_contribs}
\backmatter
\addcontentsline{toc}{section}{References}
\bibliographystyle{aa}
\bibliography{zotero_bibtex}
\end{document}

%% file: src/preamble.tex
\immediate\write18{makeindex -s nomencl.ist -o "\jobname.nls" "\jobname.nlo"} 
\documentclass[
    11pt,
    a4paper,
    titlepage,
    twoside,
    openright
]{book}
\usepackage[
    a4paper,
    twoside=true,
    textwidth=360pt,
    hmarginratio=3:2,
    vmargin=30mm
]{geometry}
\usepackage[T2A]{fontenc}
\usepackage{verse}
\usepackage{lettrine}
\usepackage{Starburst}
\usepackage{lipsum}
\usepackage{hyperref}
\usepackage{physics}
\usepackage{wasysym}
\usepackage{amsmath}
\usepackage{amsfonts}
\usepackage{amssymb}
\usepackage{layouts}
\usepackage{xfrac}
\usepackage{minted}
\usepackage{pdfpages}
\usepackage{chngcntr}
\counterwithout{footnote}{chapter}
\counterwithout{figure}{chapter}
\counterwithout{table}{chapter} 
\counterwithout{equation}{chapter}
\usepackage[object=vectorian]{pgfornament}
\newcommand{\attrib}[1]{%
\nopagebreak{\raggedleft\footnotesize #1\par}}
\usepackage{natbib}
\usepackage{bibentry}
\nobibliography* 
\newcommand{\fullcite}[1]{\bibentry{#1}}
\usepackage{xcolor}
\definecolor{grey}{HTML}{595959}
\newcommand{\clearemptydoublepage}{\newpage{\pagestyle{empty}\null\cleardoublepage}}
\usepackage{fancyhdr}
\usepackage{tikz}
\usetikzlibrary{calc,positioning,arrows.meta,decorations.pathreplacing,angles,intersections,quotes}

\usepackage{physics}
\usepackage[
    separate-uncertainty=true,
    multi-part-units=single,
    range-phrase=\mbox{--},
    range-units=single
]{siunitx}
\ifdefined\unit\else
  \ifdefined\NewCommandCopy
    \NewCommandCopy\unit\si
  \else
    \NewDocumentCommand\unit{O{}m}{\si[#1]{#2}}
  \fi
\fi
\DeclareSIUnit{\ppt}{ppt}
\DeclareSIUnit{\ppm}{ppm}
\DeclareSIUnit{\au}{au}
\DeclareSIUnit{\dex}{dex}
\DeclareSIUnit{\hjulianday}{HJD}
\DeclareSIUnit{\Jansky}{Jy}
\DeclareSIUnit{\julianday}{BJD}
\DeclareSIUnit{\mas}{mas}
\DeclareSIUnit{\mag}{mag}
\DeclareSIUnit{\parsec}{pc}
\DeclareSIUnit{\permille}{\text{\textperthousand}}
\DeclareSIUnit{\rad}{rad}
\DeclareSIUnit{\solarmass}{M\textsubscript{\(\odot\)}}
\DeclareSIUnit{\solarradius}{R\textsubscript{\(\odot\)}}
\DeclareSIUnit{\solarluminosity}{L\textsubscript{\(\odot\)}}
\DeclareSIUnit{\earthradius}{R\textsubscript{\(\oplus\)}}
\DeclareSIUnit{\earthmass}{M\textsubscript{\(\oplus\)}}
\DeclareSIUnit{\earthflux}{\text{$\Phi$}\textsubscript{\(\oplus\)}}
\DeclareSIUnit{\jovianradius}{R\textsubscript{\(\jupiter\)}}
\DeclareSIUnit{\jovianmass}{M\textsubscript{\(\jupiter\)}}
\DeclareSIUnit{\year}{yr}

\DeclareMathOperator{\ptp}{ptp}

\makeatletter
\newcommand{\thechapterwordsEN}{%
  \if@mainmatter
    \ifcase\value{chapter}%
      \or One%
      \or Two%
      \or Three%
      \or Four%
      \or Five%
      \or Six%
      \or Seven%
    \fi
  \else
    \thechapter
  \fi
}
\def\@makechapterhead#1{%
  \vspace*{20\p@}%
  {\parindent \z@ \raggedright
    \normalfont
    \ifnum \c@secnumdepth >\m@ne
      \if@mainmatter
        {\scshape
        \@chapapp\,\thechapterwordsEN 
        }
        \par\nobreak
      \fi
    \fi
    \interlinepenalty\@M
    \Huge \bfseries #1\par\nobreak
    \vskip 40\p@
  }}
\def\@schapter#1{\if@twocolumn
                   \@topnewpage[\@makeschapterhead{#1}]%
                 \else
                   \@makeschapterhead{#1}%
                   \@afterheading
                 \fi}
\def\@makeschapterhead#1{%
  \vspace*{20\p@}%
  {\parindent \z@ \raggedright
    \normalfont
    \interlinepenalty\@M
    \Huge \bfseries #1\par\nobreak
    \vskip 40\p@
  }}

\usepackage[intoc]{nomencl}
\makenomenclature

\renewcommand{\nompreamble}{The present manuscript may include some of the following notation without declaration. Unless explicitly stated, this notation shall stand for:}

\usepackage{etoolbox}
\renewcommand\nomgroup[1]{%
  \item[\bfseries
  \ifstrequal{#1}{A}{Abbreviations and acronyms}{%
  \ifstrequal{#1}{C}{Constants}{%
  \ifstrequal{#1}{D}{Physical dimensions}{%
  \ifstrequal{#1}{U}{Units}{%
  \ifstrequal{#1}{O}{Operators}{}}}}}%
]}

\newcommand{\nomunit}[1]{%
\renewcommand{\nomentryend}{\hspace*{\fill}#1}}

\usepackage{fmtcount}
\renewcommand{\chaptermark}[1]{%
  \markboth{\chaptername\ \Numberstring{chapter}:\ #1}{}%
}

%% file: src/front.tex
\thispagestyle{empty}
\large
\begin{minipage}[b]{0.46\linewidth}
{\hspace{-0.6cm} \large UNIVERSIDAD DE LA LAGUNA\\
\vspace{0.2cm}
\hspace{-0.72cm} Departamento de Astrofísica}
\end{minipage}
\hfill
\begin{minipage}[b]{0.4\linewidth}
\includegraphics[width=6cm]{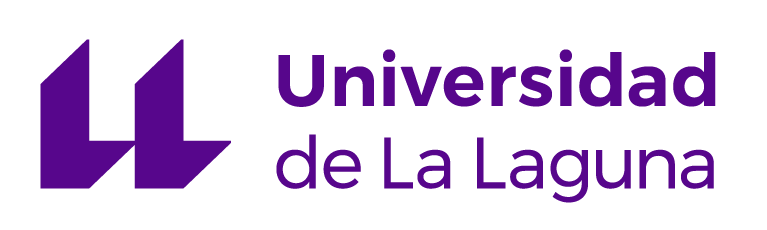}
\end{minipage}
\vfill
\begin{center}
    {\LARGE\bfseries Stellar magnetic activity and\\rocky planets in cool dwarf stars\par}
\end{center}
\vfill
\begin{center}
    {\Large Memoria que presenta\\
    Atanas Konstantinov Stefanov\\
    para optar al grado de\\
    Doctor en Astrofísica\\}
\end{center}
\vfill

\begin{center}
    \includegraphics[width=3truecm,height=3truecm]{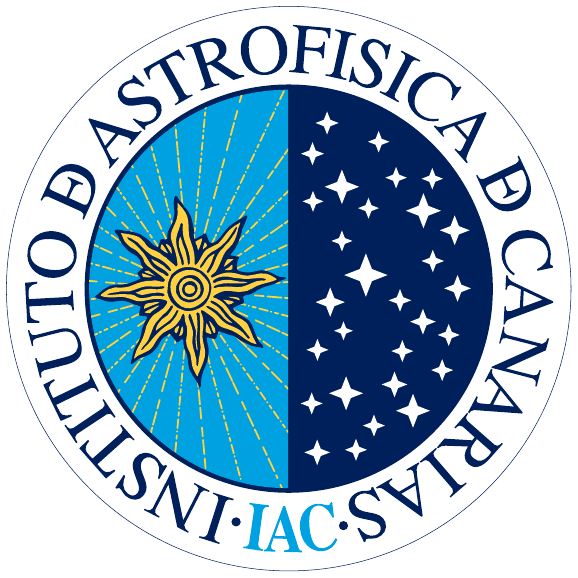}\\[2mm]
    {INSTITUTO {\fontencoding{T1}\selectfont\char208} ASTROFÍSICA {\fontencoding{T1}\selectfont\char208} CANARIAS\\[2mm]
    San Cristóbal de La Laguna\\
    septiembre de dos mil veintiséis\\}
\end{center}	
\newpage\normalsize
\thispagestyle{empty}
\ \vfill\noindent
{
\setlength{\topsep}{-\parskip}
\setlength{\partopsep}{-\parskip}
\begin{tabbing}
Programme co-supervisor:\hspace{5mm}\=Dr~Alejandro~Suárez~Mascareño \kill
Examination date: \>8 September 2026 \\[1ex]
Programme supervisor: \>Dr~Jonay~I.~González~Hernández \\
Programme co-supervisor: \>Dr~Alejandro~Suárez~Mascareño \\[1ex]
\end{tabbing}
}
Some figures and material included in the following manuscript have already been published in {\it Astronomy \& Astrophysics}.\\

%% file: src/dedication.tex
\thispagestyle{empty}
\vspace{1cm}
\begin{verse}[\versewidth]
Моторът, който пее горе,\\
е труд на моите ръце.\\
А тази песен на мотора\\
е кръв от моето сърце.
\end{verse}
\attrib{Никола Вапцаров, \textit{Двубой}, 1940}
\vspace{4mm}
\begin{verse}[\versewidth]
The motor singing high above \\
is the labour of my hands, \\
and the song that it is humming \\
is blood taken from my heart.
\end{verse}
\attrib{
Nikola Vaptsarov, \textit{А Duel}, 1940\\
}
\vspace{4mm}
\begin{verse}[\versewidth]
El motor que canta arriba \\
es trabajo de mis manos, \\
y su canto se deriva \\
de mi sangre, sin engaños.
\end{verse}
\attrib{
Nikola Vaptsarov, \textit{Duelo}, 1940\\
}

%% file: src/abstracts.tex
\chapter*{Resumen}
\addcontentsline{toc}{chapter}{Resumen}
La astronomía exoplanetaria ha entrado en su edad de oro, con la humanidad habiendo detectado más de \num{6000} exoplanetas hasta ahora. La mayor parte de estos descubrimientos se han logrado mediante técnicas de fotometría de tránsito y velocidad radial (VR). Aunque los programas de tránsito lideren en número de descubrimientos, el método de VR sigue siendo una técnica fundamental e indispensable en la ciencia exoplanetaria, especialmente entre la población estelar más prevalente: las enanas rojas. Sus masas bajas las hacen teóricamente favorables para la detección por VR,
ya que a mismo período las señales de exoplanetas son más grandes, y por lo tanto, más fáciles de detectar y medir con precisión. Además, las zonas de habitabilidad de las enanas rojas se encuentran más cerca de las estrellas, lo que hace más sencillo a priori encontrar un planeta potencialmente habitable en torno a estas estrellas. Sin embargo, la actividad estelar que presentan las enanas rojas introduce desafíos significativos, particularmente a la búsqueda de amplitudes de VR del orden de \SI{1}{\metre\per\second} o incluso menores. La actividad estelar se sabe que puede imitar o suprimir señales planetarias y sus efectos se manifiestan en todas las escalas temporales: desde señales breves provocadas por movimientos convectivos (de minutos a días), pasando por regiones activas en la superficie estelar (de días a meses), hasta ciclos magnéticos a largo plazo (de años a décadas). Estos problemas exigen metodologías robustas capaces de desentrañar con fiabilidad la variabilidad estelar de las señales verdaderamente planetarias.

Esta tesis explora estos desafíos mediante el uso de modelado con procesos gaussianos (PP.\,GG.), un marco estadístico no paramétrico que asume una relación funcional entre las covarianzas de las mediciones en función de su diferencia temporal. Presento tres publicaciones distintas, pero interconectadas, que permiten una descomposición robusta de las señales planetarias frente a la actividad estelar. En dos de ellas, se analizan datos espectroscópicos de alta precisión y fotometría mediante un enfoque unificado basado en PP.\,GG. Conjuntamente, estas tres investigaciones ilustran cómo se pueden combinar observaciones de diferentes fuentes con una maquinaria estadística adecuada con el fin de detectar exoplanetas y la posibilidad de entender la actividad estelar.

En el primer estudio, analizo datos espectroscópicos de alta precisión de HARPS-N sobre la enana roja cercana GJ~3998 y revelo al super-Tierra GJ~3998~d mediante un modelado conjunto de señales planetarias de VR, rotación estelar en velocidad radial e indicadores de actividad, así como tendencias a largo plazo en velocidad radial e indicadores de actividad. GJ~3998~d tiene un período orbital de \SI{41.78(0.05)}{\day}, una masa mínima de \mbox{$6.07^{+1.00}_{-0.96}\,\unit\earthmass$} y es una de los pocos planetas conocidos que se acerca a una flujo incidente similar al terrestre \mbox{($1.2^{+0.3}_{-0.2}\,\unit\earthflux$)}. La señal de GJ~3998~d se encontraba muy entrelazada con subproductos de la actividad estelar, lo que me llevó a realizar una serie de pruebas independientes sobre la naturaleza planetaria de GJ~3998~d. Mis esfuerzos permitieron desarrollar un esquema robusto para la detección y prueba de planetas, que he aplicado sistemáticamente a otras estrellas.

En el segundo estudio, aprovecho 26 años de espectroscopía pública de alta precisión de ESPRESSO, HIRES y HARPS sobre la enana roja cercana GJ~526, y modelo su rotación estelar en VR y sus indicadores de actividad, así como sus tendencias a largo plazo en VR e indicadores de actividad, mediante la esquema mencionada. Determino que el período de rotación de la estrella es de \mbox{$P_\text{rot}=\SI{48.7(0.3)}{\day}$} y detecto una señal tenue de tipo sinusoidal en VR, en la anchura a media altura y en el índice S de Mount Wilson. Asigno esta señal a un ciclo magnético de período \mbox{$P_\text{cyc}=1680^{+50}_{-40}\,\unit\day$}. Introduzco trayectorias en el espacio de fase entre VR e indicadores de actividad como un nuevo método para visualizar y comprender la actividad estelar. Estas trayectorias representan la evolución de VR en relación con los indicadores de actividad a lo largo del tiempo, y permiten observar cómo la actividad estelar se desarrolla como una trayectoria continua en un espacio observacional combinado. Muestro que las trayectorias en el espacio de fase revelan detalles más finos en la actividad y ofrecen una visión del desarrollo de regiones activas si dicha región está suficientemente muestreada.

En el tercer y último estudio, investigo la posibilidad de tránsitos emergentes alrededor de la estrella de Barnard en los datos fotométricos de TESS, enfocándonos en el primer y único sector que incluyó observaciones de dicha estrella. El análisis de la curva de luz de \SI{2}{\minute} de cadencia no reveló ninguna variabilidad atribuible a tránsitos de Barnard~b --o a cualquier otro cuerpo en el sistema--. Esta ausencia estableció un límite superior de $3\sigma$ en el ángulo de inclinación orbital de Barnard~b. Este análisis se llevó a cabo en un periodo breve y supuso una revisión rápida y sistemática de observaciones fotométricas.

Los resultados asociados a este programa doctoral demuestran que la actividad estelar no es simplemente otra capa de ruido aleatorio. Es una fuente dinámica, dominante, pero altamente estructurada de variabilidad. Al buscar planetas de amplitud baja, los enfoques tradicionales suelen no capturar la complejidad entera de esta actividad estelar. La aplicación contemporánea, fundamentada en modelos de PP.\,GG, parece ofrecer una estrategia más eficaz, al menos en la etapa actual de la detección de las primeras decenas de planetas con semi-amplitudes inferiores a $\unit{\metre\per\second}$. Aunque los modelos de PP.\,GG puedan perder su relevancia en el futuro, ya hemos obtenido una conclusión valiosa: la actividad estelar no se trata simplemente de ignorarla o corregirla, y ha transformado ya la interpretación de datos de VR en sistemas de enanas frías. Este cambio en la perspectiva se debe reconocer, ya que es esencial para la detección de exoplanetas de masa baja.

\let\cleardoublepage\clearpage
\chapter*{Abstract}
\addcontentsline{toc}{chapter}{Abstract}

Exoplanetary astronomy has entered its Golden Age, with humanity having found over \num{6000} exoplanets to date. The majority of these discoveries have been achieved through transit-photometry and radial-velocity (RV) techniques. While transit surveys lead in discovery numbers, the RV method persists as a vital and indispensable technique in exoplanetary science, particularly in the most prevalent stellar population -- M dwarfs. Their low masses make them the theoretical favourites for RV detections, since exoplanet signals are stronger over the same period, and are therefore easier to detect and measure accurately. Furthermore, the habitable zones of M dwarfs lie closer, which increases the likelihood of finding potentially habitable planets around them. Nevertheless, the stellar activity M dwarfs exhibit introduces significant challenges, particularly when seeking RV semi-amplitudes to the order of \SI{1}{\metre\per\second}, or lower. Stellar activity is known to mimic or suppress planetary signals, and its effects manifest across all time scales: from short-lived signals due to convective motion (lasting minutes to days), through active regions on the stellar surface (days to months), to long-term magnetic cycles (years to decades). These issues highlight the urgent need for robust methodologies that are capable of reliably disentangling stellar variability from genuine planetary signals.

This dissertation explores these challenges through the application of Gaussian-process (GP) modelling, a non-parametric statistical framework that assumes a certain functional relationship of the covariance between measurements with respect to their difference in time. I present three distinct but interconnected publications that integrate high-precision spectroscopy and photometry with a unified GP-based approach in two cases, and that enable robust disentanglement of planetary signals from stellar activity. Together, these three studies illustrate how observations of different origin can be combined with appropriate statistical machinery for the sake of detecting exoplanets, with the possibility of interpreting stellar activity.

In the first study, I analysed high-precision HARPS-N spectroscopy of the nearby early M dwarf GJ~3998, and revealed the super-Earth GJ~3998~d through joint modelling of RV planetary signals, stellar rotation in RV and activity indicators, as well as long-term trends in RV and activity indicators. GJ~3998~d has an orbital period of \SI{41.78(0.05)}{\day}, a minimum mass of \mbox{$6.07^{+1.00}_{-0.96}\,\unit\earthmass$}, and is one of the few known planets that come close to an Earth-like incident flux \mbox{($1.2^{+0.3}_{-0.2}\,\unit\earthflux$)}. The signal of GJ~3998~d appeared very well intertwined with some by-products of stellar activity, which prompted us to carry out a chain of independent tests on the planetary nature of all three planets in the system. Our cumulative efforts yielded a robust planetary-detection and planetary-testing framework, which I have since applied systematically to other stars.

In the second study, I took advantage of \SI{26}{\year} of archival high-precision ESPRESSO, HIRES, and HARPS spectroscopy of the nearby, early M dwarf GJ~526, and modelled its stellar rotation in RV and activity indicators, as well as its long-term trends in RV and activity indicators via the aforementioned framework. I determined the rotational period of the star to be \mbox{$P_\text{rot}=\SI{48.7(0.3)}{\day}$}. In addition, I detected a faint sine-like signal in radial velocity, full width at half maximum, and Mount Wilson S-index. I attributed this signal to a magnetic cycle of period \mbox{$P_\text{cyc}=1680^{+50}_{-40}\,\unit\day$}. I introduced phase-space trajectories between RV and activity indicators as a new method for visualising and understanding stellar activity. These trajectories plot the RV- against the activity-indicators solution over time, and show how activity evolves as a continuous path in a combined observational phase space. I showed that phase-space trajectories reveal finer details in activity, and that they offer a view of the activity-region evolution if the region of interest is sufficiently well-sampled.

In the third and final study, I investigated the possibility of emergent transits around Barnard's star in TESS photometric data, focusing on the first, and to date, only TESS sector that contained observations of Barnard's star. Analysis of the detrended \SI{2}{\minute} light curve revealed no discernible variability attributable to transits of Barnard~b or any other body. This absence established a $3\sigma$ upper limit on the orbital inclination of Barnard~b. This analysis was carried out in a concise timeframe, and demonstrated a rapid, yet systematic examination of photometric observations.

The results associated with this doctoral programme go on to show that stellar activity is not merely another layer of aleatoric uncertainty. It is a dynamic, dominant, yet highly structured source of variability. When searching for low-amplitude planets, traditional approaches seldom capture the full complexity of stellar activity. The current well-informed application of GPs appears to offer a more effective strategy, at least in the current stage of discovering the first few dozen planets with sub-$\unit{\metre\per\second}$ semi-amplitudes. Even if GPs lose their relevance in the future, we have already gained a valuable insight -- stellar activity 
is not simply something to be ignored or corrected for, and it has already transformed how we interpret RV data in cool-dwarf systems. This shift in perspective must be acknowledged, as it is instrumental for detecting low-mass exoplanets.

%% file: src/acknowledgements.tex
\chapter*{Acknowledgements}
\addcontentsline{toc}{chapter}{Acknowledgements}
\begin{verse}[6cm]
\noindent Adapt yourself to the things which your destiny has given you: love those with whom it is your lot to live, and love them with sincere affection.
\end{verse}
\attrib{Marcus Aurelius, \textit{Meditations}, VI.39 (G. W. Chrystal, trans.), circa AD 170--180}
\vspace{7mm}

\lettrine[nindent=0pt,findent=1pt,lraise=-0.0,loversize=-0.21]{\starburst{T}}{his} dissertation comes to me at the age of 25, a time when I am in the tremendous privilege of following my calling in a world that is far from ideal. This manuscript bears witness to my development as an astrophysicist, a process that would have been impossible without many, to whom I owe a debt of gratitude for as long as my legs keep me standing.

Thank you,
Jonay I. González Hernández,
Alejandro Suárez Mascareño,
Nicola Nari, and
Rafael Rebolo López
for the trust you put in me since the very beginning of my doctoral studies, as well as for your unreserved support over the past three years. I am grateful to you for giving me the means to pave my own way, and for passing on a craft I could now call my own. Working with you has been extremely rewarding, and despite at times demanding, it has always flowed as joyfully as a song. You lent me a hand in a place that was initially foreign and unfamiliar to me. Through your warmth and kindness, you helped me grow to love the Canaries as if they were my own motherland.

Thank you, \emph{Fundació ``la Caixa''}, for supporting me with the resources needed to realise this doctoral programme. I would have not reached this point without your vote of confidence. You welcomed me into a community of fellow \emph{becaris}, and it is within this very group that I have found both a sense of belonging and enduring motivation. Thank you,
Étienne Artigau,
Neil James Cook,
René Doyon,
Charles Cadieux,
and others at the \emph{Institut Trottier de recherche sur les exoplanètes} for giving me the opportunity to work with you during my stay in Canada. I saw in you excellent people and colleagues, with whom I look forward to continuing to work in the future.

Thank you all, with whom I walked the same path at the \emph{Instituto de Astrofísica de Canarias}:
João Pedro Verardo Benedetti,
Desmond Helmut Grossmann,
Iveth Adaena Gaspar Gorostieta,
Eduardo Albuquerque Hartmann,
Carlos Martínez-Sebastián,
and the many others who shared this stretch of the way. Your presence made these years richer, and I hold your companionship dear. I also thank my faithful and devoted friends who have been supporting me from abroad:
Aleksandar Ganchev,
Stefan Stefanov,
and others who have been a source of strength for me over the past few years. You are the people I look up to and who motivate me to be a better man.

To my past teachers, from my childhood to my final lectures:
Rositsa Primovska,
Sofia Hariskova,
Maria Robeva,
Zapryana Zapryanova,
Vasil Bazelkov,
the organisers of the Bulgarian Astronomy Olympiad,
Steve Fossey,
Giorgio Savini,
among others.
Thank you.
You are the reason I have come this far, and the better parts of this work owe much to your guidance.

Thank you, mother, for all the care you have given me selflessly since my first breath. You not only taught me to believe in and follow the good in this world, you embody it yourself. Thank you, father, for all you bore so that I could live in comfort. All I have rests on your labour, and if I could become even half the man you are, I would count myself fortunate. I love you both and I owe you more that I could even utter.

Thank you, Lord, for all I have been given. All my wins belong to You, and all my losses are mine alone.\vspace{6mm}

{\hfill Смирено ваш, Атанас{\parfillskip0pt\par}}
{\hfill Your servant, Atanas{\parfillskip0pt\par}}
{\hfill Con cariño, Atanás{\parfillskip0pt\par}}

%% file: src/nomenclature.tex
\nomenclature[A]{RV}{Radial Velocity}
\nomenclature[A]{ADU}{Analogue-to-Digital Units}
\nomenclature[A]{BJD}{Barycentric Julian Date}
\nomenclature[A]{BERV}{Barycentric Earth Radial Velocity}
\nomenclature[A]{HJD}{Heliocentric Julian Date}
\nomenclature[A]{CCF}{Cross-Correlation Function}
\nomenclature[A]{ESPRESSO}{Echelle SPectrograph for Rocky Exoplanets and Stable Spectroscopic Observations}
\nomenclature[A]{FWHM}{Full Width at Half Maximum}
\nomenclature[A]{GP}{Gaussian Process}
\nomenclature[A]{HARPS}{High Accuracy Radial Velocity Planet Searcher}
\nomenclature[A]{HARPS-N}{High Accuracy Radial Velocity Planet Searcher -- North}
\nomenclature[A]{LBL}{Line-By-Line}
\nomenclature[A]{MCMC}{Markov Chain Monte Carlo}
\nomenclature[A]{NIRPS}{Near Infra Red Planet Searcher}
\nomenclature[A]{SAP}{Simple Aperture Photometry}
\nomenclature[A]{SNR}{Signal-to-Noise ratio}
\nomenclature[A]{PDCSAP}{Pre-search Data Conditioning Simple Aperture Photometry}
\nomenclature[A]{SPIRou}{\emph{SPectropolarimètre InfraRouge}\\(Infrared Spectropolarimeter, trans.)}
\nomenclature[A]{SPOC}{Science Processing Operations Center}
\nomenclature[A]{TESS}{Transiting Exoplanet Survey Satellite}
\nomenclature[A]{CARMENES}{Calar Alto high-Resolution search for M dwarfs with Exoearths with Near-infrared and optical Echelle Spectrographs}
\nomenclature[A]{HZ, oHZ, cHZ}{Habitable Zone, optimistic Habitable Zone, conservative Habitable Zone}
\nomenclature[A]{SHO}{Simple Harmonic Oscillator}
\nomenclature[A]{SMA}{Semi-Major axis}
\nomenclature[A]{SEP}{Squared-Exponential Periodic}
\nomenclature[A]{MEP}{Matérn \sfrac{3}{2} Exponential Periodic}
\nomenclature[A]{ESP}{Exponential-Sine Periodic}
\nomenclature[A]{LTF}{Long-Term Function}
\nomenclature[A]{LMA}{Levenberg-Marquadt algorithm}
\nomenclature[A]{WF}{Window Function}
\nomenclature[A]{FAP}{False Alarm Probability}
\nomenclature[A]{FIP}{False Inclusion Probability}
\nomenclature[A]{SERVAL}{SpEctrum Radial Velocity AnaLyser}
\nomenclature[A]{H$\alpha$}{Hydrogen-alpha (line; \SI{6562.8}{\angstrom} in air wavelength)}

\nomenclature[U]{\unit\angstrom}{ångström
\nomunit{\SI{e-10}{\metre}}
}
\nomenclature[U]{\unit\ppt}
{part per thousand
\nomunit{\num{e-3}}
}
\nomenclature[U]{\unit\ppm}
{part per million
\nomunit{\num{e-6}}
}
\nomenclature[U]{\unit\au}
{astronomical unit
\nomunit{\SI[group-digits=true]{149597870700}{\metre}}
}
\nomenclature[U]{\unit\mas}{milliarcsecond
\nomunit{
approx.\footnote{Exactly $(\pi/648\,000\,000)\,\unit\rad$.}
\SI{4.8481e-9}{\rad}}
}
\nomenclature[U]{\unit\parsec}{parsec
\nomunit{
approx.\footnote{Exactly $(648\,000/\pi)\,\unit\au$.}
\SI{3.0857e16}{\metre}}
}
\nomenclature[U]{\unit\mag}{magnitude}
\nomenclature[U]{\unit\solarradius}{
Solar radius
\nomunit{\SI{6.957e8}{\metre}}
}
\nomenclature[U]{\unit\earthradius}{
Earth radius
\nomunit{\SI{6.3781e6}{\metre}}
}
\nomenclature[U]{\unit\jovianradius}{
Jovian radius
\nomunit{\SI{7.1492e7}{\metre}}
}
\nomenclature[U]{\unit\solarmass}{
Solar mass
\nomunit{
approx.\footnote{Defined through $\gamma\unit\solarmass=1.3271244\times 10^{20}\,\unit{\metre\cubed\per\second\squared}$ \citep{Prsa2016}.}
\SI{1.9884e30}{\kilo\gram}}
}
\nomenclature[U]{\unit\earthmass}{
Earth mass
\nomunit{
approx.\footnote{Defined through $\gamma\unit\earthmass=3.986004\times 10^{14}\,\unit{\metre\cubed\per\second\squared}$ \citep{Prsa2016}.
}
\SI{5.9722e24}{\kilo\gram}}
}
\nomenclature[U]{\unit\jovianmass}{
Jovian mass
\nomunit{
approx.\footnote{Defined through $\gamma\unit\jovianmass=1.2668653\times 10^{17}\,\unit{\metre\cubed\per\second\squared}$ \citep{Prsa2016}.}
\SI{1.8981e27}{\kilo\gram}}
}
\nomenclature[U]{\unit\solarluminosity}{
Solar luminosity
\nomunit{
\SI{3.828e26}{\watt}}
}
\nomenclature[U]{\unit\earthflux}{
Earth insolation flux
\nomunit{
\SI{1361}{\watt\per\metre\squared}}
}
\nomenclature[U]{\unit\year}{
year
\nomunit{
assum.\footnote{Taken to be \SI{365.2425}{\day}.}
\SI[group-digits=true]{31556952}{\second}}
}

\nomenclature[C]{$c$}{
speed of light
\nomunit{
\SI[group-digits=true]{299792458}{\metre\per\second}}
}
\nomenclature[C]{$\gamma$}{
gravitational constant
\nomunit{
\SI[group-digits=false]{6.67430e-11}{\metre\cubed\per\kilogram\per\second\squared}}
}

\nomenclature[O]{$\ptp$}{
peak-to-peak
\nomunit{
$\ptp x_i = \max x_i-\min x_i$}
}

\printnomenclature[30mm]

%% file: src/chapters/intro.tex
\chapter{Introduction}\label{sec:introduction}
\begin{verse}[62mm]
\noindent
Moreover, there is an infinite number of worlds, some like this world, others unlike it. For the atoms being infinite in number, as has just been proved, are borne ever further in their course. For the atoms out of which a world might arise, or by which a world might be formed, have not all been expended on one world or a finite number of worlds, whether like or unlike this one. Hence there will be nothing to hinder an infinity of worlds.
\end{verse}
\attrib{Epicurus, \textit{Letter to Herodotus}, IX (R. D. Hicks, trans.), circa 341--270 BCE}
\vspace{7mm}
\lettrine[nindent=4pt,findent=1pt,lraise=-0.0,loversize=-0.21]{\starburst{E}}{ver} since the dawn of time, the night sky has been a quiet, but trusted companion to humans. Across mountains, seas, and deserts, the firmaments have served us as a calendar, a map, but also as a canvas to express our strong religious and cultural convictions. Even then, millennia ago, our forefathers surely wondered whether the night sky was only ours to enjoy, or there were other worlds where foreign eyes marvel at it as much as we do.

The question of the existence of planets outside the Solar System (hereafter exoplanets) have tormented many artistic or intellectual minds. In any case, these philosophical musings ended when \citet{Wolszczan1992} discovered two planets orbiting PSR~B1257+12, a millisecond pulsar about \SI{700}{\parsec} away from the Sun. Three years later, the first confirmation of an exoplanet around a main-sequence star followed (51~Peg~b; \citealp{Mayor1995}), and humans came to understand that exoplanets are not only more common than expected, but also that some were in fact well observable with the instrumentation at the time. Within the next year or so, new discoveries were rapidly made (e.g. \citealp{Marcy1996,Butler1996,Butler1997,Cochran1997}), and so, the new field of exoplanetary science began to grow almost exponentially.

At the time of writing, there are over \num{6300} discovered exoplanets, beyond two thirds of which were found in the past ten years \citep{nasaexoplanettable2}. We find exoplanets far and wide, from the closest nooks in our cosmic vicinity (e.g. Proxima~b, d, \mbox{$d=\SI{1.3}{\parsec}$}; \citealp{Anglada-Escude2016,SuarezMascareno2020,Faria2022,SuarezMascareno2025}) to the farthest reaches of the galaxy (e.g. SWEEPS-4~b and SWEEPS-11~b, \mbox{$d=\SI{8500}{\parsec}$}; \citealp{Sahu2006}). Some exoplanets are enormous in size (e.g. HAT-P-67~b, \mbox{$R=\SI{24}{\earthradius}$}; \citealp{Zhou2017}), while others are minute (e.g. Kepler-37~b, \mbox{$R=\SI{0.31}{\earthradius}$}; \citealp{Barclay2013}). We have been managing to uncover the most extraordinary worlds, such as some with incredible planetary rotation rates (e.g. $\beta$~Pic~b, \mbox{$v\sin i=\SI{19.9(1.0)}{\kilo\metre\per\second}$}; \citealp{Lagrange2009,Landman2024}), extreme instellations (e.g. KELT-9~b, \mbox{$T_\text{eq}=\SI{4600}{\kelvin}$}; \citealp{Gaudi2017}), or at a tender age (e.g. TIDYE-1~b, \SI{3}{\mega\year}; \citealp{Barber2024}). At the same time, we have also learnt that exoplanets often come with company. We know of other stars that host as many as eight planets (e.g. Kepler-90; \citealp{Cabrera2014,Shallue2018}), as well as exoplanets that are shared by more than one star (e.g. Kepler-64~b; \citealp{Schwamb2013}). These are just some examples that reflect the vast diversity of the exoplanets we have discovered until now. This diversity charts a broad area in the parameter space, and it helps us conjecture how exoplanets form and evolve.

Until this day, we have been carefully assembling a colourful bouquet of exoplanets, which we continue to fill with new blooms each month. Nevertheless, we are yet to discover an Earth analogue -- that is, an Earth-like exoplanet orbiting a Sun-like star on a similar orbit.\footnote{In this work, `Earth-like' and `Sun-like' pertain solely to the masses of said bodies. In particular, `Earth-like' shall suggest a mass near \SI{1}{\earthmass}, and no other similarity such as: planetary radius, planetary composition, atmospheric composition, and so on.}
One reason for our lack of success is a natural observational bias against such detections. An Earth analogue would orbit its star with a period of about \SI{1}{\year}. Most detection methods rely on observing multiple orbital repetitions in order to guarantee an unambiguous detection.\footnote{This does not apply to the direct imaging method and gravitational microlensing method, but they come with their own shortcomings. Direct imaging favours young, well-separated, self-luminous planets -- conditions, none of which an Earth analogue satisfies. Gravitational microlensing is sensitive to low-mass planets, but arises from single-instance events. This implies that follow-up spectroscopy is often not possible, which is central to characterisation beyond detection.}
This comes to the benefit of short-period planets, which often can be discovered in a single observing season -- at the time of writing, 67\% of discovered exoplanets have $P_\text{orb}<\SI{30}{\day}$ \citep{nasaexoplanettable2}. For Earth analogues, however, one needs continuous monitoring of the same star that spans several years.

This has been addressed on a photometric level by the Kepler space telescope \citep{kepler}, which continuously observed a \SI{115}{\deg\squared} portion of the sky for four years. The Kepler survey was dedicated to the search of Earth-like planets through the transit method (Sect.~\ref{sec:transit_method}). Indeed, it found some temperate Earth-sized planets, but around faint stars, where follow-up mass measurements through the RV method (Sect.~\ref{sec:rv_method}) proved to be challenging.\footnote{
Currently, there are only two discovered exoplanets with a radius within \SIrange{0.5}{1.5}{\earthradius}, an SMA within \SIrange{0.5}{1.5}{\au}, and orbit a \SIrange{0.5}{1.5}{\solarmass} star: Kepler-62~f \citep{Borucki2013} and Kepler-1126~c \citep{Valizadegan2023}. Both were found through the Kepler facility.}
As far as the RV method is concerned, an Earth analogue would induce an RV sinusoidal with a semi-amplitude of \SI{9}{\centi\metre\per\second} and a period of \SI{1}{\year}. Such RV precisions can be now achieved by ESPRESSO, a high-resolution ultra-stable échelle spectrograph at the ESO Very Large Telescope, Paranal Observatory, Chile \citep{espresso_early,GonzalezHernandez2018,espresso}. ESPRESSO already achieves an RV precision of \SI{25}{\centi\metre\per\second} over the course of a night and \SI{50}{\centi\metre\per\second} over several months. In fact, \citet{espresso} suggested that the instrumental precision of ESPRESSO is \SI{10}{\centi\metre\per\second} if one leaves photon noise and stellar jitter aside. Modern works have begun to utilise data with typical RV uncertainty below \SI{10}{\centi\metre\per\second} (e.g. \citealp{GonzalezHernandez2024,Nari2025,SuarezMascareno2025}). Overall, it appears that we are progressing towards the necessary instrumental performance to detect Earth analogues with the RV method, and that we have got to the point where stellar activity becomes the main impediment (Sect.~\ref{sec:cool_dwarf_activity}).

\section{The transit method}\label{sec:transit_method}
Planetary systems form along diverse spatial orientations. This leads us to assume that they show no particular preference towards a specific alignment within the Galaxy. Let us consider one such system, comprised of a planet of radius $R$ and mass $M$ that orbits a star of radius $R_\star$ and mass $M_\star$. Let the planet complete its orbit over a period $P$. Although the orbital plane is randomly oriented, there is a non-zero chance that a distant observer lies sufficiently near it to witness the planet crossing the stellar disc. The observer would consequently note a periodic dimming of stellar apparent brightness. This phenomenon, occurring at a period $P$, is known as a transit.

Should our observer be drawn to measure the observed stellar flux $\Phi(t)$ during a subsequent transit, they would find that $\Phi(t)$ behaves similarly to a rectangular function. At first, prior to the transit, $\Phi(t)$ stays at a particular level, usually normalised to unity. As soon as as the planet begins to cross the luminous stellar disc, the effective luminous area of the latter diminishes, and $\Phi(t)$ monotonically decreases until the planet enters the disc in its entirety. This stage of the transit is called an ingress.\footnote{From Latin \emph{ingressus}, `an advance, an entrance'.}
While the planet is fully within the disc, $\Phi(t)$ stabilises at a constant $\Phi_\text{transit}<1$, which is often parametrised as \mbox{$\Phi_\text{transit}=1-D$}, where $D$ is known as the transit depth. The observer may relate $D$ to the radii ratio of the star and the planet by virtue of elementary geometry:
\begin{equation}
    D = \frac{R^2}{R_\star^2}.
    \label{eq:transit_depth}
\end{equation}
Of course, the event is time-symmetric: the planet exits the stellar disc gradually, and $\Phi(t)$ consequently rises back to unity. The time-inverse equivalent of an ingress is known as an egress.\footnote{From Latin \emph{egressus}, `an exit'.}
During an egress, the planet gradually leaves the stellar disc, and so, the transit comes to an end.

There are four distinct moments in time that define a transit. They are sometimes referred to as contacts. The first and the second contact define the start and end of an ingress; similarly, the third and fourth contact define those of an egress. If the planet is on a circular orbit with semi-major axis $a$, our observer would measure a time interval between first and fourth contact
\begin{equation}
    T_{14} =
    \frac{PR_\star}{a\pi}
    \sqrt{
    \left(1+\frac{R}{R_\star}\right)^2-b^2
    },
    \label{eq:transit_t14}
\end{equation}
where $b$ is the observed projected distance between the centre of the stellar disc and the planetary-transit chord, in units of $R_\star$. This quantity is known as the impact parameter, and can be related to the orbital inclination of the planet $i_o$ through
\begin{equation}
    b=\frac{a\cos i_\text{o}}{R_\star}.
    \label{eq:transit_impact_parameter}
\end{equation}
$T_{14}$ is often referred to as the total transit duration, and is sometimes published plainly as the `transit duration'. The duration between second and third contact, during which the planet is fully embedded in the stellar disc, also carries an analytical form:
\begin{equation}
    T_{23} =
    \frac{PR_\star}{a\pi}
    \sqrt{
    \left(1-\frac{R}{R_\star}\right)^2-b^2
    }.
    \label{eq:transit_t23}
\end{equation}

All of these equations go on to suggest that the observer can derive many useful quantities from the shape of $\Phi(t)$ alone. For example, a measure of the transit depth $D$ already gives the planet-to-star radii ratio through \eqref{eq:transit_depth}. As another example, one could divide \eqref{eq:transit_t23} by \eqref{eq:transit_t14}, and then substitute in \eqref{eq:transit_depth} to obtain 
\begin{equation}
    \frac{T_{23}}{T_{14}}
    =
    \frac{
    \sqrt{\left(1-\sqrt{D}\right)^2-b^2}
    }{
    \sqrt{\left(1+\sqrt{D}\right)^2-b^2}
    },
\end{equation}
an equation that underlines how we could get to the impact parameter $b$ from the three direct observables $D$, $T_{14}$, and $T_{23}$. Since $\Phi(t)$ is directly related to the overlap of two discs, it can be solved analytically. From this analytic treatment, several other parameters can be derived, including the scaled semi-major axis $a/R_\star$ or the scaled orbital velocity $v/R_\star$ (e.g. \citealp{Mandel2002,Seager2003}).

What we initially presented is an elegant and tractable problem: the transit of a single planet orbiting a featureless, tame star on a circular orbit. In reality, Nature has found countless ways to complicate the matter. What follows is only a brief overview of some of the many circumstances that twist this simple case, together with some detail on their numeric modelling and whether they yield further information about the planet or the star.

\subsection{Stellar-disc heterogeneities}
\subsubsection{Limb darkening}\label{sec:limb_darkening}
Limb darkening is an effect where areas near the centre of the stellar disc appear brighter than their edges (limbs). This is generally because different lines of sight probe different stellar depths. Lines of sight near the centre of the stellar disc probe deeper layers in the star, at depths where the atmosphere is hotter and brighter. Conversely, lines of sight near the limb only skim through the outermost layers of the stellar atmosphere. This tends to result in lower perceived intensity in areas near the limb.

Limb darkening mostly influences the shape of $\Phi(t)$ during ingress and egress, and is usually modelled through parametrisations that reflect the stellar physics of the parent star in some way. Most parametrisations are known as limb-darkening laws, and describe the total emergent intensity $I$ and the angle between the line of sight and the direction of emergent radiation $\vartheta$. Some of the popular limb-darkening laws include:
\begin{align}
    \frac{I(\vartheta)}{I(\vartheta=\SI{0}{\degree})}
    &= 1-a(1-\cos\vartheta) \\
    \frac{I(\vartheta)}{I(\vartheta=\SI{0}{\degree})}
    &= 1-u_1(1-\cos\vartheta)-u_2(1-\cos\vartheta)^2 \label{eq:quadratic_limb_darkening}\\
    \frac{I(\vartheta)}{I(\vartheta=\SI{0}{\degree})}
    &= 1-\sum_{n=1}^4 c_n\left[1-(\cos\vartheta)^{n/2}\right],
\end{align}
and are referred to as the linear law \citep{Schwarzschild1906}, the quadratic law \citep{Kopal1950}, and the four-terms law \citep{Claret2000}, respectively. There are other parametrisations of higher complexity, e.g. a logarithmic law \citep{Klinglesmith1970} and an exponential law \citep{Claret2003}, among others (see \citealp{Espinoza2015} for review). The first few limb-darkening laws have analytic solutions \citep{Mandel2002}, and therefore come with low computational overheads.
Most other limb-darkening laws require numeric integration over the overlap between the stellar and the planetary discs (\citealp{jktebop,pytransit,batman,starry}, among others). In this regard, the quadratic limb-darkening law \eqref{eq:quadratic_limb_darkening} has been a popular choice in modelling, as it is simple enough for analytic treatment, yet flexible enough to fit through most observational data \citep{Kipping2013,Espinoza2016}.

The emissivity in the stellar atmosphere does not necessarily increase with optical depth. In fact, the upper layers of the atmosphere can be more emissive, which reverses the limb-darkening effect, and causes the disc centre to appear darker than its limbs (e.g. \citealp{Schlawin2010}). This effect is known as limb brightening.

\subsubsection{Gravity darkening}

Main-sequence stars tend to undergo a change of internal structure somewhere near \SIrange{6000}{6300}{\kelvin}, a boundary known as the Kraft break \citep{Kraft1967}. Stars hotter than the Kraft break tend to have a convective interior and a radiative envelope, and vice versa.\footnote{There are low-mass main-sequence and pre-main-sequence stars that are fully convective. For main-sequence stars, this appears to be the case for masses below \SI{0.35}{\solarmass} \citep{Chabrier1997}, such as Proxima.}
In convective envelopes, magnetic winds emerge, and decrease the stellar angular momentum of cooler stars over time. This process is sometimes referred to as `magnetic braking', and it makes stars below the Kraft break particularly sensitive to angular-momentum losses. On the other hand, stars above the Kraft break often retain their rotational angular momentum. As a result, they have a wide range of rotational velocities available to them, and can attain projected rotational velocities to the order of
\mbox{$v\sin i_\star\sim\SI{e2}{\kilo\metre\per\second}$} (e.g. \citealp{Zorec2012,Albrecht2022}). Rotational velocities of such magnitude are so large that the centrifugal motion of these stars can cause them to become oblate. This gives rise to a surface-gravity gradient along their photospheres, which translates to an effective-temperature gradient. This suggests that the stellar equator of a fast rotator is characterised by a lower effective temperature $T_\text{eff}$ because it has a smaller surface gravity $g$ due to its oblateness \citep{VonZeipel1924}. That is to say, the equator is darkened relative to the stellar poles. This effect is termed gravity darkening, and has been observed to lead to $T_\text{eff}$ differences that reach nearly \SI{1000}{\kelvin} \citep{Ahlers2020a}.

\begin{figure}[t]
    \centering
    \includegraphics[width=\linewidth]{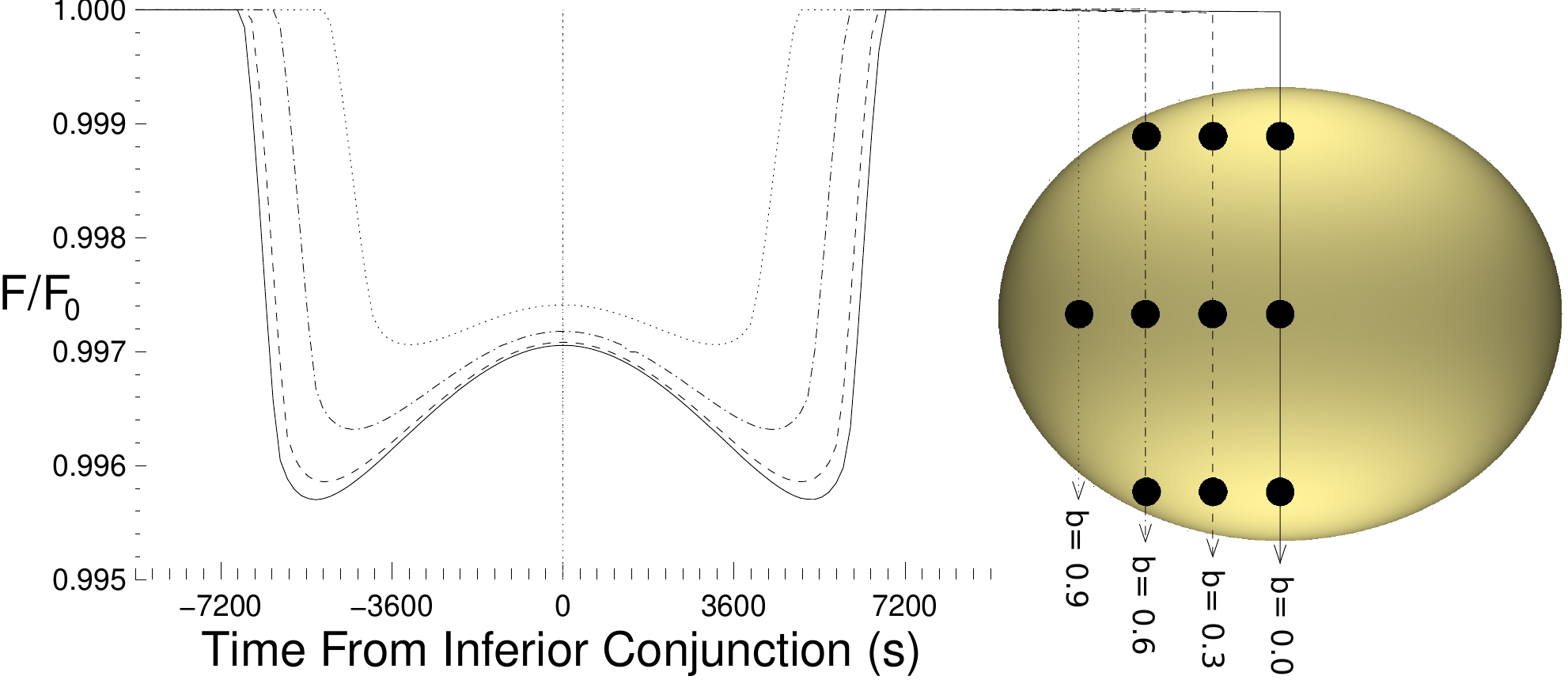}
    \caption{
    Synthetic $\Phi(t)$ for a \SI{1}{\jovianradius} body on a polar \SI{0.05}{\au} orbit around an Altair-like star. The system orientation is described by
    a sky-projected obliquity \mbox{$\lambda=\SI{-90}{\deg}$},
    a stellar inclination of \mbox{$i_\star=\SI{90}{\deg}$}, and
    a series of planetary inclinations near \mbox{$i=\SI{90}{\deg}$}, with impact parameters of
    \mbox{$b=0$} (solid black line),
    \mbox{$b\approx 0.3$} (dashed black line),
    \mbox{$b\approx 0.6$} (dash-dotted black line),
    \mbox{$b\approx 0.9$} (dotted black line).
    This is a significantly misaligned orientation, which results in an unusual, non-rectangular $\Phi(t)$.
    Reproduced from \citet{Barnes2009}, Figure~5, with permission from the lead author and AAS Publishing.
    }
    \label{fig:Barnes2009_fig5}
\end{figure}

If the angular-momentum vectors of stellar rotation and planetary motion happen to coincide, planetary transit chords are guaranteed to have the same stellar latitude $\theta$. In such a case, the planet occults stellar regions of the same $g$, and therefore of the same $T_\text{eff}$ -- making the light curve indistinguishable from ordinary. However, in cases where there is a misalignment between the two vectors (hereafter spin-orbit misalignment), the planet occults parts of the stellar disc that are subject to variable amounts of gravity darkening. In those cases, gravity darkening perturbs the light curve (see \citealp{Barnes2009} for review). A good example is a planet on a polar orbit (Figure~\ref{fig:Barnes2009_fig5}). The planet firstly obscures the hot, bright stellar pole. It then perpendicularly crosses the stellar equator, which is subject to the largest degree of gravity darkening. Shadowing a darker region removes less from the total flux, which the observer would note as a bump in the centre of the light curve. Such polar orbits have indeed been observed, where gravity darkening plays a significant role -- though typically secondary to limb darkening -- and must be 
modelled in tandem (e.g. \citealp{Ahlers2020a,Ahlers2020,Deline2022}).

The first systems with measured spin-orbit orientations were found to be aligned (e.g. \citealp{Winn2006}), but later, planets on polar (e.g. \citealp{Hebrard2008}) and even retrograde orbits (e.g. \citealp{Anderson2010}) were discovered.
To comment on spin-orbit (mis)alignment, however, one needs to measure the stellar inclination of the star $i_\star$. This parameter cannot be probed photometrically unless there is a stellar-surface heterogeneity that the observer can exploit. In this light, gravity-darkened stars offer a very favourable opportunity.

\subsubsection{Starspots}\label{sec:starspots}

\begin{figure}
    \centering
    \includegraphics[width=\linewidth]{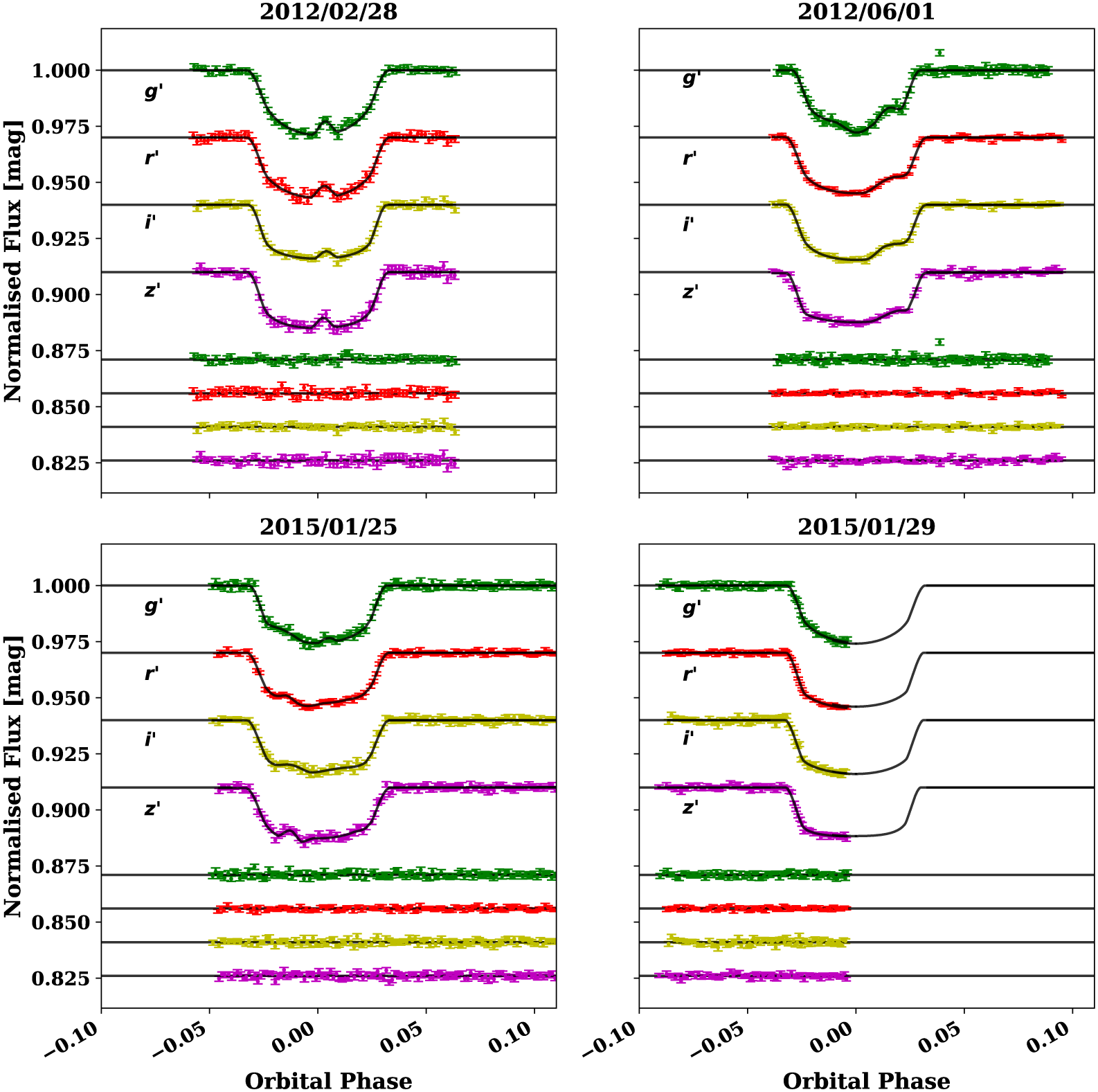}
    \caption{
    Observed $\Phi(t)$ during transits of four HATS-2~b transits, simultaneously observed in four optical bands: Sloan g', r', i', and z'. We observe starspot crossings in the first three datasets, which manifest as abrupt bumps in the transit curve. Observational data comes from \citet{Mohler-Fischer2013} and \citet{Biagiotti2024}. Modelled $\Phi(t)$ are given in solid black lines, and the residual time series after the fit are shown under each transit panel.
    Reproduced from \citet{Biagiotti2024}, Figure~2, under a CC-BY 4.0 licence.
    }
    \label{fig:Biagiotti2024_fig2}
\end{figure}

One of the signs of stellar activity is the presence of starspots, which are dark features in the stellar photosphere. If a planet happens to transit across a starspot, the former would occult a less luminous region than average, as the starspot temperature is lower than its surroundings. This reduces the amount of flux taken away by the planet, which results in a small bump in the transit light curve (Fig.~\ref{fig:Biagiotti2024_fig2}), sometimes referred to as a `spot crossing'.

Spot crossings can be used to infer the stellar inclination $i_\star$, and thereby, the spin-orbit misalignment in a similar manner to gravity darkening. It bears mentioning that both methods aid the study of stellar physics and misalignment on the same principle -- rather than viewing transits as simply removing light from $\Phi(x)$, they inform of the stellar intensity that lies behind the transit chord.\footnote{
There is an intuitive real-life analogy to this: a letterpress roller passing over a page. The roller picks up varying amounts of ink as it rolls, according to what is being printed. Using $\Phi(t)$ to interpret the stellar-disc structure behind the planetary chord is akin to using the amount of lost ink over time to infer what was being printed.
}
Despite this, measuring $i_\star$ through spot crossings is considerably more challenging. Unlike gravity darkening, where each stellar latitude is assumed to have its own local, invariant temperature, in this case we have to rely on sporadic spot crossings, which in fact may not necessarily correspond to the same spot at all. This makes $i_\star$ solutions strongly degenerate if an insufficient number of transits are observed. Yet, it has been demonstrated that long-term analyses of spot crossings can indirectly constrain $i_\star$ (e.g. \citealp{Desert2011,Sanchis-Ojeda2011}).

\subsection{Eccentric orbits}
The initial posing of the problem assumed that the planet lies on a circular orbit, and this may not necessarily be the case. In their definition, eccentric orbits come with additional parameters beyond the semi-major axis $a$, such as the eccentricity $e$ and the argument of periastron $\omega$.\footnote{The longitude of the ascending node $\Omega$ is also required for a complete description of an eccentric orbit. However, it cannot be probed by $\Phi(t)$ alone, and is beyond our scope.} 
Moreover, in eccentric orbits, planets no longer orbit with a constant velocity $v$. Instead, their velocity is governed by
\begin{equation}
    v(r) = \sqrt{
    \gamma M_\star \left(
    \frac{2}{r}-\frac{1}{a}
    \right)
    },
    \label{eq:vis_viva}
\end{equation}
where $\gamma$ is the gravitational constant, and $r$ is the instantaneous distance between the planet and the star. This result is known as the vis-viva equation,
and shows that the planet moves faster near its periastron and slower near its apoastron. That is to say, depending on the orientation of the elliptic orbit, the observer may measure shorter or longer transit durations, irrespective of the transit chord and its impact parameter $b$.

Within an eccentric-orbit context, the impact-parameter equation \eqref{eq:transit_impact_parameter} can be generalised to
\begin{equation}
    b = \frac{a\cos i_\text{o}}{R_\star}
    \frac{1-e^2}{1+e\sin\omega},
    \label{eq:eccentric_transit_impact_parameter}
\end{equation}
and the transit-duration equations \eqref{eq:transit_t14} and \eqref{eq:transit_t23} can be likewise expressed as
\begin{align}
    T_{14} &=
    \frac{PR_\star}{a\pi}
    \frac{\sqrt{1-e^2}}{1+e\sin\omega}
    \sqrt{
    \left(1+\frac{R}{R_\star}\right)^2-b^2
    }
    \label{eq:eccentric_transit_t14} \\[1ex]
    T_{23} &=
    \frac{PR_\star}{a\pi}
    \frac{\sqrt{1-e^2}}{1+e\sin\omega}
    \sqrt{
    \left(1-\frac{R}{R_\star}\right)^2-b^2,
    }
    \label{eq:eccentric_transit_t23}
\end{align}
following \citet{Winn2010a}. That is to say, for non-circular orbits, we can think of $R_\star$ to be effectively scaled with $e$ and $\sin\omega$. We remark that in \eqref{eq:eccentric_transit_impact_parameter}, \eqref{eq:eccentric_transit_t14}, and \eqref{eq:eccentric_transit_t23}, $\omega$ always appears in a product with $e$. This suggests that $e$ and $\omega$ cannot be separated from transit information alone. We can use transits together with secondary eclipses,\footnote{Secondary eclipses occur when an exoplanet passes behind its star as seen from Earth, and are sensitive to $e\cos\omega$. Their discussion is beyond our scope; we refer to \citet{Perryman2018} for a more detailed account.} or radial-velocity curve fitting to retrieve $e$ and $\omega$ (Sect.~\ref{sec:rv_method}).

Beyond its effect on transit durations, \eqref{eq:vis_viva} alludes to a more subtle effect: $v(r)$ may change ever so slightly between ingress and egress, which could lead to an asymmetric curve. If the planet transits past its periastron but before its apoastron, the time between first and second contact will be shorter than the time between third and fourth contact -- and vice versa \citep{Barnes2007,Kipping2008}. \citet{Ford2008} showed that this difference between ingress- and egress duration can be approximated to
\begin{equation}
\begin{split}
    T_{12}-T_{34} \approx~&\SI{4}{\second}
    \left(\frac{P}{\unit\year}\right)^{-1/3}
    \left(\frac{R}{\unit\earthradius}\right)
    \left(\frac{R_\star}{\unit\solarradius}\right)
    \left(\frac{M_\star}{\unit\solarmass}\right)^{-2/3} \\[1ex]
    &\times\left[
    \frac{(1+\mu)^{-2/3}e\sin\Psi}{(1-b^2)(1+e\cos\Psi)(1-e^2)^{3/2}}
    \right],
\end{split}
\end{equation}
where $\mu=M/M_\star$ is the planet-to-star mass ratio, and $\Psi$ is the true anomaly at the time of mid-transit. The authors evaluate this difference to be near \SI{3.4}{\minute} for a Jupiter-like planet on a \SI{4}{\day} orbit around a Sun-like star, and discuss that this asymmetry tends to exert a rather weak constraint on $e$.

\subsection{Influences by other bodies}
We finalise our transit-method review by focusing less on the complexities that rise in modelling and more on the indirect detection methods that are largely driven by transits, such as transit-timing- and transit-duration variations (TTVs and TDVs). These methods are often treated as distinct and separate techniques. However, in the interest of this dissertation and given that they remain artisanal and yet to be fully understood, we have grouped them under a single umbrella. We now touch upon their fundamental principles and assess their current fruitfulness.

\subsubsection{Transit-timing variations (TTVs)}\label{sec:ttvs}
\begin{figure}[t]
    \centering
    \includegraphics[width=0.8\linewidth]{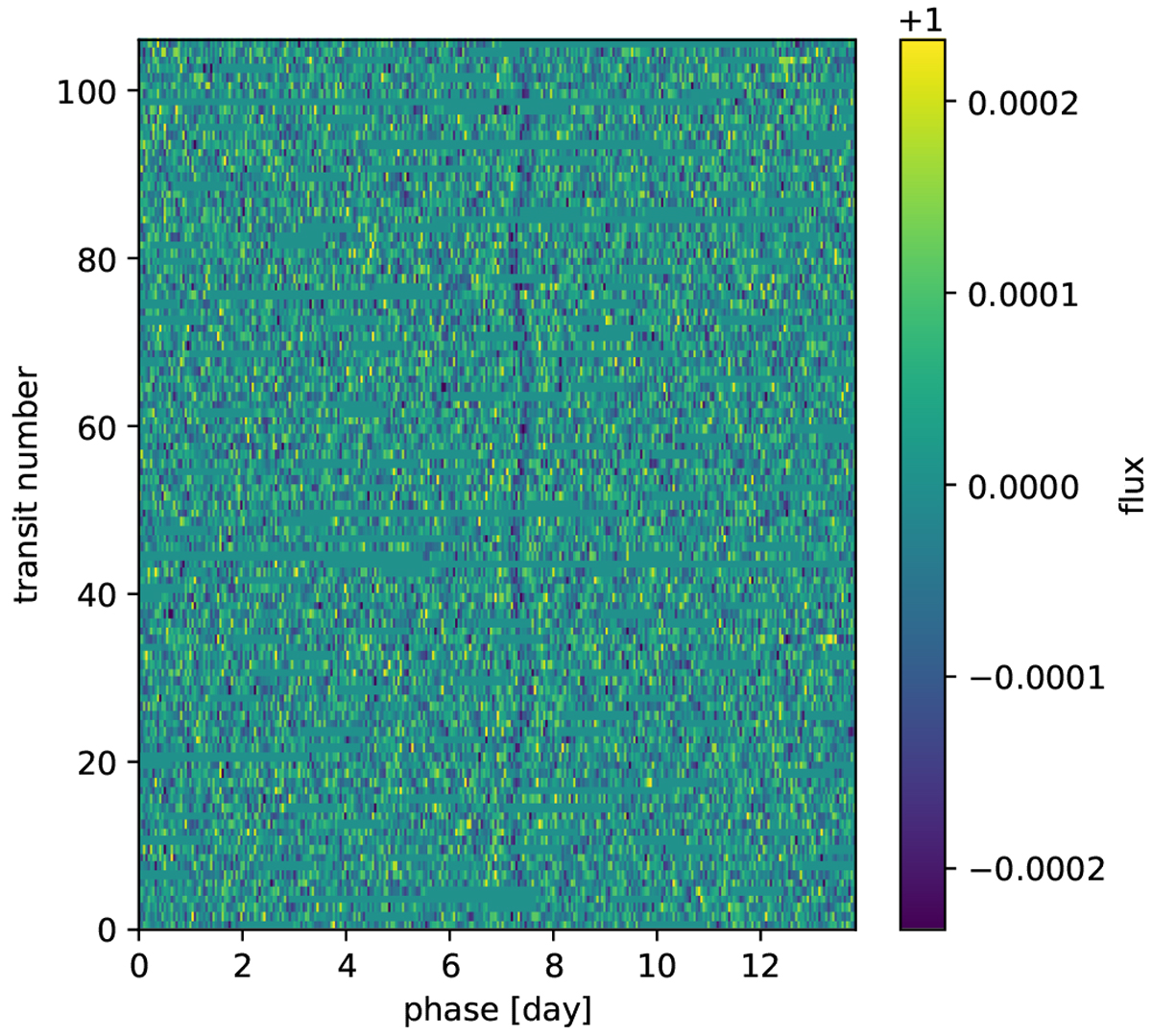}
    \caption{
    Visualisation of Kepler-36~b TTVs through a riverplot. Each row depicts one $\Phi(t)$ portion, folded at an average transit period $P$. Rows are arranged chronologically from the earliest (bottom) to the most recent transits (top). Columns indicate the observation time modulo $P$ (i.e. non-normalised phase). Coloured pixels show relative flux deviations from $\Phi(t)$ (blue to yellow).
    Reproduced from \citet{Leleu2021a}, Figure~2, under a CC-BY 4.0 licence.
    }
    \label{fig:Leleu2021a_Fig2}
\end{figure}

The transit ephemerides $t_i$ of a lone planet around a star are predictable and determined solely by the orbital period $P$:
\begin{equation}
    t_i=t_0+nP,
    \label{eq:linear_ephemerides}
\end{equation}
where \mbox{$t_0$} is the time of any mid-transit and \mbox{$n\in\mathbb{Z}$}. However, as soon as a third body is present in the system, the planetary orbit is no longer Keplerian, and $t_i$ begin to deviate from their linear solution in \eqref{eq:linear_ephemerides}. These deviations are termed transit-timing variations (TTVs). While all multi-planetary systems in theory produce TTVs, the vast majority do so in a way that generates minute effects, much smaller than any instrumental precision. TTVs can only be instrumentally detected under specific configuration conditions \citep{Agol2005,Holman2005}. Figure~\ref{fig:Leleu2021a_Fig2} demonstrates this concept through what is referred to as a riverplot -- an ordered array that arranges continuous $\Phi(t)$ photometry in rows, depending on $n$ in \eqref{eq:linear_ephemerides}; and in columns, depending on the instantaneous time of the measurement, modulo $P$. Before us stand 106 transits of Kepler-36~b \citep{Carter2012}, which are represented by the congregation of dark pixels of phases near \SI{7}{\day} in Fig.~\ref{fig:Leleu2021a_Fig2}. If these transits were perfectly periodic, they would align to form a uniform blue vertical stripe. Instead, we observe a river-like oscillatory feature, which is a clear signature of TTVs.

The planetary systems that have been identified to produce substantial TTVs can be broadly classified in four distinct configurations \citep{Agol2005,Nesvorny2014}:
(i)~an inner component with a significantly shorter $P_1$;
(ii)~a distant, eccentric outer component with a large $P_2$;
(iii)~two transiting stars on circular orbit and a ratio $P_1/P_2$ that is arbitrary but not in resonance; and
(iv)~planets initially on circular orbits that become locked into a mean-motion resonance (see \citealp{Perryman2018} and discussion therein). At the time of writing, there are 40 planets discovered through TTVs, and research in this area is still undergoing development.

TTVs typically cannot be described in an analytical way. There exist approximations that can be applied to planets close to a first-order mean-motion resonance -- however, there is a known degeneracy between their masses and their eccentricities \citep{Lithwick2012}. TTVs tend to be addressed via two standard approaches. The first involves fitting a series of transits on observational data using fixed transit parameters but allowing their mid-times $t_i$ to vary freely (e.g. \citealp{Nesvorny2012}). This yields a TTV time series that can be subsequently analysed through closed-form models or N-body simulations. This method fundamentally treats photometry and orbital dynamics as two decoupled steps, but is computationally faster, and \citet{Nesvorny2014a} argued that it offers greater understanding of how various dynamical parameters influence the observed TTVs. The second approach, known as photo-dynamical modelling, simultaneously integrates the system with an N-body model, generates the full light curve, and fits a model $\Phi(t)$ onto photometric data (e.g. \citealp{Carter2012}). Photo-dynamical modelling naturally captures the evolution of orbital parameters and other phenomena such as transit-duration variations (TDVs; Sect.~\ref{sec:tdvs}), offers a more physically informed treatment, and requires fewer free parameters. However, it comes at the cost of significant computational expenses.

\subsubsection{Transit-duration variations (TDVs)}\label{sec:tdvs}
Not only may transits vary in timing, but they may also exhibit changes in duration. These variations, known as transit-duration variations (TDVs), may stem from the gradual shift in the orbit orientation, such as apsidal or nodal precession (e.g. \citealp{Damiani2011}). Alternatively, TDVs could come from the gravitational influence of other planets in the system (e.g. \citealp{Becker2016,Boley2020}) or even exomoons \citep{Kipping2009,Kipping2009a,Martin2019}.
So far, we have found only a few systems adorned with the crown of TDVs. These include:
TrES-2 \citep{Mislis2009},
Kepler-13 \citep{Szabo2012,Mazeh2013},
Kepler-88 \citep{Nesvorny2013},
Kepler-108 \citep{Mills2017}, and
KOI-134~b \citep{Nabbie2025}.
TDVs complement TTVs by probing a different aspect of the orbital geometry of the system. While TTVs are sensitive to planetary masses and eccentricities, these two parameters tend to be degenerate with one another. On the other hand, TDVs are very sensitive to orbital-inclination variations and nodal precession. This allows a clearer separation between mass and eccentricity, and provides a more complete picture of the true three-dimensional orbital structure we find in a system.

\subsection{Observational and instrumental effects}
\subsubsection{Finite exposure time}
So far, we have introduced analytic expressions of the observed normalised flux $\Phi(t)$ at a particular instantaneous time $t$. Observations, however, rely on the integration of light for a finite interval during which the camera shutter is open. As such, what we measure is not instantaneous $\Phi(t)$, but
\begin{equation}
    \Phi_\text{obs}(t) = \int^{t+E/2}_{t-E/2}\Phi(t)\dd t,
    \label{eq:flux_finite_exposure}
\end{equation}
where $E$ is the finite exposure time. Typically, $E$ ranges from \SIrange{e0}{e3}{\second}, with the longer end of 
this range predominantly occupied by observations from space-based surveys.

Such high values of $E$ may cause substantial smearing of $\Phi(t)$, and effectively attenuate high-frequency features such as the events of ingress and egress. \citet{Kipping2010a} has demonstrated analytically that long exposures could introduce significant systematic errors in the derived physical parameters. This work demonstrates that finite exposure times artificially elongate the ingress and egress phases and reduce the observable limb darkening during transits. These distortions individually cause an overestimation of the impact parameter $b$, and may underestimate stellar densities, which runs into the risk of rejecting valid planetary candidates on account of unphysical parameters.

There appear to be no established analytic treatments of finite exposure times at the present moment. The most common and widely adopted approach seems to be to oversample $\Phi$ at times with temporal difference $\Delta t\ll E$, and then to numerically integrate \eqref{eq:flux_finite_exposure}. We refer the reader to \citet{Kipping2010a} for a more detailed account of this approach and its comparison to other numerical techniques.

\subsubsection{Instrumental uncertainty}\label{sec:instrument_uncertainty}

All measurements are fundamentally constrained by the capabilities and the characteristics of the used instrumentarium. Every measurement -- whether photometric or 
otherwise -- has a finite level of precision. Photometric measurements in particular are 
affected by various instrumental factors such as detector noise, readout errors, 
pixel-to-pixel non-uniformities, and instrumental drift, among others. These circumstances introduce uncertainties that propagate through analysis and eventually make their way to inferred physical parameters.

Sources of uncertainty can be broadly divided into two groups: aleatoric\footnote{From Latin \emph{alea}, `dice'.} and systematic. Aleatoric uncertainties are inherently probabilistic and can be reduced through longer exposures, repeated observations, or any other process that increases or extracts more information. An example of an aleatoric uncertainty is the one associated with photon noise. Systematic uncertainties, by contrast, stem from intrinsic biases in instruments, from implicit assumptions in our own analysis, or from correlated noise that cannot be eliminated trivially by any form of averaging. As such, they introduce consistent deviations, which, if uncorrected, question the validity of our results and conclusions.

Measurement theory warrants that observational data are interpreted within a framework that acknowledges the presence of uncertainty. Generally, uncertainty should not be regarded as an indication of failure; rather, it is an intrinsic and unavoidable component of the measurement process. We should be striving to reduce the systematic uncertainty, for it is the only component we could control to an extent. At the same time, any scientific inference of ours must be accompanied by a clear measure of uncertainty. The distinction between reporting a planetary radius as \SI{1(0.05)}{\earthradius} or \SI{1(0.5)}{\earthradius} is significant -- just as there is a  difference between stating that the author of this dissertation is two metres tall to the nearest centimetre, or to the nearest metre.

\section{The RV method}\label{sec:rv_method}

It is commonly believed that planets orbit their stars in a simple, isolated manner. In fact, all bodies in a planetary system orbit their common centre of mass (hereafter barycentre). This includes the star itself, which is affected by the gravitational interaction from its planets. The presence of planets causes periodic perturbations of another direct observable property of the star -- not its apparent brightness, but its motion along the line of sight.

\begin{figure}[t]
    \centering
    \includegraphics[width=3.5in]{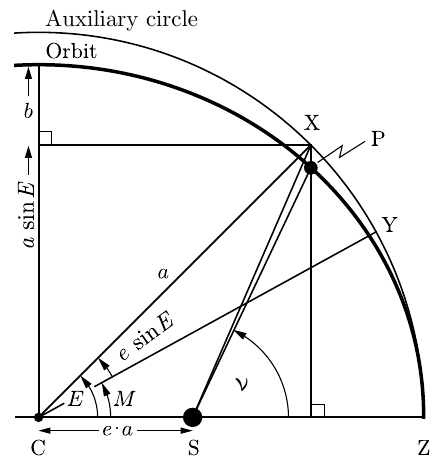}
    \caption{The orbital parameters of Keplerian orbits in its two-dimensional plane. Adapted from AndrewBuck and René Schwarz via Wikimedia Commons, under a CC BY-SA 3.0 licence. This derivation includes English labels, and replaces $T$ with $\nu$.}
    \label{fig:2d_keplerian}
\end{figure}

Let us consider again a system that is comprised of a planet of mass $M$ that orbits a star of mass $M_\star$, in a similar fashion to Sect.~\ref{sec:transit_method}. Under Newton's law of universal gravitation, both the planet and the star move in closed elliptical orbits, with the barycentre located at one focus of each ellipse. These orbits are described by
\begin{equation}
    \frac{x^2}{a^2}+\frac{y^2}{b^2} = 1,
    \label{eq:ellipse_cartesian}
\end{equation}
where $x$ and $y$ define the Cartesian coordinates of the orbital plane. The parameters $a$ and $b$ determine the size and shape of the ellipse, and are known as the semi-major and the semi-minor axis, respectively. The time-variant distance between the orbiting body and the barycentre is given by
\begin{equation}
    r(t)=\frac{a\left(1-e^2\right)}{1+e\cos\nu(t)},
\end{equation}
where the eccentricity $e$ is the fundamental parameter that characterises the shape of the orbit:
\begin{equation}
    e=\sqrt{1-\frac{b^2}{a^2}},
    \label{eq:eccentricity}
\end{equation}
and the true anomaly $\nu(t)$ is the angle between the direction of the closest point on the ellipse to the barycentre (the periapsis) and the current position of the body, measured from the barycentre. This is rather well illustrated in Figure~\ref{fig:2d_keplerian}.
This figure shows one quadrant of an elliptical orbit, with its centre $C$, the barycentre $S$, and the periapsis $Z$. The instantaneous position of the orbiting body is $P$. Thus, the true anomaly $\nu(t)$ is the angle $\angle ZSP$.

Two quantities complement the true anomaly $\nu(t)$: the eccentric anomaly $E(t)$ and the mean anomaly $M(t)$. They are defined by
\begin{align}
    \cos\nu(t)&=\frac{\cos E(t)-e}{1-e\cos E(t)}
    \label{eq:eccentric_anomaly} \\[1ex]
    M(t)&=\frac{2\pi}{P}\left(t-t_p\right),
\end{align}
where $t_p$ is the time after periapsis passage. Geometrically, $E(t)$ is measured from the ellipse centre and determines the position of the orbiting body via a projection of the orbit onto an auxiliary circle of radius \mbox{$r=a$}. A perpendicular from the major axis $CZ$ to $P$ intersects the auxiliary circle at point $X$. The angle $\angle ZCX$ corresponds exactly to $E(t)$. The mean anomaly $M(t)$ is a hypothetical quantity that increases linearly with time. It represents the fraction of the orbital period that has elapsed since periapsis if the body were moving at a constant, mean motion. After passing through periapsis $Z$, the radius vector of the orbiting body has swept out a certain area within the closed region $[ZSP]$. There exists a auxiliary-circle sector with the same area, $[ZCY]$. The angle $\angle ZCY$ is precisely $M(t)$.

All of these orbital parameters describe the elliptical orbit in its own two-dimensional plane. This orbit can be arbitrarily oriented in three-dimensional space, much as was discussed in Sect.~\ref{sec:transit_method}.
\begin{figure}[t]
    \centering
    \includegraphics[width=\linewidth]{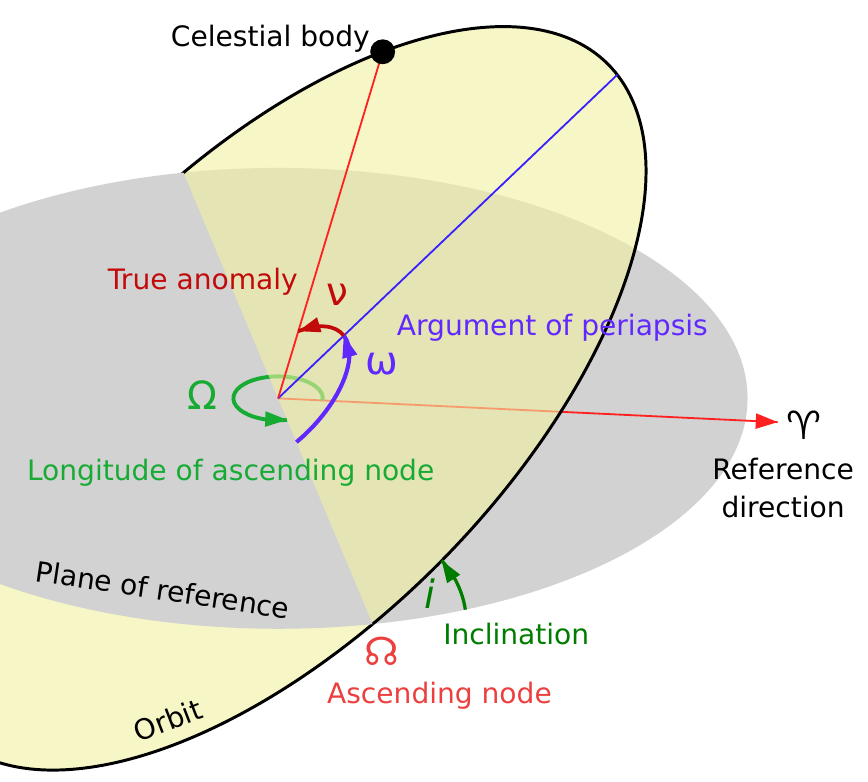}
    \caption{The orbital parameters of Keplerian orbits in three dimensions, relative to an arbitrary plane of reference. Note that our definition of orbital inclination $i_\text{o}$ in Sect.~\ref{sec:transit_method} matches $i$ if the plane of reference is the sky plane. Adapted from Lasunncty, via Wikimedia Commons, under a CC BY-SA 3.0 licence.}
    \label{fig:3d_keplerian}
\end{figure}
Figure~\ref{fig:3d_keplerian} aims to visualise the additional parameters that come with the three-dimensional orientation. The orbit is inclined at some angle relative to an arbitrary reference plane. Let this be the sky plane, i.e. the plane perpendicular to the line of the sight. Let us situate our observer in Fig.~\ref{fig:3d_keplerian} so that they observe the orbit from atop. For the observer, the orbit appears inclined exactly at the orbital inclination $i_\text{o}$, as defined earlier in \eqref{eq:transit_impact_parameter}. The orbit intersects the sky plane in two points: the ascending and the descending node. An additional parameter, termed the argument of periapsis $\omega$, determines the orientation of ellipse within its plane, by measuring the angle from the ascending node to the periapsis.

The introduction of $\omega$ allows us to write a complete expression for the observed radial velocity of the star:
\begin{equation}
    v_\text{r}(t)=
    \frac{2\pi}{P}\frac{a_\star\sin i_\text{o}}{\sqrt{1-e^2}}
    \left\{
    \cos\left[\omega+\nu(t)\right]+
    e\cos\omega
    \right\},
    \label{eq:rv_formula}
\end{equation}
where $a_\star$ is the semi-major axis of the star in its orbit around the barycentre. In practice, researchers fit the simpler function
\begin{equation}
    v(t)=
    k_\text{rv}
    \left\{
    \cos\left[\omega+\nu(t)\right]+
    e\cos\omega
    \right\},
    \label{eq:rv_formula_simple}
\end{equation}
where $k_\text{rv}$ is the RV semi-amplitude. The RV semi-amplitude encodes not only the mathematical orientation of the ellipse, but also some key physical parameters of the system. \citet{Perryman2018} remarked that for a system of a lone planet and a lone star, Kepler's third law can be substituted in \eqref{eq:rv_formula} to our implicit definition of $k_\text{rv}$ to give
\begin{equation}
    k_\text{rv} = \sqrt{
    \frac{\gamma M^3}{(M_\star+M)^2}
    \frac{\sin^2i_\text{o}}{a_\star(1-e^2)}.
    }
\end{equation}
In this form, it becomes evident that $k_\text{rv}$ reflects the planet-to-star mass ratio $M/M_\star$, particularly when $M\ll M_\star$. If the stellar mass $M_\star$ is known, then the planetary mass $M$ can be determined up to a factor of $\sin i_\text{o}$. Indeed, for any observed modulation of the RV signal, we remain fundamentally uncertain whether it comes from a small, nearly-transiting planet \mbox{($\sin i_\text{o}\to 1$)}; or from a massive planet at a nearly-perpendicular orbit \mbox{($\sin i_\text{o}\to 0$)}. This is a core limitation of the RV method. While alternative techniques can resolve this (e.g. transits, TTVs, or astrometry), they are not always feasible. As a result, researchers typically report $M\sin i_\text{o}$, while acknowledging that the true mass of the planet cannot be determined without additional information.

There is another subtlety in \eqref{eq:rv_formula_simple} that bears attention. As shown in the geometric interpretation of Fig.~\ref{fig:2d_keplerian}, $M(t)$ is the only anomaly that is linear in time -- while the true anomaly $\nu(t)$, which appears in \eqref{eq:rv_formula_simple}, is not. To obtain $\nu(t)$, we use a convenient result from orbital mechanics:
\begin{equation}
    M(t)=E(t)-e\sin E(t),
    \label{eq:kepler_equation}
\end{equation}
termed Kepler's equation. In this way, we can go from the linear $M(t)$ to $E(t)$ through \eqref{eq:kepler_equation}, and then from $E(t)$ to the desired $\nu(t)$ through \eqref{eq:eccentric_anomaly}. Kepler's equation is transcendental, and must be solved numerically. This is typically done through iterative methods, with the Newton-Raphson method being among the most commonly employed. Nevertheless, the performance and the convergence of such methods can be very sensitive to the initial guess. This issue has been long recognised in the literature, and is thoroughly discussed in \citet{Odell1986}. We recommend reading their full account of the problem. The works associated with this dissertation utilised the EKEPL1 procedure by the same authors, which typically achieves precision in $E(t)$ better than \SI{e-12}{\rad} after just five iterations.

And so, we have derived the usual recipe that models $v_\text{r}(t)$ through five parameters: $k_\text{rv}$, $e$, $\omega$, the orbital period $P$, and the normalised phase $\varphi$. Each of these parameters has its own unique role:
$k_\text{rv}$ determines the extent of modulation of $v_\text{r}(t)$ and points to the projected planetary mass; $e$ and $\omega$ determine the non-circular `complexity' in the functional shape of $v_\text{r}(t)$; the orbital period $P$ scales $v_\text{r}(t)$ in the time dimension, somewhat similar to $k_\text{rv}$; and $\varphi$ helps to determine the location of the orbiting planet for a given ephemeris.

In the works that comprise this dissertation, $\varphi$ is defined such that the most recent measurement has a normalised phase $\varphi$ relative to the last inferior conjunction.\footnote{The definition of the last inferior conjunction follows the definition of \citet{Kane2009} under the name `midpoint of planetary transit'.} As a result, inferior-conjunction ephemerides are defined through
\begin{equation}
    \varepsilon = (n-\varphi)P,
    \label{eq:ephemerides}
\end{equation}
where \mbox{$n\in\mathbb{Z}$}. This approach entails several modelling decisions, which we will now briefly address. First, $\varphi$ is defined with respect to the most recent measurement. While any point of reference within the time series can be used, the most recent measurement minimises the time interval $\Delta t$ to a hypothetical future transit follow-up. A smaller $\Delta t$ reduces the propagation error into the derived $\varepsilon$, and would be therefore more helpful for teams that attempt to detect transits. Second, $\varphi$ is defined relative to inferior conjunctions, and not to periapsis passages, which are more commonly used in orbital modelling. This is also for the sake of convenience of potential transit follow-ups. Finally, instead of directly fitting $\varepsilon$, we sample for a normalised phase. This approach avoids expected degeneracies between $P$ and $\varepsilon$ for large periods, as well as multiple solutions in the $\varepsilon$ prior for small periods. Despite these advantages, the definition of $\varphi$ is based on $P$. If $P$ is poorly constrained or is very large, uncertainties in $P$ can affect the precision of the derived $\varphi$.

For circular planets, the eccentricity vanishes: \mbox{$e=0$}. In this case, the argument of periapsis $\omega$ becomes irrelevant, and all anomalies equal the mean anomaly $M(t)$. \eqref{eq:rv_formula_simple} then simplifies to
\begin{equation}
    v_\text{r}(t)=k_\text{rv}\sin\left[
    \frac{2\pi(t+P\varphi)}{P}
    \right].
    \label{eq:rv_formula_simple_circle}
\end{equation}
\citet{Stefanov2025a} referred to this form as the planetary-sine approximation, or the circular-orbit approximation. This form is often favoured over the full expression in \eqref{eq:rv_formula_simple}, because it involves only three free parameters and does not require solving Kepler's equation. As such, it is much faster to compute. Note that the $P\varphi$ term has undergone a sign change compared to \eqref{eq:ephemerides}. This maintains the consistency in the physical interpretation of $\varphi$.

The planetary-sine approximation is an appropriate first step in uninformed searches of planets around late dwarfs. Most planets on close-in orbits tend to be circularised due to tidal interactions between the planet and the host star, as well as their energy dissipation within the planetary interior (e.g. \citealp{Rasio1996,Marcy1997}). In addition, given the frequent sparsity and the non-uniformity of spectroscopic observations, fitting for additional parameters such as eccentricity is often not justified. Moreover, eccentricity is defined in a non-intuitive way. From \eqref{eq:eccentricity}, an eccentricity of $e=0.1$ implies the ratio \mbox{$b/a\approx 0.995$}, giving an RV curve that is barely distinguishable from a circular orbit. We should keep in mind that uninformed searches of planets prioritise detection. At first, the focus is on identifying clear, periodic signals. Eccentricity can be introduced in a more refined analysis after a signal has been detected, or following the acquisition of additional data.

The RV method has long been a rival of the transit method in terms of discovery yield. It was the first to produce a detection of a planet around a main-sequence star (51~Peg~b; \citealp{Mayor1995}), and in fact, over 70 more RV detections of exoplanets were made before the first planet discovered by the transit method (OGLE-TR-56~b; \citealp{Udalski2002,Udalski2002a,Konacki2003}). Unlike the transit method, which requires a precise alignment for a planet to pass across the stellar disc, the RV method is much less selective in terms of orbital geometry. Additionally, RV surveys do not require continuous monitoring, as transit surveys do. These factors likely made the RV method the most fruitful detection technique until 2014, when it was surpassed by the transit method, primarily due to the success of the Kepler Space Telescope mission \citep{kepler}. With the launch of other space-based photometric missions, the transit method became exceptionally effective for `planet harvesting'. Automated space telescopes can monitor large sky fields continuously, and detect very small planets due to their photometric precision. The RV method, on the other hand, relies on spectroscopy that is much more resource-intensive. This method requires medium-to-large ground-based telescopes, stable spectrographs, and high signal-to-noise spectra. And so, despite the high precision of RV measurements, they ultimately could not compete with the discovery volume produced by transit missions.

Transits, however, do not measure planetary masses -- a fact that made the RV method fundamentally indispensable. Once both the mass and radius of a planet are known, we can move from simply detecting exoplanets to understanding their internal structures. This in turn offers insights into their formation processes. As a result, although being second in terms of fruitfulness, the RV method did not become obsolete. Instead, it has taken on a crucial secondary role -- to confirm transit detections and measure their masses. And so, the two methods evolved from direct competitors to complementary techniques, which are now used together in tandem.

The probability of an exoplanet transiting its host star decreases with increasing orbital distance. As a result, planets in the habitable zones of their host stars are unlikely to transit from our point of view on Earth. This circumstance makes the RV method particularly advantageous for detecting exoplanets around nearby stars. For instance, an Earth analogue has a transit probability of nearly 0.5 per cent. This means that we would find one such planet around only one of the nearest 200 Sun-like stars, given the very optimistic assumption that all of those stars harbour an Earth analogue.

\subsection{RV measurement techniques}\label{sec:rv_measurement}
Measurements of stellar RVs are derived from the small and consistent shifts in wavelength that are observed in the spectral lines of their light. These shifts arise due to the Doppler effect, which causes a change in the observed wavelength of electromagnetic radiation when a source moves relative to an observer. In particular, if said source is moving with velocity $v$ at an angle $\theta$ relative to the line of sight of the observer, the measured information at wavelength $\lambda_1$ is related to that at emitted wavelength $\lambda_0$ in the rest frame of the source by:
\begin{equation}
    \lambda_1 = \lambda_0 \frac{1+\beta\cos\theta}{\sqrt{1-\beta^2}},
    \label{eq:doppler_shift}
\end{equation}
where $\beta=v/c$. For non-relativistic velocities and small $\theta$, \eqref{eq:doppler_shift} simplifies to
\begin{equation}
    v\cos\theta \equiv v_\text{r} = \frac{\lambda_1-\lambda_0}{\lambda_0}\cdot c,
    \label{eq:doppler_shift_simple}
\end{equation}
where $v_\text{r}$ is the RV component along the line of sight. By convention, $v_\text{r}>0$ for receding sources. The following practice is then to identify a well-known spectral feature (e.g. a strong absorption line from a species) and measure the position of that line in the observed spectrum, $\lambda_1$. Using the measured $\lambda_1$ and the known rest wavelength $\lambda_0$, we can compute $v_\text{r}$ through \eqref{eq:doppler_shift_simple} -- and then build an RV time series that are later modelled via \eqref{eq:rv_formula}.

From this point onwards, we assume that we are working with a set of one-dimensional spectra $\{A\}$, from which we would like to measure an RV time series. In addition, we assume that each spectrum is represented in pixel space, and that we have access to the flux value, its associated uncertainty and the wavelength solution of each pixel. This intermediate data product arises through a sequence of non-trivial procedures, such as:
(i)~the acquisition of two-dimensional images of dispersed starlight,
(ii)~the removal of detector artefacts and instrumental effects,
(iii)~the extraction of individual spectral orders from the two-dimensional image,
(iv)~the application of wavelength calibration,
(v)~the subtraction of sky background and scattered light,
(vi)~the correction for terrestrial motion around the Solar-System barycentre.
For a comprehensive introduction and detailed explanation of these procedures, along with many others, we recommend that the reader consult \citet{Perryman2018} and \citet{Burt2025}. While these steps are essential in practice, they are not central to the core arguments of this dissertation, and we proceed without detailing them.

\subsubsection{CCF methods}\label{sec:ccf_methods}
One of the most intuitive methods to make an RV measurement from a spectrum $A$ is to cross-correlate it with a weighted mask $M$ that reflects the line structure of $A$. This mask weights the spectrum to emphasise line features that are expected in the spectra of the associated stellar type. Then, the RV shift of the spectrum is the value that minimises this cross-correlation:
\begin{equation}
    v_\text{r} = \operatorname{argmin}\int^{\infty}_{-\infty}A(v'_\text{r})M(v'_\text{r}-v_\text{r})\dd v'_\text{r},
\end{equation}
where both $A$ and $M$ are expressible in velocity space. This is now known as the cross-correlation function (CCF) technique. It was first proposed by \citet{Fellgett1955}, demonstrated experimentally by \citet{Griffin1967}, and extended to échelle spectroscopy by \citet{Baranne1979}. In the absence of systematic line asymmetries, the shape of the CCF is well approximated by a Gaussian function. Real absorption lines, however, are typically asymmetric due to stellar-rotation activity, as well as stellar convection and its associated net blueshift (see Sect.~\ref{sec:activity_starspots}). The CCF integrates information from hundreds to thousands of spectral lines, and assumes that any asymmetries should cancel out on average.

For many RV applications, fitting the CCF to a Gaussian profile provides an reliable RV measure, along with useful metrics such as the contrast and the full width at half maximum (FWHM; Sect.~\ref{sec:fwhm_indicator}). CCF methods perform best when the spectrum contains a moderate number of strong, well-separated spectral lines that can be represented by a weighted mask. While the mask $M$ should closely resemble the spectrum of the observed star, a perfect match is not necessary. A small set of carefully selected reference masks can suffice to cover late-F to early-M dwarfs \citep{Burt2025}. For later spectral types, however, spectra become increasingly crowded due to blended molecular features. In such cases, the CCF method discards most of the available spectral information, relies only on a few selected lines, and is less attractive. Furthermore, a Gaussian fit on the CCF fundamentally assumes symmetry. If the CCF is asymmetric, the measured centroid in velocity space does not correspond to the true velocity; and even minor asymmetries can introduce biases in the measured RV that can be as small as a few \unit{\metre\per\second}, which already challenges sub-\unit{\metre\per\second} searches for planets.

There is another issue that affects all methods of RV measurement from spectra, and is particularly pronounced in CCF methods. Stellar spectra are never measured in isolation -- they also contain lines from our own atmosphere, primarily due to molecules such as oxygen, water, carbon dioxide, and methane, among others. These features are known as telluric lines. They are fixed in the terrestrial frame, whereas the stellar spectrum shifts due to the orbital motion of the Earth (approx. $\pm\SI{30}{\kilo\metre\per\second}$) and the intrinsic radial velocity of the star in space. As a result, the same stellar line can appear clean in one observation and overlap a telluric feature in another, depending on the observational epoch. This overlap can shift the apparent centroid of a spectral line and distort the CCF, leading to a biased Gaussian fit and an erroneous RV measurement.

\subsubsection{Template-matching methods}
We find a solution of M-dwarf velocimetry in \citet{Bouchy2001}, that extended the method originally proposed by \citet{Connes1985} that makes use of the full spectral information in each pixel of the observed spectrum. \citet{Bouchy2001} considered the shift in wavelength space $\delta\lambda(i)$ between an observed spectrum $A(i)$ and its (assumed) noiseless reference template $A_0(i)$ at a given pixel $i$:
\begin{equation}
    A(i)-A_0(i)=\pdv{A_0(i)}{\lambda(i)}\delta\lambda(i),
    \label{eq:bouchy_primitive}
\end{equation}
where $\pdv*{A_0(i)}{\lambda(i)}$ is the wavelength gradient of the template at pixel $i$. Combining this with \eqref{eq:doppler_shift_simple} yields
\begin{equation}
    v_\text{r} =
    \frac{A(i)-A_0(i)}{\lambda(i)}
    \frac{1}{\pdv*{A_0(i)}{\lambda(i)}}\cdot c.
    \label{eq:bouchy}
\end{equation}
This equation captures a key idea -- RV is encoded in the rate of change of observed flux with wavelength. If a high-SNR reference template $A_0(i)$ is available, even small RV shifts can result in detectable changes in flux at individual pixels. This method has since been implemented in various forms, including least-squares minimisation (e.g. \citealp{terra,Astudillo-Defru2015,serval}) and Bayesian frameworks (e.g. \citealp{sbart}).

\subsubsection{Line-by-line algorithms}
Template matching has long been recognised for its ability to utilise the full spectral information, and made way to a new and increasingly relevant approach. Rather than applying a single template across the entire spectrum, \eqref{eq:bouchy} can be applied line-by-line (LBL), and return a set of individual (albeit less precise) RV measurements for each spectral line. Then, these line-specific velocities can be combined to produce a robust, averaged RV measurement (e.g. \citealp{Dumusque2018,Cretignier2020,Artigau2022,Lafarga2023}). By isolating the contribution of each line, the LBL method can be used to detect lines corrupted by tellurics, instrumental effects, or other anomalies that would deviate from the overall distribution. LBL methods helped to reveal families of lines that are especially sensitive to stellar activity (\citealp{Dumusque2018,Cretignier2020,Moutou2020,AlMoulla2022,Lafarga2023,GomesDaSilva2025,Larue2025}; Stefanov et al., in prep), and these findings may ultimately contribute to a deeper understanding of how stellar activity manifests and evolves (Sect.~\ref{sec:cool_dwarf_activity}).

\section{Cool dwarfs and their stellar activity in RV}\label{sec:cool_dwarf_activity}
A major challenge in identifying planetary RV signals around cool (GKM) dwarfs is the influence of stellar activity at the $\unit{\metre\per\second}$ level \citep{Saar1997,Santos2000a,SuarezMascareno2017}. These stars often obscure planetary signals across a range of timescales through various physical processes, such as
stellar oscillations (e.g. \citealp{Arentoft2008,Bazot2012}),
granulation (e.g. \citealp{Meunier2017,Meunier2017a}),
and magnetic activity on both short and long timescales (e.g. \citealp{Borgniet2015} and \citealp{Dumusque2011a}, respectively). Stellar activity often generates velocity patterns that closely resemble genuine planets. Such patterns have, on numerous occasions, deceived the scientific community. For example, some planets that had already been discovered were subsequently ruled out by stellar-activity analyses (e.g. \citealp{Queloz2001,Huelamo2008,Carolo2014,Rajpaul2015a,Bortle2021}). Less frequently, signals that were claimed to be stellar-activity by-products were later shown to be of planetary nature after surviving many rigorous tests (e.g. GJ~411~c, \citealp{Hurt2022}; GJ~3998~d, \citealp{Stefanov2025a}). Sometimes, the stellar activity of the same star can produce pseudo-planetary RV signals, while suppressing signatures of actual planets (e.g. Barnard's star; \citealp{Lubin2021,GonzalezHernandez2024}). In other cases, the same planet can be detected, then declared a false positive on account of stellar activity, only to be later resuscitated (HD~192263~b; \citealp{Santos2000,Henry2002,Santos2003}).

The mischievousness of cool dwarfs is well illustrated by the peculiar case of GJ~581. Early discoveries by the HARPS team in this system (GJ~581~b, \citealp{Bonfils2005}; GJ~581~c,d; \citealp{Udry2007}) relied predominantly on conventional techniques at the time: periodograms, sequential Keplerian fitting, FAP-informed thresholds (see Sect.~\ref{sec:pgram_tests}), among others. Shortly after, the published planets around this system became four (GJ~581~e; \citealp{Mayor2009}), prompting an increased interest in the system and leading to additional RV monitoring. \citet{Vogt2010a} reanalysed the accumulated HARPS data and combined it with about as many HIRES measurements to confirm these four planets, and detect two more -- GJ~581~f and GJ~581~g -- and suddenly, GJ~581 found itself with six planets. In particular, GJ~581~g generated considerable excitement within the scientific community, as it would have been the first rocky exoplanet within the habitable zone of its star.

However, as these low-amplitude signals have been emerging, analyses by \citet{Gregory2011} and \citet{Tuomi2012} introduced fully Bayesian-motivated approaches and explicit model comparison through Bayes factors. These works concluded that observational evidence suggested that GJ~581 harboured five or four planets, respectively, depending on the chosen detection thresholds and prior distributions for model parameters. Subsequently, \citet{Robertson2014} and \citet{Hatzes2016} demonstrated that the signal attributed to GJ~581~d had strong correlations with the H$\alpha$ activity indicator of the star, and this planet was also eliminated as a false positive. And so, the current consensus recognises only three confirmed planets in the system: GJ~581~b, GJ~581~c, and GJ~581~e.

This was only a brief outline on the broader scientific debate around GJ~581, which attracted a sustained and ongoing discussion (see \citealp{VonStauffenberg2024} for a comprehensive timeline). The controversy surrounding GJ~581 catalysed a widespread recognition that models must incorporate a realistic stochastic representation of stellar variability.

\subsection{Understanding stellar activity through starspots}\label{sec:activity_starspots}

We illustrate stellar activity by virtue of a simple and an already brought-up example: starspots (Sect.~\ref{sec:starspots}). Consider a rotating, rigid-body star with no limb darkening, and dark spots on its surface. As the star rotates, the visible projected areas of these regions vary. This leads to a fractional flux modulation, which can be understood as the real-time extent of active regions on the stellar surface. However, every active region available to the observer also influences the measured RV through two separate physical processes that occur simultaneously.

\begin{figure}
    \centering
    \includegraphics[width=\linewidth]{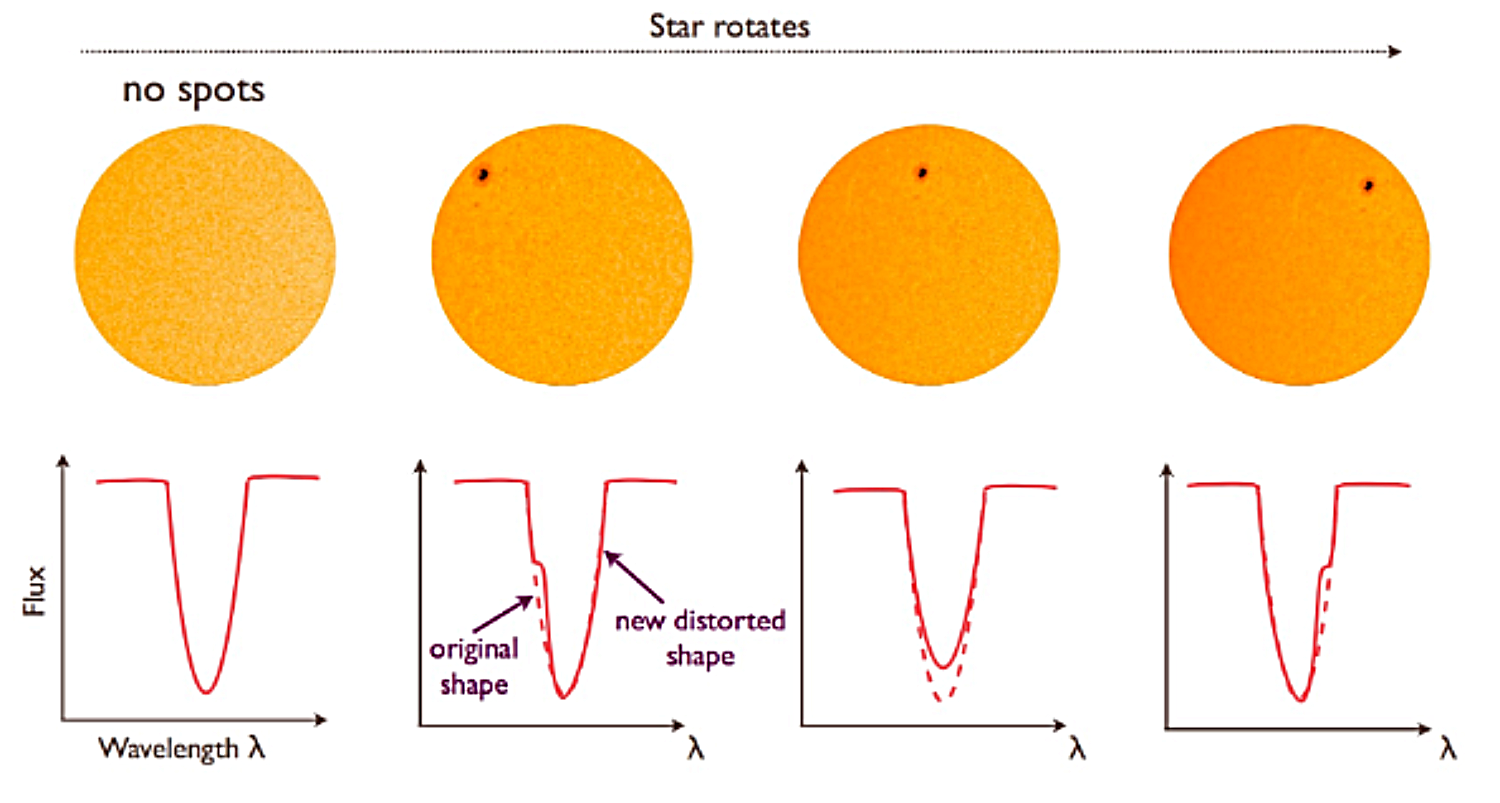}
    \caption{Illustration of how the flux effect causes an uneven absorption in the average line in the stellar spectrum, which in turn changes the observed RV. Adapted from \citet{Haywood2015}, Figure 1.4, under a CC-BY 4.0 licence.}
    \label{fig:haywood_spots}
\end{figure}

The first mechanism, hereafter the `flux effect', is geometric and stems from rotational Doppler imbalance. Different parts of the stellar disc supply with information along different line-of-sight velocities, which, in a sense, defines a Doppler velocity map of the star. As the spot moves across the stellar disc, it reduces the flux somewhere on the Doppler map, and thereby perturbs the average Doppler-broadened symmetric line. A naïve Gaussian fit on the distorted line would therefore measure an apparent RV shift, which is not due to an actual stellar motion (Fig.~\ref{fig:haywood_spots}).

The second mechanism is thermodynamic in origin and arises from the suppression of convective blueshift in active regions. In a quiet photosphere, convection proceeds normally: hotter, more luminous plasma rise from the stellar depths to the surface, and blueshift spectral information. At the surface, cooler, less luminous parcels give way to their hotter counterparts by sinking, and produce a redshift. Because hotter, rising parcels are more luminous and cover a larger surface area, the net effect is a measurable blueshift. In active regions, however, magnetic fields suppress these convective motions. This reduces the upwards flow of hot plasma and diminishes the associated blueshift. This effect is known as `inhibition of convection'.

These are only two of the many ways in which late-type dwarfs have exhibited persistent variability in RV searches for exoplanets. In response, the professional community has turned to the use of ancillary time-series observables that are solely driven by stellar activity. Such metrics are referred to as activity indicators. For example, if we revisit the flux-effect illustration of Fig.~\ref{fig:haywood_spots}, we see that spot-induced perturbations of the aggregate line would not only influence its centraal wavelength (and consequently RVs), but its width as well. This effect, for example, motivates the use of CCF FWHM as such an activity indicator (Sect.~\ref{sec:fwhm_indicator}).

Because they reflect variations in stellar-activity, activity indicators are routinely examined and used alongside RV data. As a result, activity-indicator time series are given equal scientific weight as RVs when assessing potential exoplanet candidates. A good activity indicator should satisfy the following criteria:
(i)~it has a solid physical basis,
(ii)~it can be straightforwardly related to RV,
(iii)~it is unaffected by the Doppler shifts caused by planetary motion,
(iv)~it is sufficiently informative of physical phenomena within the limits of the instrumental precision,
(v)~it is computationally feasible to calculate.
We now briefly review some activity indicators that have been employed in the studies that form this dissertation.

\subsection{A handbook of activity indicators}\label{sec:activity_handbook}
\subsubsection{Photometry}
\begin{figure}[t]
    \centering
    \includegraphics[width=\linewidth]{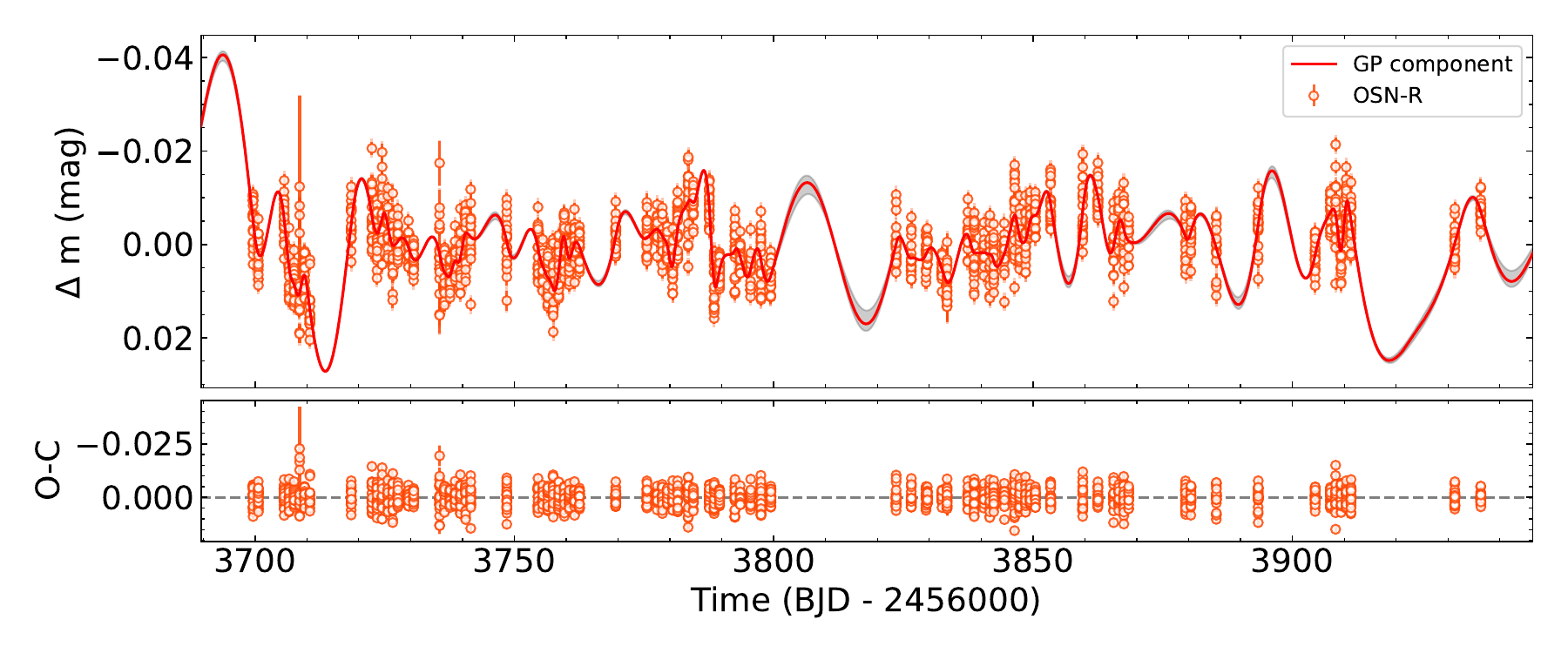}
    \includegraphics[width=\linewidth]{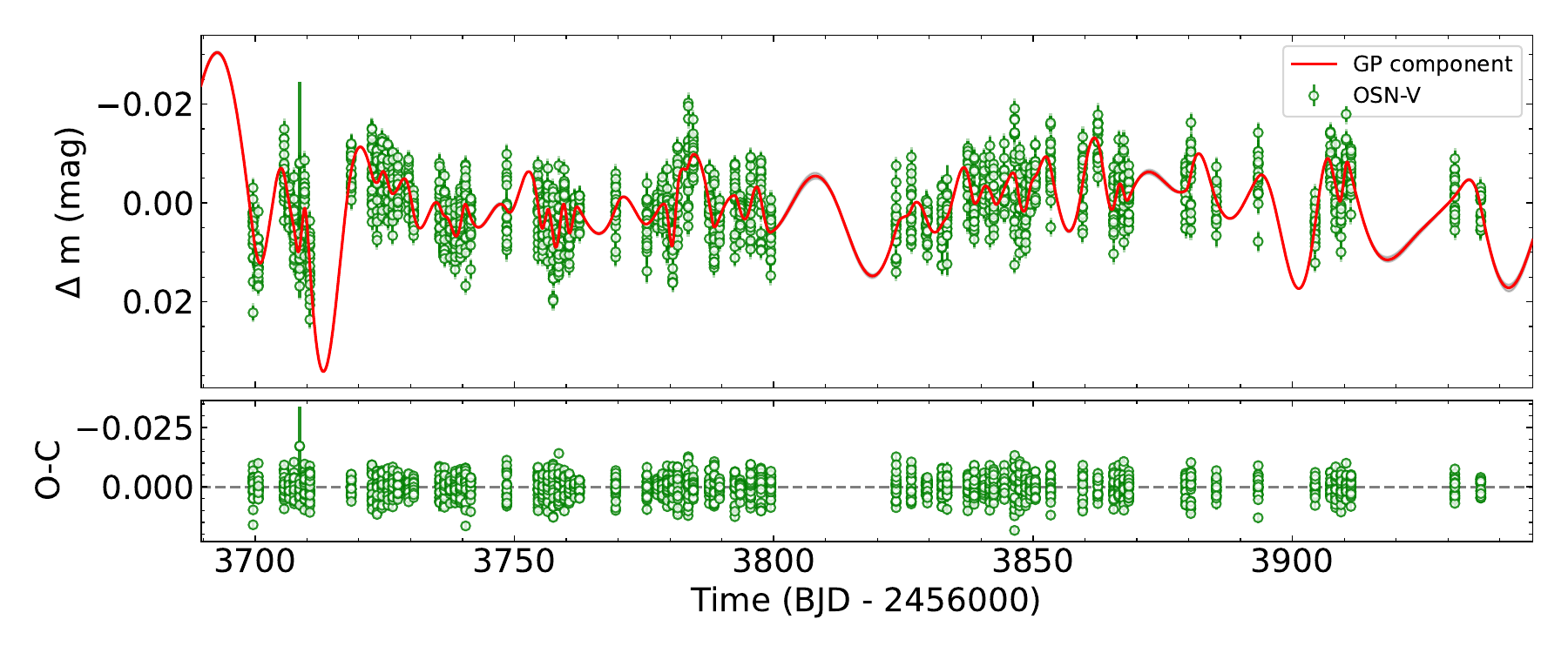}
    \caption{
    Photometric light curves of TOI-2093 from \textit{Observatorio de Sierra Nevada}, in Johnson V (top) and Johnson R (bottom), with a GP model that was used to account for stellar activity (solid red line). Adapted from \citet{Sanz-Forcada2025}, Figure A.4, under a CC-BY 4.0 licence.
    }
    \label{fig:photometry_demo}
\end{figure}

Activity regions are often believed to alter the apparent brightness of stars. The RV effects from such activity regions are typically explained through statistical re-weighting of photometric information, much like our flux-effect discussion. As a result, photometry itself has been an intuitive and accessible indicator that formed the basis of many stellar-activity models (e.g. Fig.~\ref{fig:photometry_demo}; see also \citealp{Aigrain2012,soap,Rajpaul2015,fenrir} for theoretical work). Compared to high-precision spectroscopy, photometry requires fewer observational resources, and this makes it an attractive activity indicator (e.g. \citealp{Haywood2014,Cegla2019,Barragan2024}). While contemporaneous photometric measurements tend to be more challenging to obtain, they can directly trace brightness homogeneities caused by active regions such as spots. These inhomogeneities can then be linked to activity-induced RV variations \citep{Meunier2010,Aigrain2012}.

\subsubsection{CCF FWHM}\label{sec:fwhm_indicator}
Alternatively, activity indicators can be derived from the same spectra that are used to measure RVs. Some of those indicators are based on properties of the CCF profile. One such indicator is the FWHM of the CCF profile, which has been known to be a useful proxy for activity-induced RV variations in slowly rotating stars (e.g. \citealp{Dumusque2014,Santos2014,SuarezMascareno2020,GonzalezHernandez2024,Stefanov2025a}). It has been also shown to help reveal long-term activity cycles of K- and M dwarfs (\citealp{Santos2010} and \citealp{GomesDaSilva2012}, respectively).

The CCF FWHM is fundamentally based on the CCF method, and therefore inherits all its strengths and limitations. This activity indicator can be sensitive to instrumental effects (e.g. \citealp{Zechmeister2013,Dumusque2018,Costes2021,Stoeva2026}), variations in observing conditions, and the choice of CCF mask \citep{Malavolta2017}. In M dwarfs, the presence of many overlapping spectral lines causes CCF profiles to deviate from a Gaussian shape (Fig.~\ref{fig:haywood_spots}; Sect.~\ref{sec:ccf_methods}). This non-Gaussianity understandably reduces the sensitivity of the CCF FWHM to track activity-related changes. Past works have addressed this by combining a Gaussian profile with a polynomial component (e.g. \citealp{SuarezMascareno2017,SuarezMascareno2017a}). Despite these limitations, the CCF FWHM remains an attractive indicator, as it is frequently obtained automatically during RV extraction. However, its interpretation tends to be less direct than other chromospheric diagnostics such as Ca~II H\&K or H$\alpha$ (see Sections to follow), because it traces a geometric effect on the CCF line, rather than actual physical processes. The idea of CCF FWHMa has been extended to template-matching methods, including the differential line width (dLW) in \textsc{SERVAL} \citep{serval} and the second-derivative velocity (d2v) in the \textsc{LBL} implementation of \citet{Artigau2022}.

\subsubsection{Chromospheric activity indicators}
\begin{figure}
    \centering
    \includegraphics[width=\linewidth]{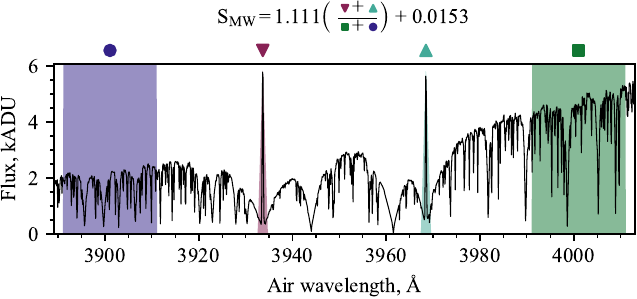}
    \caption{An illustration of how the Mount Wilson S-index is calculated on one of the ESPRESSO spectra of GJ~526 in \citet{Stefanov2025b}.}
    \label{fig:s_index_demo}
\end{figure}

Activity indicators can be derived from the flux measurements of specific chromospheric lines. For example, the intensity of the emission in the cores of the Ca~II H\&K lines (\SI{3933}{\angstrom}, \SI{3968}{\angstrom}) is closely related to the strength of the magnetic fields of FGKM stars. In turn, this magnetic activity is strongly linked to their stellar rotational periods. As active regions move across the stellar surface, the emission intensity of these lines varies, and this allows us to trace the stellar rotation. The S-index is a well-established measure of chromospheric activity, and is derived from the fluxes of the Ca~II H\&K lines. The index traces its origins to long-term monitoring at the Mount Wilson Observatory, and is defined as
\begin{equation}
    S_\text{MW} = 1.111\times\frac{\Tilde{N}_\text{H}+\Tilde{N}_\text{K}}{\Tilde{N}_\text{R}+\Tilde{N}_\text{V}}+0.0153,
    \label{eq:s_index}
\end{equation}
where $\Tilde{N}_\text{H}$ and $\Tilde{N}_\text{K}$ are triangular passbands with centers at \SI{3968.470}{\angstrom} and \SI{3933.664}{\angstrom}, respectively, each with a FWHM of \SI{1.09}{\angstrom}; while $\Tilde{N}_\text{R}$ and $\Tilde{N}_\text{V}$ are rectangular passbands centred at \SI{4001.07}{\angstrom} and \SI{3901.07}{\angstrom}, each with a width of \SI{20}{\angstrom} \citep{Vaughan1978}. Figure~\ref{fig:s_index_demo} shows how the Mount Wilson S-index is computed from an ESPRESSO spectrum of GJ~526 in the study of \citet{Stefanov2025b}.
Measurements of the S-index for certain stars, including the Sun, have been carried out since the 1960s. These data have been widely used to study a variety of stellar characteristics. A detailed review of these lines of work is given in \citet{Hall2008}.

\begin{figure}[t]
    \centering
    \includegraphics[width=\linewidth]{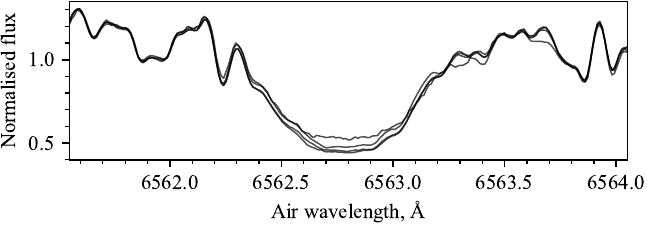}
    \caption{Five ESPRESSO spectra of GJ~526 around the H$\alpha$ line.}
    \label{fig:h_alpha_demo}
\end{figure}

Although Ca II H\&K lines have long been utilised for measuring chromospheric activity, they suffer from some limitations when applied to late-type stars. These stars are redder, and as a result, the blue calcium lines become fainter compared to other spectral features in the visible range \citep{Cincunegui2007,Robertson2013,IbanezBustos2023}. Like the Ca II H\&K lines, the emission in the core of the H$\alpha$ (\SI{6563}{\angstrom}) line is associated to the magnetic-field strength and the presence of active regions on the stellar surface (Fig.~\ref{fig:h_alpha_demo}).
For this reason, H$\alpha$ has also been a popular choice for an activity indicators. The works associated with this dissertation \citep{Stefanov2025b,Stefanov2025a} computed H$\alpha$ following \citet{GomesDaSilva2011}.

\begin{figure}[t]
    \centering
    \includegraphics[width=\linewidth]{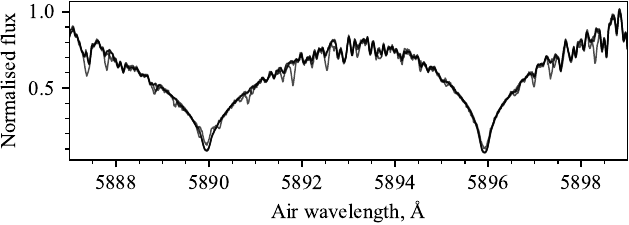}
    \caption{Five ESPRESSO spectra of GJ~526 around the sodium doublet.}
    \label{fig:na_i_demo}
\end{figure}

Another family of indicators relies on flux ratios associated with the sodium doublet (\SI{5889}{\angstrom}, \SI{5895}{\angstrom}), for this doublet has also been shown to respond to magnetic activity in cool stars \citep{Hall2008}. The works associated with this dissertation \citep{Stefanov2025b,Stefanov2025a} computed the Na~I activity index following \citet{Diaz2007}. Figure~\ref{fig:na_i_demo} shows several ESPRESSO spectra of GJ~526, centred near the sodium doublet.

All of these indices share a common design choice -- they measure the flux in narrow windows centred on activity-sensitive line cores, which is then normalised by nearby pseudo-continuum regions. Variations in those ratios reflect chromospheric activity. Since these indicators directly probe the magnetic chromosphere, they complement other activity indicators, such as CCF metrics (e.g. the FWHM), which mainly reflect changes in the shape and width of spectral lines. Some activity events produce strong chromospheric signatures, while others mainly affect line profiles. In practice, analysing both types of indicators together tends to provide a more complete picture of the activity of a star.

\subsubsection{Differential temperature}
Although we introduced \eqref{eq:bouchy_primitive} in the context of RV measurement, this equation leaves a rather interesting idea. Rather than limiting ourselves to wavelength space $\lambda(i)$, we could measure a per-spectrum shift in any observable $\mu$, as long as there exists a way to compute $\pdv*{A_0(i)}{\mu(i)}$. \citet{Artigau2024}, applied this idea to the temperature derivative of the template $\pdv*{A_0(i)}{T(i)}$, which could be obtained from a spectral library of objects with known temperatures (in their case, from \citealp{gaiaDR3}). This allowed the authors to determine the temperature shift $\delta T$ at a given pixel $i$:
\begin{equation}
    \delta T(i) = \frac{A(i)-A_0(i)}{\pdv*{A_0(i)}{T(i)}}.
    \label{eq:delta_temperature}
\end{equation}
This temperature shift gives the per-spectrum change of disc-averaged temperature $\Delta T$. This quantity is referred to as differential temperature for short, and has been shown to be a promising activity indicator for cooler stars (e.g. \citealp{Cadieux2025,Lamontagne2026,Osborn2026}).

\subsection{Gaussian-process (GP) frameworks}\label{sec:gp_frameworks}
Richard Feynman famously drew an analogy between scientific understanding and a observing the game of chess. By examining the board and its evolution of state, an observer can gradually determine the rules of chess by recognising patterns -- for example, they can reason how the Bishop moves by noticing how it moves only along squares of a single colour. In the same way, by making careful scientific observations, we gradually form a consistent set of laws that describe reality \citep{Feynman1963}.

I dare say the situation with stellar activity is no different -- but instead of chess, we observe the game of backgammon. Backgammon is a highly variant game, driven by random rolls of dice. Nevertheless, there are deterministic rules of play: a stone may be hit and borne off the board, only to return later if the dice are kind. To an uninvolved observer that does not see the dice, a pattern is revealed. This pattern is highly variable, yet structured, much like the stochastic emergence and decay of active regions on stellar surfaces. Thus, just as a backgammon spectator, who sees only the board and the movement of stones, but tries to infer the rules of the game, so too do we seek to deduce the rules of the `game' we play with stellar activity. 

The fundamental contribution of this dissertation rests on Gaussian processes (GPs) -- a type of stochastic processes the community uses to model and interpret stellar activity. \citet{Rasmussen2006} formally define a GP as `a collection of random variables, any finite number of which have a joint Gaussian distribution'. GPs began to be applied in exoplanet RV analyses in the early 2010s. Among the first works to use GPs for modelling temporally correlated stellar nuisance signals were \citet{Hou2014}, who developed a GP framework to model stochastic stellar oscillations in RV data, and \citet{Barclay2015}, who employed GPs to model granulation signals in Kepler-91 photometry. The earliest study that introduced GPs into RV activity mitigation, however, was \citet{Haywood2014}. In this work, the authors used a quasi-periodic GP kernel, trained it on photometric data, and used its inferred covariance properties to model the available RV time series at hand. A second early example of this line of work is \citet{Grunblatt2015}, which applied a similar strategy to the Kepler-78 system. Later, \citet{Rajpaul2015} developed the multivariate GP framework that became a standard reference, 
and provided a principle means for separating planetary signals from activity-induced variability.

The main advantage that was recognised in GPs is that they enable the fitting of time-series data, given a specific closed-form expression that describes the covariance between observations. Many early GP studies in the field (e.g. \citealp{Haywood2014,Grunblatt2015}) re-analysed already known systems with the primary aim of refining orbital parameters and improving mass estimates. This application remains relevant today (e.g. \citealp{Burt2024}), but more recently, GPs have proven particularly useful in detecting low-amplitude planets around late dwarfs with appreciable stellar activity. Many of these detections now include a GP component as part of the baseline model (e.g.~\citealp{Amado2021,Damasso2022,Kossakowski2023,SuarezMascareno2023,Dreizler2024,GonzalezHernandez2024,SuarezMascareno2025}), and without such a component, some of those would likely not have been identified.

\subsubsection{The squared exponential periodic kernel}\label{sec:sep_kernel}

What would be the typical assumptions that motivate such closed-form expressions? For instance, regardless of the physical event that is being monitored, we typically assume that measurements taken with a smaller time separation $\Delta t$ would show greater correlations than those separated by a large $\Delta t$. By way of illustration, we would make such an assumption onto the temperature fluctuations in a room. Without additional knowledge, we could easily imagine that two temperature readings in the room a few minutes apart would be more correlated than a third measurement on a later day. The same can be said about the temperature, the wind speed, or the solar irradiance at a fixed location. These vary with time of day, season, and weather, but their short-term fluctuations are more predictable than long-term ones. And finally, by extension, we could argue that this holds for active regions too. Stars have magnetic fields that are generated by dynamo processes deep within their interiors. These fields evolve over time, but should have some characteristic time scale. Although an analytic solution probably does not exist, we can assume that the covariance of two stellar-activity measurements of any kind should somehow depend exponentially on their time separation. This motivates the covariance-function component
\begin{equation}
    k(\Delta t)\propto\exp\left[-\frac{(\Delta t)^2}{2\tau^2}\right],
    \label{eq:gp_exponential_kernel}
\end{equation}
where $\tau$ is some characteristic timescale we associate with their fluctuations, and bears a unit of time, typically days. This covariance function is called the squared exponential kernel, the Radial Basis Function kernel, or the Gaussian kernel. The squares in $\Delta t$ and $\tau$ are not arbitrary -- they arise from underlying assumptions of smoothness. The $(\Delta t)^2$ term ensures infinitely differentiable, symmetric functions with a natural Euclidean-distance interpretation. The factor of two in the denominator connects this exponential terms to the Gaussian distribution, and gives a clean interpretation of $\tau$ as some `functional inertia' of the stochastic process. Larger $\tau$ lead to smoother and more time-stable functions, and vice versa. Extrapolation beyond $\tau$ from a collection of points is generally unreliable.

Some stochastic processes are driven by environmental factors that act at specific periodicities. Let us return to the example of measuring temperature at a fixed location. Beyond the short-term dependencies that we expect by nearby environmental conditions, we know that temperatures tend to rise during the day and fall at night, as they are driven by periodic solar radiation. That is to say, we expect some recurring pattern at a particular period -- in this case, \SI{1}{\day}. If we go back to stellar activity, we can easily see that stellar rotation drives periodic variations in a very similar manner. As a star rotates around itself, it manifests different regions of its surface to the observer. Our covariance function should therefore reflect the rotational period of the star:
\begin{equation}
    k(\Delta t)\propto\exp\left[
    -\frac{\sin^2(\pi\Delta t/P)}{
    2\eta^2
    }
    \right],
    \label{eq:gp_periodic_kernel}
\end{equation}
where $P$ determines the characteristic period of covariance, and $\eta$ determines the `functional inertia' of the sine term akin to $\tau$ in \eqref{eq:gp_exponential_kernel}. This covariance function is called the periodic kernel. The more inquisitive will note that the periodic kernel does not simply scale with a sine term. Instead, the term shown on the right-hand side of \eqref{eq:gp_periodic_kernel} is analogous to the one on the right-hand side of \eqref{eq:gp_exponential_kernel}, under the additional transformation \mbox{$\Delta t\to\sin(\pi\Delta t/P)$}. In other words, the periodic kernel acts as an exponential kernel, but by using a periodic notion of difference, as if $\Delta t$ were the phase distance on a unit circle. As such, the periodic-kernel component is maximised when \mbox{$\Delta t=nP$}, \mbox{$n\in\mathbb{Z}$}. While $P$ carries a unit of time, $\eta$ is dimensionless, as it is relative to P \citep{Rajpaul2017}.

The quantity $\eta$ has been subject to various parametrisations in the literature. It has been referred to as:
`the harmonic complexity' $\Gamma=1/\sqrt{2\eta^2}$ \citep{Angus2018}
or $\Gamma=1/2\eta^2$ \citep{Lakeland2026},
`the GP smoothing parameter' $\eta'=\sqrt{2}\eta$ (e.g. \citealp{Lam2024,Rescigno2024a}),
and occasionally, the `inverse complexity harmonic
parameter' $\eta_4=2\eta$ (\citealp{Camacho2023}; see also \citealp{Sulis2024,Nari2025}). At the same time, many works share our definition of $\eta$ (e.g. \citealp{SuarezMascareno2018a,Amado2021,Perger2021}), but occasionally with different names:
`the harmonic complexity' (e.g. \citealp{Damasso2024a,Barragan2026}), `the inverse harmonic complexity' (e.g. \citealp{Ahrer2021,Barragan2021,Klein2024}), or `the smoothing factor' (e.g. \citealp{Cortes-Zuleta2023}). \citet{Stefanov2025a} proposed to refer to $\eta$ as the `sinescale', on account of its algebraic analogy to $\tau$ in \eqref{eq:gp_exponential_kernel}. We adopt this naming, for it allows a natural analogy to the timescale $\tau$, but by acting in unit-circle space. Large sinescales indicate stochastic processes that are less sensitive to phase differences, and vice versa.

\begin{figure}[t]
    \centering
    \includegraphics[width=\linewidth]{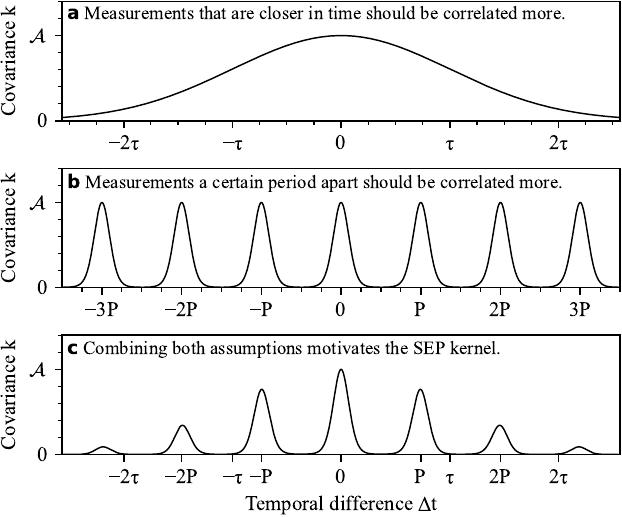}
    \caption{
    The intuition behind the SEP kernel. We take one of our measurements, and consider how it correlates with the remaining measurements as a function of the temporal difference $\Delta t$. The combination of two general assumptions gives us a complex form of covariance, which we try to solve for $\mathcal{A}$, $\tau$, $P$, and $\eta$.
    }
    \label{fig:sep_kernel_intuition}
\end{figure}

So far, we have examined two general assumptions -- that measurements should be correlated more when they are: (i)~closer in time, and (ii)~a certain period apart. The combination of both assumptions takes an algebraic form in the product of equations \eqref{eq:gp_exponential_kernel} and \eqref{eq:gp_periodic_kernel}:
\begin{equation}
    k(\Delta t)=\mathcal{A}^2\cdot\exp\left[
    -\frac{(\Delta t)^2}{2\tau^2}
    -\frac{\sin^2(\pi\Delta t/P)}{
    2\eta^2
    }
    \right],
    \label{eq:gp_sep_kernel}
\end{equation}
and is known as the quasi-periodic kernel or the squared exponential periodic (SEP) kernel. $\mathcal{A}$ is known as the GP amplitude, and carries the units of the physical quantity in question, e.g. \unit{\metre\per\second} in RV. Figure~\ref{fig:sep_kernel_intuition} visualises how the combination of our two general assumptions leads to \eqref{eq:gp_sep_kernel}.

The SEP kernel has remained a consistent choice in exoplanet-detection studies that employ Gaussian processes for stellar-activity 
modelling over the years, and has increasingly established itself as a de facto standard in the field (e.g. \citealp{Affer2016,Yu2017,Mortier2018,Barragan2019,Benatti2020,Amado2021,Damasso2022,Desidera2023,Rescigno2024a,Lamontagne2026}). What makes SEP kernels particularly useful in exploratory analyses is that \eqref{eq:gp_sep_kernel} collapses to the squared exponential kernel for large $\eta$, or to the periodic kernel for large $\tau$. In this way, if we begin modelling with the SEP kernel, we can quickly assess whether a simpler model is sufficient to describe the stellar activity in question.

The quantities \mbox{$\mathcal{A},\tau,P,\eta$} are not parameters in the classic sense of the term, and they should not be referred as such. Parametric functions have determinate algebraic forms that are defined by a finite set of parameters, chosen to capture the overall functional behaviour. Such functions include polynomials, exponentials, trigonometric functions, and more generally, closed-form expressions. GPs, however, are non-parametric. They do not assume a pre-defined functional form, and instead allow the data to determine the shape of the stochastic process (see \citealp{Haywood2015} for discussion). \mbox{$\mathcal{A},\tau,P,\eta$} cannot specify the functional values of the stochastic process without reference to the data. Instead, they govern the structure of the considered covariance kernel. Consequently, \mbox{$\mathcal{A},\tau,P,\eta$} are collectively referred to as `hyperparameters', with the prefix `hyper-' indicating their position beyond data itself. Table~\ref{tab:gp_sep_hyperparameters} reviews the hyperparameters of the SEP kernel, along with typical values derived from the practical experience that came along with the three-year preparation of this dissertation. These values should be taken as a starting point, before any more rigorous analysis is undertaken.

\begin{table}
\centering
\caption{Hyperparameters of the squared exponential periodic (SEP) kernel, along with their common notation, units, and typical range of solutions.}
\vspace{4pt}
\label{tab:gp_sep_hyperparameters}
\begin{tabular}{clcl}
\hline\hline
\multicolumn{1}{c}{Symbol} &
\multicolumn{1}{c}{Name} &
\multicolumn{1}{c}{Unit} &
\multicolumn{1}{c}{Typical range} \\ \hline
$\mathcal{A},\mathcal{B}\ldots$ &
Amplitude &
- &
$[0,\ptp y_i]$ \\
$\tau$ &
Timescale &
\unit\day &
$[2P,+\infty)^\dagger$ \\
$P$ &
Period &
\unit\day &
$[2,200]^\ddagger$ \\
$\eta$ &
Sinescale &
1 &
$[10^{-2},10^1]$ \\
\hline
\multicolumn{4}{l}{
\footnotesize
$^\dagger$informed by \citet{Giles2017},
$^\ddagger$informed by \citet{SuarezMascareno2017}.
}
\end{tabular}
\end{table}

Empirical evidence suggests that conditioning GP kernels solely on RV data can lead the model to misattribute planetary signals to stellar activity, thereby reducing successful planet detections. An emerging solution has been the fitting of several GP models simultaneously to both RVs and stellar-activity indicators. By incorporating stellar-activity information into the GP kernel structure, the model can gain a more accurate understanding of the underlying signal behaviour. This enables better distinction between planetary and non-planetary separation in RV data (e.g. \citealp{Ahrer2021,Rajpaul2021}). There are several main ways to assign GPs to several physical quantities, as outlined in \citet{Tran2023}. While a detailed account is provided therein, we now briefly recap the key points.

\subsubsection{Serial GPs}\label{sec:gp_serial}
The simplest way to constrain GP models on stellar-activity and RV time series is to first fit a GP to an activity-indicator time series, and obtain the posterior distributions of hyperparameters. These posteriors can be then used as prior distributions when modelling RV data under the same kernel structure. In this manner, the same GP kernel is informed sequentially by two time series: first by an activity indicator, then by RVs. This sequence motivates the name of this approach -- the serial approach. The first step of the serial approach is meant to act as a protective blanket against overfitting RV observations, as it incorporates an independent dataset that should reflect stellar activity alone. The serial approach has been among the early methods to incorporate stellar activity in the GP analysis, and has been usually carried out in tandem with photometry (e.g. \citealp{Haywood2014,Grunblatt2015,KunovacHodzic2021}).

The serial approach is a good step towards the use of multi-dimensional time-series data, but is not always advisable. \citet{Nicholson2022} identified two main potential problems, which we now briefly address. First, there is observational evidence that SEP hyperparameters may vary significantly over the magnetic cycle of the star \citep{Kosiarek2020}. Therefore, training a GP on activity indicators may result in misleading priors for the RV fit, especially when the two time series are not contemporaneous to one another. Second, it has been observationally found that RV- and activity-indicator time series may have consistent $P$ and $\tau$, but significant differences in $\eta$. This suggests distinct patterns in RV and activity (\citealp{Kosiarek2020,Nicholson2022}; also confirmed by \citealp{Barragan2021} with simulated data). Indeed, as it was alluded to in Sect.~\ref{sec:activity_handbook}, each activity indicator captures a distinct aspect of stellar activity. Therefore, it is unreasonable to expect them to reflect the same physical processes, or exhibit identical complexity.
 
\subsubsection{Semi-independent GPs}\label{sec:gp_semishared}

One way to combat the problem of different sinescales in RV and activity indicators is to take the time series $y_j$, measured at times $t_j$ -- with $j$ indexing different physical quantities -- and construct $M$ semi-independent GP kernels $G$ that share $P$ and $\tau$, but have different $\mathcal{A}_j$ and $\eta_j$. Then, we could condition every kernel on a separate set of measurements $y_j$. This implies solving the system of equations
\begin{equation}
\begin{array}{|l}
y_j = G(t_j;\mathcal{A}_j,\tau,P,\eta_j) \\
\vdots\\
\end{array}.
\end{equation}
Finally, the log-likelihood function is obtained from the sum of log-likelihoods of said kernels. We refer to this as the semi-independent regime. This regime has also been used in stellar activity modelling (e.g. \citealp{Bortle2021,Lamontagne2026}), where in certain cases $\eta_j$ was even treated as a shared parameter across dimensions (e.g. in earlier studies of Proxima Centauri; \citealp{Faria2022}).

Regardless of the question posed on the potential difference in $\eta$, the semi-shared approach eliminates the need to sequentially run models and transfer the output probability density function from one training sample to another -- a procedure that inherently discards information along the way \citep{Osborn2021}. In addition, the semi-shared approach is more aligned with the simultaneous time-series modelling of RV and activity -- they are now treated on the same footing for the first time in our narrative, and unlike previously discussed approaches.

\subsubsection{$FF'$ GPs}\label{sec:ff_gps}
Earlier, we justified the use of SEP kernel (and GPs in general) via two elementary assumptions: larger covariance for smaller $\Delta t$, and covariance that exhibits sensitivity at periodic $\Delta t$. At this stage, however, our rationale did not include any context relevant to the physics of stellar activity. As such, a SEP-kernel approach on its own lacks the means to distinguish physical from unphysical solutions. It has become evident to the community that incorporating more physically motivated priors would enhance the robustness of GP applications. An emerging GP regime in recent years is grounded in the $FF'$ formalism as described by \citet{Aigrain2012} and \citet{Rajpaul2015}, which we will now briefly describe. We recommend that readers consult these two publications for a more detailed account.

In the beginning of Sect.~\ref{sec:activity_starspots}, we introduced the two physical phenomena in starspots that influence the measured RV: the flux effect and the inhibition of convection. A key insight is that the flux-effect RV perturbation does not scale with $F(t)$, but with its time derivative $\dv*{F(t)}{t}$. To illustrate why, we revisit two key configurations of a spot on the disc. When the spot is in the centre of the disc, the flux deficit is greatest, so $F(t)$ is large. Yet, the rotational symmetry between the approaching at the receding half of the disc is preserved, leading to a \mbox{$\Delta\text{RV}\to0$}. On the other hand, when the spot is near the limb, $F(t)$ is smaller due to projection and, yet the approaching-receding half asymmetry is maximised.\footnote{Limb darkening also reduces $F(t)$, but as a separate effect that does not motivate the scaling with $\dv*{F(t)}{t}$.}
On the other hand, the inhibition-of-convection RV perturbation is insensitive to the precise location of active regions on the disc (beyond the effects of limb darkening), and is influenced primarily by their total area, i.e. by $F(t)$. And so, RV perturbations stemming from the inhibition of convection scale with $F(t)$, while those driven by a flux effect scale with $\dv*{F(t)}{t}$.

Nevertheless, we must account for an important boundary condition -- all activity induced effects must vanish in the absence of activity regions. That is to say, all RV perturbations must, at lowest order, be proportional to the extent of active regions, i.e. $F(t)$. This leads us to
\begin{equation}
    \Delta\text{RV}(t)=\alpha F^2(t)+\beta F(t)\dv{F(t)}{t},
    \label{eq:ff_formalism_primitive}
\end{equation}
where $\alpha$ and $\beta$ scale the effects of inhibition of convection and the flux effect, respectively (see \citealp{Aigrain2012} for a formal treatment). It is namely through \eqref{eq:ff_formalism_primitive} that the FF' formalism derives its name.

We remark that \citet{Aigrain2012} derived \eqref{eq:ff_formalism_primitive} for a lone active region. Yet, this equation can be applied to account for the collective influence of all active regions across the visible half of the star. \citet{Rajpaul2015} extended this framework by seeking similar expressions to activity indicators beyond photometry. As such, they generalised \eqref{eq:ff_formalism_primitive} to the system of equations
\begin{equation}
    \begin{array}{|l}
    y_j=\mathcal{A}_j\,G(t_{j})+\mathcal{B}_j\,(\dv*{G}{t})_{t=t_{j}}\\
    \vdots
    \end{array}
    \label{eq:ff_formalism}
\end{equation}
where \mbox{$G(t_j)\equiv G(t_j;\tau,P,\eta)$} is a latent Gaussian-process kernel that is evaluated with the same hyperparameters for all time series. \citet{Rajpaul2015} discussed that some activity indicators, e.g. $\log R'_\text{HK}$, scaled similarly with photometry, in that they should be proportional to $G$ only, and not to $\dv*{G}{t}$. In such cases, \mbox{$\mathcal{B}_j=0$}. In contrast, other indicators should scale with the linear combination of $G$ and $\dv*{G}{t}$, e.g. the inverse slope of the bisector of the CCF (BIS; \citealp{Queloz2001}). This distinction classified activity indicators into two main groups. \citet{Barragan2021} termed these two categories `photometry-like' (i.e. $\mathcal{B}_j=0$) and `RV-like' (i.e. $\mathcal{B}_j\neq 0$) activity indicators, and we will be using the same expressions from now on.

The $FF'$ formalism has been gaining popularity for the disentanglement of stellar activity of late-type dwarfs in recent years, with examples including
mass measurements of multi-planet systems (e.g. \citealp{Passegger2024}), as well as discoveries of planets around active stars
(e.g. \citealp{Barragan2022}),
nearly ultra-short period planets (e.g. \citealp{Deeg2023}),
temperate planets (e.g. \citealp{SuarezMascareno2023,Stefanov2025a}),
and most notably, the candidate sub-Earth planetary system around Barnard's star \citep{GonzalezHernandez2024}. The $FF'$ approach proves particularly effective in breaking degeneracies in cases of strong correlations between RVs and activity indicators (e.g. \citealp{SuarezMascareno2020,SuarezMascareno2025}).

As the system of equations associated with the $FF'$ framework fits a single latent kernel $G$ on all quantities, these quantities may be positively or negatively correlated with one another. Consequently, $\mathcal{A}_j$ and $\mathcal{B}_j$ can be positive or negative. There are two key circumstances pertinent to this modelling process. First, there must be at least one photometry-like activity indicator (at least to leading order) in \eqref{eq:ff_formalism}, in order to break the free mixing of $G$ and $G'$. Second, among these photometry-like indicators, at least one must have its single free parameter $\mathcal{A}_j$ constrained to be non-negative. This eliminates a two-fold sign degeneracy that arises in \eqref{eq:ff_formalism}.

\section{A practical guide for planet hunting in RV}\label{sec:hunter_guide}
We often associate `planet hunting' with the grand quest of discovering new worlds. However, we should not forget that hunting is equally a process of elimination -- and some planetary candidates are slain not by a bullet, but by subtle, misleading stellar activity.

So far, our discussion has largely focused on how stellar activity manifests in RV measurements. Indeed, stellar 
activity poses a significant challenge not only in RV measurements but also in observables from other methodologies: active regions on the surface can bias transit photometric parameters (e.g. \citealp{Martioli2021}), or introduce non-trivial astrometric jitter by shifting the apparent-centroid position \citep{Eriksson2007}. Nevertheless, it seems to me that planetary RV signals are more vulnerable to 
systematic contamination than transit-photometry ones. This comes from the fundamental nature of planetary signals themselves. Transits tend to have sharp ingress- and egress features, and yield well-defined durations, depths, and orbital periods. Their direct, box-like signatures are more distinctive, and subsequent stellar-activity modelling follows more readily by virtue of their strong constraints in period and phase \citep{Burt2025}.\footnote{At the same time, isolating an RV signal with a very narrow prior on the period comes with its caveats; see \citet{Rajpaul2024}.}
On the other hand, RV signatures of planets are continuous and smooth modulations in velocity, and are therefore inherently less distinguishable.

This should not dismay us, planet hunters in the RV realm. Our community has been remarkably ingenious in developing many tests of planetary nature, that are grounded in two main axioms: (i)~that planets induce Doppler shifts in all spectral lines alike, and (ii)~that the parameters describing these shifts are time-invariant (e.g. $k_\text{rv}$; Sect.~\ref{sec:apodisation}). Each new test adds another layer to the armour of procedures we employ when detecting and confirming planets under the heavy veils of stellar activity. Naturally, every individual planet must pass a sufficient amount of such tests, and it is inevitable that some older detections that fail to meet all criteria must be disregarded as false positives.

We now discuss a suite of the planetary-nature tests that were used in most works associated with this dissertation \citep{Stefanov2025b,Stefanov2025a}, as well as in other works where I have had some contribution \citep{SuarezMascareno2024,GonzalezHernandez2024,Nari2025,SuarezMascareno2025,SuarezMascareno2025a,Nari2026,Nari2026a,Lamontagne2026,Stoeva2026}. All tests in the suite rely on comparison between RV and some activity indicator. We remark that the purpose of these tests is not to infer stellar-activity mechanisms, but rather to minimise the number of false positives in exoplanetary detection. As such, the tests are data-oriented, and do not require a deep understanding of the particular stellar physics associated with the indicators of choice. In fact, only in this context, the definition of `activity indicators' can be relaxed to include all observable proxies of instrumental or pipeline behaviour that are suspected to introduce false RV signals (such as incomplete BERV correction; e.g. \citealp{Wright2014}).

\subsection{Periodogram tests}\label{sec:pgram_tests}
The first family of tests involves the use of periodograms and the peaks they reveal in observed time series. Periodograms are a way of analysing a time series in order to find periodic signals within those time series. They are fundamentally based on the Fourier transform
\begin{equation}
    \hat{f}(\xi) = \int^{\infty}_{-\infty}
    f(x)\cdot\exp\left(
    -2\pi i \xi x
    \right)\dd x
\end{equation}
and its associated power spectrum
\begin{equation}
    P(\xi) = \left|\hat{f}(\xi)\right|^2,
\end{equation}
but are generalised to account for:
(i)~measurements at discrete timestamps,
(ii)~non-uniform sampling,
(iii)~measurement uncertainties,
(iv)~measurements from several instruments, and so on. Periodograms have been widely employed since the early days of RV detection works (e.g. \citealp{Lovis2005}), and some of its variants such as the generalised Lomb-Scargle periodogram (GLSP; \citealp{Lomb1976,Scargle1982,Zechmeister2009,VanderPlas2015}) or the Bayesian Lomb-Scargle periodogram (BLSP; \citealp{Mortier2015}) are standard tools in the repertoire of the professional community. An introduction to GLSPs is beyond the scope of this dissertation; we refer the reader to \citet{VanderPlas2018} for a comprehensive discussion of GLSPs and best practices in their interpretation.

Periodograms, though valuable and well understood, are inherently limited in their ability to fully capture the complexities of real data (e.g. \citealp{Dumusque2017,Hara2022a,Hara2023}). Periodograms by design assume non-correlated (`white') noise between measurements, which is often violated in real measurements. Additionally, they search for one signal at a time, but in reality, strong peaks can interfere with weaker ones. The spectroscopic analysis of the K dwarf GJ~1137 presented in \citet{Stoeva2026} is a good anecdotal example. Current evidence suggests that GJ~1337 harbours two planets: a temperate Saturn at a period near \SI{144.7}{\day}, and a hot super-Earth at a period near \SI{9.6}{\day}. In addition, \cite{Stoeva2026} found GJ~1137 to be spectroscopically active in RV and two activity indicators: FWHM and $R'_\text{HK}$. Then, the authors measured the rotational period of GJ~1137 to be near \SI{32.3}{\day}, and detected a magnetic cycle near \SI{5900}{\day}.

\begin{figure}[t]
    \centering
    \includegraphics[width=\linewidth]{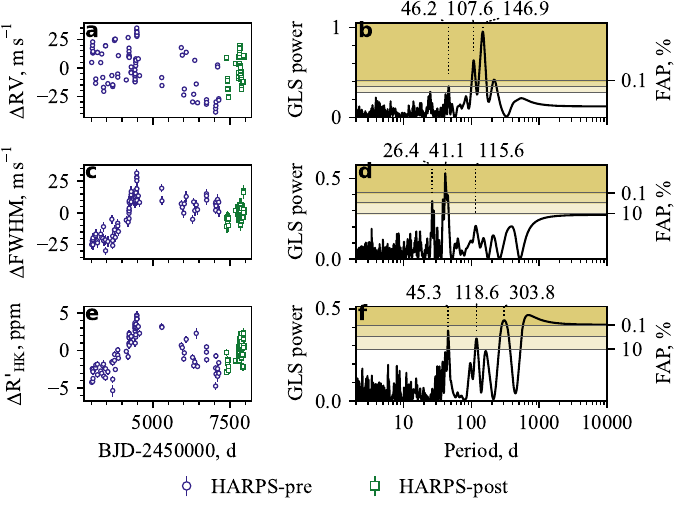}
    \caption{
    HARPS time series of GJ~1137 of mean-subtracted:
    (a)~RV,
    (c)~FWHM,
    (e)~R$'_\text{HK}$.
    Data points are colour-coded depending on whether they were from before or after the optical fibre upgrade of HARPS in May 2015 \citep{LoCurto2015}. (b,d,f) Associated wide-period GLSPs of RV and activity indicators. These periodograms are segmented by three FAP levels: 10\%, 1\%, and 0.1\%, each corresponding to a distinct colour band. The most significant peaks in each GLSP are individually highlighted. Adapted from \citet{Stoeva2026}, Figure 3, under a CC-BY 4.0 licence.
    }
    \label{fig:pgram_misleading}
\end{figure}

All of this suggests a cocktail of four signals, near: \SI{9.6}{\day}, \SI{32.3}{\day}, \SI{144.7}{\day}, and \SI{5900}{\day}; all at different timescales. All four signals manifest in RV, and only the second and the fourth should appear in activity indicators. The GSLPs of their time series, however, display a highly anomalous picture (Fig.~\ref{fig:pgram_misleading}). Even though the \SI{5900}{\day} cycle is unmistakably visible in the time series, it only manifests at around \SI{700}{\day} in the R$'_\text{HK}$ GLSP (Fig.~\ref{fig:pgram_misleading}f), and no significant powers are detected beyond \SI{200}{\day} in either RV or FWHM (Fig.~\ref{fig:pgram_misleading}b,d). Such discrepancies reinforce that all periodogram variants should be looked to solely as forms of exploratory data analysis, and not to draw solid conclusions.

\subsubsection{False alarm probability (FAP)}

In Sect.~\ref{sec:instrument_uncertainty}, we referred to an important aspect of measurement theory -- that scientific inference must be accompanied by a formal measure and propagation of uncertainty. In the case of periods identified in GLSPs, however, uncertainties on peak periods are typically not meaningful from the functional form of GLSP signatures, and have to be indirectly evaluated. \citet{VanderPlas2018} provides a detailed account on different ways to assess GLSP uncertainties, which we will now briefly go over.

A basic idea to assess the significance of a peak is to consider its power (or `height') relative to the those of spurious background peaks that arise in the periodogram. These background peaks should become comparable to the size of the sought peak in cases where time series have few or low-SNR observations. Therefore, this measure of significance should depend both on the number of measurements, as well as their signal-to-noise ratio. A typical metric that tracks GLSP-peak significance is the False Alarm Probability (FAP), which estimates the chance that a peak of similar magnitude would arise in a dataset with no underlying signal due to random fluctuations. \citet{Scargle1982} demonstrated that, for 
Gaussian noise, the non-normalised periodogram values follow a $\chi^2$ distribution with two degrees of freedom. Therefore, at any frequency $f_0$, the cumulative probability of observing a periodogram power smaller than $z$ is
\begin{equation}
    p_\text{single}(z)=1-\exp(-z)
\end{equation}
in data consisting only of Gaussian noise.

In uninformed RV searches of planets, however, we tend to be concerned not with the significance of some randomly selected $f_0$, but with the significance of the highest peak in the periodogram. This is a fundamentally different statistical scenario, for which \citet{VanderPlas2018} gives a good analogy with dice: the probability of rolling a high value increases significantly with the number of rolls, even if the probability is low for a single roll of a die. In this analogy, rolls are independent -- whereas GLSP powers at frequencies are correlated, so such a simplistic treatment is impossible. Various analytical and simulation-based methods have been developed to refine this (e.g. \citealp{Horne1986,Cumming2004}), though they all remain approximate.

The works associated with this dissertation have assumed the upper limit of the FAP, following \citet{Baluev2008}. The authors derived analytically that
\begin{equation}
    \text{FAP}(z)\lesssim
    1-p_\text{single(z)}\exp\left[
    -W(1-z)^{(N-4)/2}\sqrt{z}
    \right],
    \label{eq:fap_baluev}
\end{equation}
where \mbox{$W=f_\text{max}\sqrt{4\pi\operatorname{var}t}$} is an effective width of the observing window in units of the maximum chosen frequency $f_\text{max}$, and $N$ is the number of measurements. This is a useful analytic result that assumes alias-free periodograms (i.e. when window functions are weak; Sect.~\ref{sec:window_functions}), but it tends to hold well even in cases of realistic, non-uniform sampling of data.

Periodogram signals are typically classified by FAP into three categories: below 10\%, below 1\%, and below 0.1\%.\footnote{Note that this goes in increasing significance. More significant signals stand `taller' in GLSP power space, while at the same time decreasing in FAP space.}
Very generally, signals with FAP below 10\% may indicate structure and should not be dismissed, those below 1\% warrant close attention, and those below 0.1\% suggest a strong rationale for investigation and discussion. These thresholds suggest their relative urgency for analysis. They are not absolute criteria, and should be interpreted with caution. Finally, we conclude that FAP is not a test of planetary nature in itself, but rather, a convenient power-spectrum transformation that we use to quantify the significance of periodogram peaks in a way that inherently accounts for the number of observations and their SNR.

\subsubsection{Aliases and window functions}
\label{sec:window_functions}

It is common for observational data to be non-uniformly sampled, and instead shaped by practical constraints such as limited observing windows. Stars are typically only observable during a finite number of hours over a night, and often during specific months over a year. This introduces a sampling pattern that is sensitive to the inverse frequencies of \SI{1}{\day} and \SI{1}{\year}.\footnote{There may be other constraints that manifest subtly into the sampling patterns. Stars close to the ecliptic may be unobservable in cases of their lunar separation, which will unavoidably perturb the sampling in a systematic manner. Some instruments may be occasionally disconnected to their optical systems, which will too perturb sampling.}

\begin{figure}[t]
    \centering
    \includegraphics[width=\linewidth]{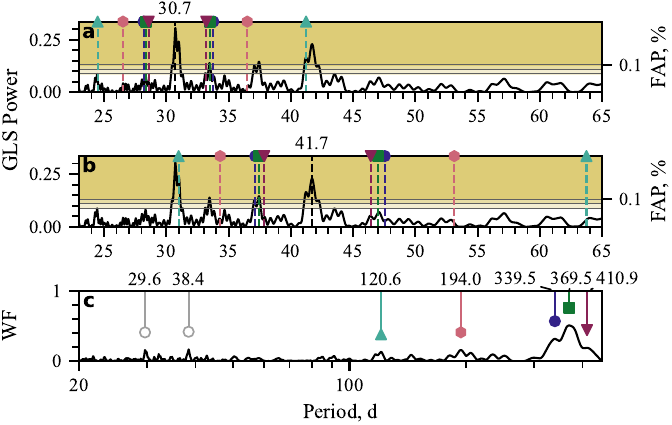}
    \caption{
    Alias analysis of the long-term detrended HARPS-N RVs of GJ~3998.
    We consider the aliases of the peaks at:
    (a)~\SI{30.7}{\day}, associated with $P_\text{rot}$,
    (b)~\SI{41.7}{\day}, associated with the planet GJ~3998~d.
    (c)~The WF reveals five prominent peaks at:
    \SI{120.6}{\day} (upward teal triangle),
    \SI{194.0}{\day} (pink hexagon),
    \SI{339.5}{\day} (blue circle),
    \SI{369.6}{\day} (green square), and
    \SI{410.9}{\day} (downward red triangle). Relevant aliases in data periodograms are plotted with dashed lines of a corresponding colour and marker. Several other strong peaks appear in WF, but they do not generate aliases within the defined period range (hollow grey circles). Adapted from \citet{Stefanov2025a}, Figure 2, under a CC-BY 4.0 licence.
    }
    \label{fig:alias_analysis}
\end{figure}

\citet{Dawson2010} explored in detail how aliases arise when a continuous signal is sampled at discrete intervals. Mathematically, the sampled signal is equivalent to the product of the original continuous function and a Dirac comb constructed by the non-uniform times of measurement (`the sampling function'). The periodogram of this sampled signal is therefore the convolution of the periodogram of the original signal and the periodogram of the sampling function (`the window function'). We therefore expect that every alias that appears in our periodograms is a common harmonic of a true signal $P_\text{true}$ and a window-function peak $P_\text{wf}$:
\begin{equation}
    P^{-1}_\text{alias} = \left|P^{-1}_\text{true}\pm P^{-1}_\text{wf}\right|.
    \label{eq:alias_equation}
\end{equation}
This allows us to check whether some of the significant peaks in our periodograms are aliases of one another, by computing \eqref{eq:alias_equation} with the period values of the most prominent WF peaks. Aliases are a good initial step to establish relationships between the strongest peaks in a periodogram. It is more straightforward to work with `families' of peaks that are known to be correlated, than with a collection of individual peaks which we a priori treat as independent. A very clear example of this is the signal structure of the RV time series of GJ~3998 between \SIrange{20}{60}{\day} (Fig.~\ref{fig:alias_analysis}). A previously identified \SI{41.7}{\day} RV signal had been interpreted as a \sfrac{1}{3}\,\unit{\year} alias of the rotational period \mbox{$P_\text{rot}\approx\SI{30.7}{\day}$} \citep{Affer2016}. However, following several additional years of observational coverage, \citet{Stefanov2025a} reported a more complex signal structure. It was subsequently revealed that the \SI{30.7}{\day} and the \SI{41.7}{\day} signals were independent, yet they generated aliases near each other through the common \sfrac{1}{3}\,\unit{\year} WF peak (Fig.~\ref{fig:alias_analysis}a,b; upward teal triangle). Figures~\ref{fig:alias_analysis}a and~\ref{fig:alias_analysis}b clearly illustrate the two distinct families of aliases that are superimposed within the aggregated GLSP, thereby simplifying the overall signal structure by a great extent. The independence of the \SI{41.7}{\day} signal from the known stellar-rotation period at \SI{30.7}{\day} prompted the analysis of the former signal, which ultimately led to its identification as a third planet in the GJ~3998 system \citep{Stefanov2025a}. Another illustrative example is the YZ~Ceti multi-planetary system. The orbital periods of its planets were a point of contention in the literature due to strong aliasing in the RV time series (e.g. \citealp{Astudillo-Defru2017,Robertson2018,Pichierri2019}). \citet{Stock2020} re-analysed the YZ~Ceti system and resolved the alias issues using a similar treatment, following \citet{Dawson2010}.

Like most periodograms, GLSPs are sensitive to the measurement uncertainty of points. Precise measurements carry greater statistical weight that imprecise ones; however, a precise RV measurement does not necessarily imply a precise measurement of activity. Consequently, aliases may manifest only in some physical dimensions. In the context of reducing false positives in detection, we should exercise caution when interpreting RV peaks at periods $P$ near a strong alias with the rotational period of the star -- which we may know from activity indicators -- even in the absence of a corresponding activity peak at $P$.

Finally, we remark that long-term trends (typically low-order polynomials) are decomposed in Fourier space as a series of significant powers at periods $P$ comparable to the baseline
\begin{equation}
    P_\text{baseline}=\operatorname{ptp}(t_i)
\end{equation}
of our measurements. These periods are usually so large that $P^{-1}_\text{true}$ approaches zero, and \eqref{eq:alias_equation} collapses to $P_\text{alias}=P_\text{wf}$. That is to say, a long-term trend is likely to inject a GLSP signal near a WF peak. At the same time, GLSP peaks near known WF peaks tend to be difficult to interpret in the absence of apparent trends in the time series. To determine whether such a signal truly exissts in the original continuous function, the best course of action, when feasible, would be to change the observational strategy, and examine how the structure of GLSP signals changes under the new WF.

\subsubsection{Stacked periodograms}
\label{sec:stacked_periodograms}
Stacked periodograms are a two-di\-men\-sion\-al generalisation of the one-di\-men\-sion\-al periodogram, in which the first dimension remains in inverse-frequency space, while the second dimension indicates the first $N$ points of the time series in the construction of the periodogram \citep{Mortier2017}. Extending from a small number of $N$ (typically a few dozen) to the full dataset builds an incremental GLSP-like structure, which reveals how the strength of a signal evolves as measurements accumulate.

From \eqref{eq:fap_baluev}, we can reason that the FAP as defined by \citet{Baluev2008} would decrease exponentially with $N$ for persistent signals. Since planets are expected to induce persistent planetary signals, their FAP should be monotonically decreasing with $N$ to first order. This leads to our first test of planetary nature -- if an RV signal has a FAP evolution driven primarily by a small subset of points in a stacked periodogram, it is not persistent and therefore unlikely to be of planetary origin.

\begin{figure}[t]
    \centering
    \includegraphics[width=\linewidth]{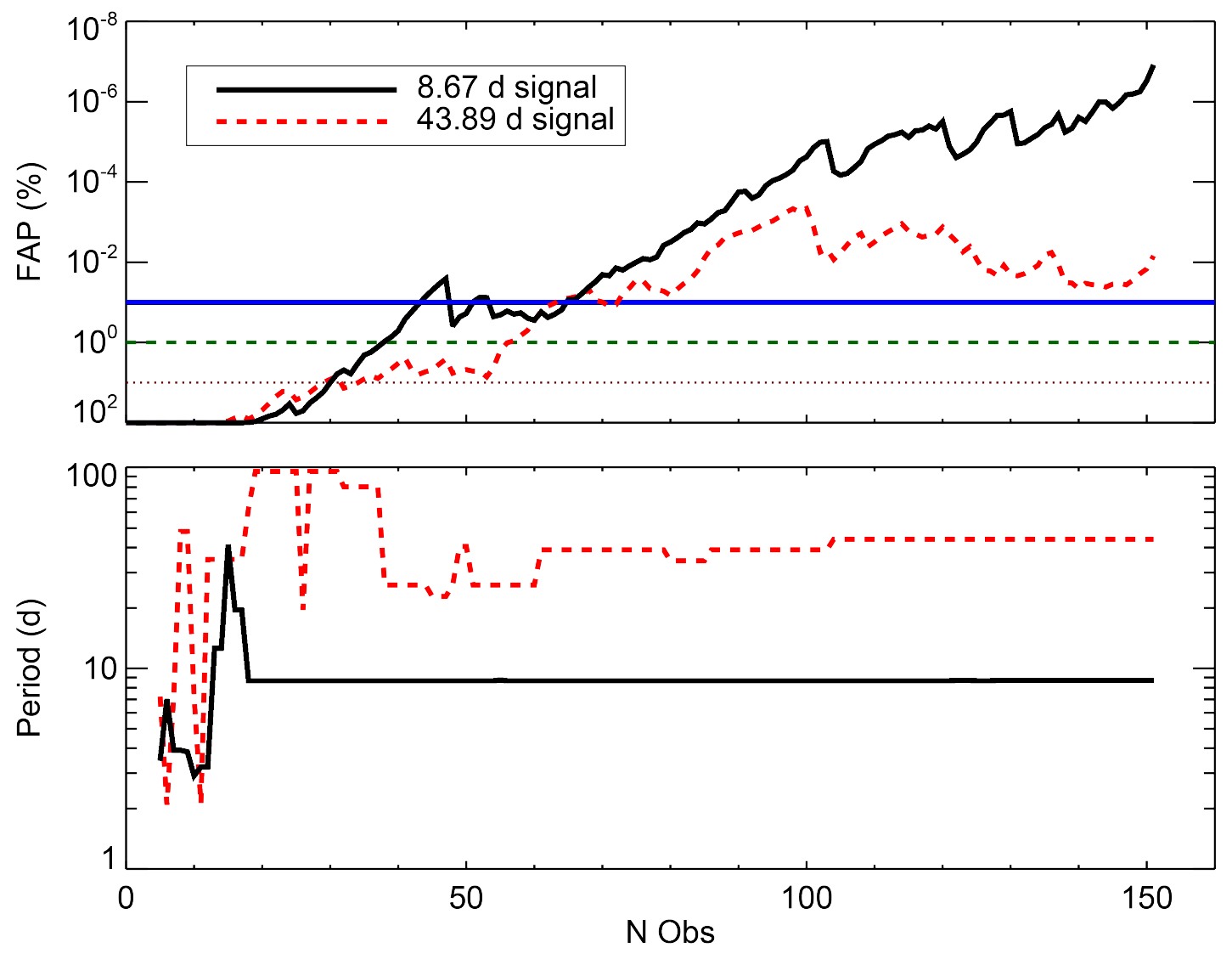}
    \caption{
    Top: FAP evolution of the \SI{8.67}{\day} (solid black; GJ~536~b) and the \SI{43.89}{\day} (red dashed; stellar-rotation period) signal in the HARPS and HARPS-N spectroscopy of GJ~536 in \citet{SuarezMascareno2017a}.
    Bottom: the best periodogram solution in terms of period.
    Adapted from \citet{SuarezMascareno2017a}, Figure 12, under a CC-BY 4.0 licence.
    }
    \label{fig:stacked_pgram}
\end{figure}

To illustrate this test, we revisit \citet{SuarezMascareno2017a} that discovered the first planet in the nearby multi-planetary system GJ~536. The authors identified two peaks in the RV time series: one at \SI{8.7}{\day}, corresponding to the soon-to-be identified GJ~536~b; and another at \SI{43.9}{\day}, which matched their measured stellar-rotation period
\mbox{$P_\text{rot}=\SI{45.39(1.33)}{\day}$}. To assess whether the \SI{8.7}{\day} signal was genuinely planetary, the authors examined how the FAP evolved over time for both isolated signals, as well as for the most prominent period when each signal was considered individually. They found that the \SI{8.7}{\day} signal increased steadily with time and remained stable in period once it became the most significant in the periodogram. In contrast, the \SI{43.9}{\day} signal exhibited more erratic behaviour and gradually lost significance in the final part of the test (Fig.~\ref{fig:stacked_pgram}).
This is an unambiguous example of a positive planetary result of a stacked-periodogram test.

\subsubsection{Rolling-window periodograms}
Stacked periodograms have been employed in many recent detection efforts (e.g. \citealp{SuarezMascareno2017a,Stock2020,Barnes2023,Cortes-Zuleta2025}), yet they inherently favour early measurements by definition. For instance, Fig.~\ref{fig:stacked_pgram} demonstrates how the strongest 
\SIrange{41.5}{41.9}{\day} RV signal evolves in FAP from the first 50 data points onwards. However, these first 50 points were always included in all subsquent samples that were assessed. Should any systematic effect be present in these early 
measurements, the stacked periodogram could run the risk of leading to an incorrect interpretation.

A solution of this issue is to replace the time-series subset, defined by the first $N$ points, with a window that encompasses all data points within a specified temporal interval. By progressively advancing this interval in time -- what we term a rolling window -- the subset changes, and allows us to 
detect shifts in persistent peaks and the emergence of new ones. This generalisation is referred to as a rolling-window periodogram. We remark that it bears a strong conceptual resemblance to the discrete short-term Fourier transform, as both aim to track the sinusoidal frequency of local measurement neighbourhoods, as they evolve over time.

\begin{figure}[t]
    \centering
    \includegraphics[width=\linewidth]{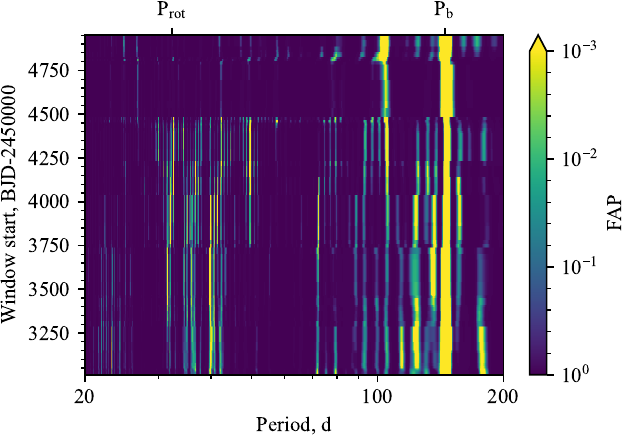}
    \caption{
    Rolling-window GLSP of the mean-subtracted HARPS RV time series of GJ~1137. A \SI{3000}{\day} window is advanced in steps of \SI{10}{\day}, beginning at the earliest available measurement and ending when the window encompasses the most recent observation. Time progresses upwards along the plot. The measured values of \citet{Stoeva2026} on the stellar-rotation period ($P_\text{rot}$; \mbox{$32.3^{+1.2}_{-1.3}\,\unit\day$}) and the orbital period of GJ~1137~b ($P_\text{b}$; \mbox{$\SI{144.720(0.029)}{\day}$}) are marked at the top.
    }
    \label{fig:rolling_pgram}
\end{figure}

Figure~\ref{fig:rolling_pgram} displays the rolling-periodogram of HARPS-N RV time series of GJ~1137 that was contemporaneous to \citet{Stoeva2026}.
This rolling-window GLSP spans over a period range of \SIrange{20}{200}{\day}, has a window length of \SI{3000}{\day}, and uses a step of \SI{10}{\day}. The start of the first window did not precede the oldest measurement point; neither did the end of the last window succeed the most recent measurement point.

GJ~1137~b exhibits a semi-amplitude of \SI{19.8(0.4)}{\metre\per\second} \citep{Stoeva2026}, significantly larger from the stellar-activity contribution in RV. As such, Figure~\ref{fig:rolling_pgram} is not particularly representative in the context of most searches among stellar activity, though it reveals a key insight: just as the RV signal of GJ~1137~b remains stable across all windows, any genuine planetary signal is expected to persistent over time. In contrast, stellar-activity signatures tend to have subtle drifts or oscillations (see $P_\text{rot}$ in Fig.~\ref{fig:rolling_pgram}). And so, we come to another test of planetary nature -- if an RV signal displays systematic variation in periodicity or lacks persistence across a rolling-window periodogram, it is unlikely to be of planetary origin.

Rolling-window periodograms benefit from relatively uniform sampling, and their performance degrades when gaps between observations exceed the chosen window width. Space-based photometric missions such as Kepler \citep{kepler}, TESS \citep{tess}, or PLATO \citep{plato} are particularly well suited to meet these requirements, as they tend to provide continuous and largely uninterrupted sampling.

\subsection{Bayesian-inference tests}\label{sec:bayesian_tests}
The family of periodogram tests, some of which were covered in Sect.~\ref{sec:pgram_tests}, was very relevant during the early decades of exoplanetary astronomy. The famous forerunner in RV detection, 51~Peg~b, was initially measured to have a semi-amplitude of \mbox{$k_\text{rv}=\SI{59(3)}{\metre\per\second}$}, a value that was exceptionally large even for the spectroscopic precision available at the time (\SI{13.5}{\metre\per\second}; \citealp{Mayor1995}).
A circular-planet signal of such strength is rather readily detectable using a periodogram approach. In fact, it would take another decade before an RV-discovered exoplanet with \mbox{$k_\text{rv}<\SI{10}{\metre\per\second}$} appeared for the first time (55~Cnc~e, \mbox{$k_\text{rv}=\SI{6.665(0.81)}{\metre\per\second}$}; \citealp{McArthur2004}).\footnote{
The earlier discovery of 47~UMa~c by \citet{Fischer2002} is now known to have broken the \SI{10}{\metre\per\second} barrier (\mbox{$k_\text{rv}=7.87^{+0.64}_{-0.57}\,\unit{\metre\per\second}$}; \citealp{Harada2025}), but was initially published with $k_\text{rv}=\SI{11.1(1.1)}{\metre\per\second}$.
}
The first planets with RV semi-amplitudes below \SI{1}{\metre\per\second} emerged another decade later, when \citet{Pepe2011} discovered
HD~20794~b and
HD~20794~d
(with $k_\text{rv}$ values of
\SI{0.83(0.09)}{\metre\per\second} and
\SI{0.85(0.10)}{\metre\per\second}, respectively).\footnote{\citet{Pepe2011} also published HD~20794~c (\mbox{$k_\text{rv}=\SI{0.56(0.10)}{\metre\per\second}$}), which was subsequently questioned by \citet{Feng2017} on account of insufficient Bayesian evidence.}
Today, in our decade, we have reached only 17 RV detections with sub-\unit{\metre\per\second} semi-amplitudes that have not been challenged. All of them reside in multi-planetary systems,\footnote{These are:
Barnard's star,
GJ~887,
HD~134606,
HD~20781,
HD~20794,
HD~215152,
HD~96700,
Proxima, and
Teegarden's star.}
where we have to mind the presence of other planets; and 12 of them orbit around late-type dwarfs, where we know stellar activity is an issue at this level (Sect.~\ref{sec:cool_dwarf_activity}). With such needles as thin as the straws in the haystack, periodograms can no longer serve as a robust analytical tool and should be regarded as merely an initial, exploratory part of the analysis.

The professional community recognised that a fundamentally new methodology was required to formally detect and characterise such weak 
signals, and turned an eye to Bayesian inference. This approach 
builds from the ordinary decomposition of sines that periodograms enable, and extends it to modelling of arbitrary complexity and parameter dimensionality, provided only that a scalar likelihood function over the parameter space exists. What is more, Bayesian inference permits the incorporation of different prior knowledge, quantifies parameter uncertainties, and enables hypothesis testing through the notion of Bayesian evidence. Several early works in the 2000s developed MCMC methods for the analysis of RVs and the search for periodicities (e.g. \citealp{Gregory2007}). In these approaches, the number of planets in the model became another variable, and marginal likelihoods began to be treated as central quantities. The advent of high-precision instruments, most notably HARPS \citep{harps} and HIRES \citep{hires}, exposed the community to RV signals of much smaller semi-amplitudes, and naturally called for model-comparison approaches. As a result, many of the first Bayesian re-analyses of known exoplanetary systems were carried out during this period (e.g. \citealp{Tuomi2009,Feroz2011a,Feroz2011,Tuomi2011}). This marked a gradual shift that went beyond a mere technical improvement -- it represented a fundamental change of philosophy in how we interpret observational data. Exoplanetary detection ceased to be seen as a binary outcome -- instead, it became an event of quantifiable confidence, which we could assess through the Bayes factor (Sect.~\ref{sec:bayes_factor}).

We refer to \citet{Trotta2008} for a comprehensive overview of Bayesian inference in astrophysics. We will now introduce the idea of nested sampling, which has been central to our implementation of Bayesian inference in the works of this dissertation \citep{Stefanov2025,Stefanov2025b,Stefanov2025a}.
Then, we will turn to the central idea of the Bayes factor, and assert its pre-eminence as the `king' of signal-detection tests. Finally, we will cover briefly other useful tests that are enabled by the Bayesian-inference framework.

\subsubsection{Nested sampling}\label{sec:nested_sampling}
The Bayes' theorem states that the probability of some hypothesis $\mathcal{H}$ to hold, given some evidence $\mathcal{R}$, is the probability of $\mathcal{H}$ to hold, multiplied by the probability of observing $\mathcal{E}$, given $\mathcal{H}$, and normalised by the probability of observing $\mathcal{E}$ whatsoever:
\begin{equation}
    p(\mathcal{H}|\mathcal{E}) = \frac{p(\mathcal{H})p(\mathcal{E}|\mathcal{H})}{p(\mathcal{E})}.
    \label{eq:bayes_theorem}
\end{equation}
If $\mathcal{H}$ can be mathematically described by a model that is parametrised by a vector $\vb*{\theta}$ within a parameter space $\Theta$, the Bayes' theorem takes the form
\begin{equation}
    \mathcal{P}(\vb*{\theta}) = \frac{\pi(\vb*{\theta})\mathcal{L}(\vb*{\theta})}{\mathcal{Z}},
    \label{eq:bayes_model}
\end{equation}
where
$\pi(\vb*{\theta})$ and $\mathcal{P}(\vb*{\theta})$ are the prior and posterior distribution of parameters, respectively;
$\mathcal{L}(\vb*{\theta})$ is the likelihood function that maps $\vb*{\theta}$ onto a scalar space; and $\mathcal{Z}$ is the probability of observing the evidence:
\begin{equation}
    \mathcal{Z} = \int_\Theta \pi(\vb*{\theta})\mathcal{L}(\vb*{\theta})\dd\vb*{\theta},
    \label{eq:bayes_evidence}
\end{equation}
and is typically referred to as `the evidence' for brevity.

The usual approach to evaluating $\mathcal{P}(\vb*{\theta})$ in \eqref{eq:bayes_model} is to treat the product $\pi(\vb*{\theta})\mathcal{L}(\vb*{\theta})$ as a scalar function defined over $\Theta$. This function is referred to as the prior-transformed likelihood, and is denoted $\mathcal{L}_\pi(\vb*{\theta})$. From here, one could try to sample from $\mathcal{L}_\pi(\vb*{\theta})$ directly. This is the foundational principle behind all Markov Chain Monte Carlo (MCMC) methods, including the Metropolis-Hastings algorithm \citep{Metropolis1953,Hastings1970}, Gibbs sampling \citep{Geman1984}, and Hamiltonian Monte Carlo \citep{Duane1987}. However, navigating through $\mathcal{L}_\pi(\vb*{\theta})$ is difficult, and sampling directly from it becomes progressively more computationally expensive.

\citet{Skilling2004} observed that the uniform sampling within the region
\mbox{$\mathcal{L}_\pi(\vb*{\theta}) > k$} for a given threshold $k$ is a much lighter task than sampling from $\mathcal{L}_\pi(\vb*{\theta})$. Moreover, by iteratively increasing $k$ until some convergence criterion, one can effectively approximate the posterior distribution. This idea constitutes the core of nested sampling. Rather than following a single trajectory along $\mathcal{L}_\pi(\vb*{\theta})$, nested sampling partitions $\Theta$ into many discrete regions, defined by local $k$. In doing so, this algorithm solves a sequence of many simpler problems, and enables the exploration of multi-modal distributions of arbitrarily complex shapes. The latter circumstance makes nested sampling particularly advantageous over traditional MCMC methods.

\subsubsection{Bayes factors}\label{sec:bayes_factor}
A further benefit of nested sampling is its ability to numerically evaluate the evidence $\mathcal{Z}$ during the sampling process. This is especially valuable for the purposes of exoplanet detection, as illustrated below. Consider a model that includes a planetary signal, and that this model is characterised by a parameter vector $\vb*{\theta}_1$, a prior distribution $\pi_1$, and a likelihood function $\mathcal{L}$. Against that model, consider a planet-free model that is characterised by
$\vb*{\theta}_0 \subset \vb*{\theta}_1$,
$\pi_0 \subset \pi_1$, and the same $\mathcal{L}$.\footnote{This assumption is realised in RV detection through setting \mbox{$k_\text{rv}=0$} and excluding all relevant orbital parameters from $\Theta$.} The Bayes factor is defined as
\begin{equation}
    K = \frac{\mathcal{Z}_1}{\mathcal{Z}_0},
    \label{eq:bayes_factor}
\end{equation}
where $\mathcal{Z}_1$ and $\mathcal{Z}_0$ are the evidences for the models that include and exclude, respectively, the planetary signal. The Bayes factor $K$ quantifies by how much the observational data favour one model over the other. As such, $K$ provides a direct means of testing a null hypothesis \citep{Kass1995,Trotta2008,Morey2016}.

This makes $K$ a fundamental benchmark for exoplanetary detection. Unlike other tests that rely on arbitrary model specifications (such as FAP-informed thresholds, where the very definition of FAP may vary), $K$ compares model frameworks themselves. Bayes factors account for uncertainties in parameter estimation through the nested-sampling process, and inherently penalise overparametrisation, thereby mitigating risks of overfitting. By  formally comparing a model with a planetary signal to its planet-free counterpart (i.e. the null hypothesis), $K$ directly quantifies the evidence for a presence of a planetary signal, as suggested by observations. Although our rationale for $K$ assumed RV measurements, this dependency is not essential -- the mathematical model that describes $\mathcal{H}$ can encompass RVs, transit photometry, other detection techniques, or their combination -- provided that the likelihood function $\mathcal{L}(\vb*{\theta})$ is sufficiently well-behaved. As a result, the Bayes factor $K$ remains a very robust and scientifically defensible metric for asserting signal detection. In this light, it is reasonable to regard the Bayes factor as the `king' metric among signal-detection tests. We recommend \citet{Morey2016}, which provides a valuable philosophical perspective on the Bayes factor, and largely influenced the intuitive understanding of the inference approaches used in this dissertation.

In astrophysics, nested-sampling software often integrates \eqref{eq:bayes_evidence} over enormous parameter spaces $\Theta$. As such, $\mathcal{L}$ and $\mathcal{Z}$ are often minute, and nested-sampling approaches often report their natural logarithms to alleviate numerical issues (e.g. \citealp{multinest,dynesty,ultranest}). Consequently, the Bayes factor is also typically given under a natural-logarithm transform $\ln K$ (or $\Delta\ln\mathcal{Z}$ in the literature). In exoplanetary science in particular, the $\ln K$ threshold that constitutes an RV detection is not derived from first principles. \citet{Jeffreys1939} proposed interpretive categories for Bayes factor, which we summarise in Table~\ref{tab:bayes_factor}.
\begin{table}
\centering
\caption{
Interpretation of the Bayes factor $K$ by \citet{Jeffreys1939}, but rearranged so that positive $\ln K$ challenges the null hypothesis. The values of $\ln K$ intervals are approximate.
}
\vspace{4pt}
\label{tab:bayes_factor}
\begin{tabular}{ccl}
\hline\hline
\multicolumn{1}{c}{$\ln K$} &
\multicolumn{1}{c}{Strength of evidence} \\ \hline
$(-\infty,0]$ & Negative (supports null hypothesis) \\
$(0,1.15]$ & Not worth more than a mention \\
$(1.15,2.30]$ & Substantial \\
$(2.30,3.45]$ & Strong \\
$(3.45,4.61]$ & Very strong \\
$(4.61,+\infty)$ & Decisive \\ \hline
\end{tabular}
\end{table}
These were heuristic guidelines that were later adapted to serve the purposes of the RV-detection field. Typically, \mbox{$\ln K > 5$} is accepted as a robust detection, and would correspond to a near 150-fold preference for the planetary model over the non-planetary one.\footnote{This comes from $\exp(5)\approx 150$.}
This criterion is continuously being re-affirmed in the literature, though variations with other thresholds \mbox{$\ln K>\alpha$} occasionally occur across research consortia or even within individual scientific group. There is now a growing recognition that $\ln K$ is highly sensitive to the choice of priors (e.g. \citealp{Jenkins2011,Tuomi2012a,Hara2022,LLorente2023}), and that the numerical evaluations of $\ln Z$ usually have a non-negligible uncertainty (e.g. \citealp{Jenkins2011,Nelson2020}). This aleatoric uncertainty, combined with the prior dependence, reinforces that \mbox{$\ln K>\alpha$} should not be seen as a universal law for any $\alpha$, and thus the role of scientific judgement remains central.

Finally, we must be very candid about three fundamental details in the application of Bayes factors. First, $\ln K$ enables only a comparative assessment between two models. This quantity provides no information about the absolute strength of evidence of any single model. Instead, it expresses the relative support for one model over another, given the observed evidence. Second, $\ln K$ is only indicative when both models are fitted to the same dataset, and explore the same domain $\Theta$. Every new data point perturbs the likelihood function $\mathcal{L}$, and consequently alters the evaluation of the integral in \eqref{eq:bayes_evidence}, thereby influencing $\ln K$. In this sense, comparing models fitted on different datasets is meaningless. Third, we have been using the term `planetary model' implicitly to denote a model that includes a planetary-like signal in RVs. A high $\ln K$ in favour of such a model implies that such a signal is present in the data, not that it is of planetary origin. This distinction is crucial: \mbox{$\ln K >\alpha$} is a necessary, but insufficient criterion for planetary detection. Formal confirmation of planetary nature must go beyond $\ln K$ and include other independent tests to ensure that the observed signal is not a by-product of noise, aliasing, or alternative astrophysical phenomena. The next few sections are dedicated to some planetary-nature tests that Bayesian inference enables.

\subsubsection{Model-grid searches}

Indeed, \eqref{eq:bayes_factor} shows that we cannot find the best model globally, but only the better one between a pair. What we can do, however, is to evaluate a large collection of models, and, by virtue of comparison, focus on those with the most positive $\ln \mathcal{Z}$. Naïvely, examining many models rather than just two increases the probability of selecting the `ground-truth' model. There is another good reason to consider a collection of models -- by introducing small variations in each model (e.g. going from circular to eccentric orbits), we can see whether observational evidence supports these particular complexities. This provides additional insight into the properties of the planetary signal, should there be one, or into the stellar-activity behaviour.

Usually, many modelling pathways come to us when we are faced with a given time series. For example, as previously noted, planetary signals can be assumed circular, or non-circular. For compact late-dwarf systems, dynamical planet-planet interactions may also be relevant, and may be accounted for. Stellar activity may not necessarily be well captured by a quasi-periodic covariance structure -- instead, its covariance might be mostly periodic, mostly exponential, or have another functional form. In fact, the SEP kernel itself, as described in \eqref{eq:gp_sep_kernel}, can be approximated by different kernels, each of which can be tested. All these methodological variations build a model grid in $N$-dimensional space. After the inference each model in this grid, all values of $\ln\mathcal{Z}$ can be explored to identify the most appropriate model for the given scientific case.

\begin{figure}
    \centering
    \includegraphics[width=0.8\linewidth]{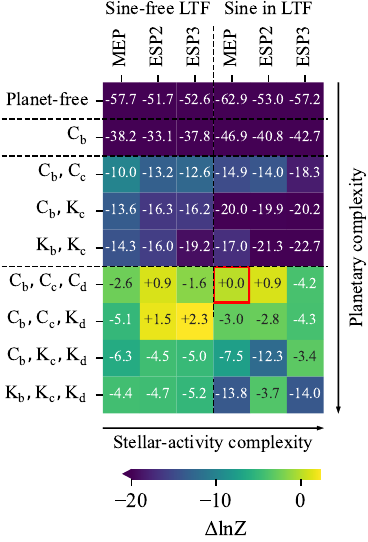}
    \caption{
    Bayesian-evidence comparison in the model grid of GJ~3998. The grid explores different planetary configurations and stellar-activity kernels, as well as for a potential presence of a sine component in the long-term function. Planetary configurations include planets with circular (C; \mbox{$e=0$}) or Keplerian (K; \mbox{$e\geq 0$}) orbits. Stellar-activity configurations include different approximations of the SEP kernel: the Matérn \sfrac{3}{2} exponential periodic kernel (MEP) and the exponential-sine periodic kernel (ESP) with two or three harmonics \citep{spleaf1,spleaf2}. The model that was eventually chosen for analysis assumed a sine in the LTF, three circular orbits and a MEP kernel (bordered in red; \mbox{$\ln\mathcal{Z} = -1114.4$}). We give the Bayes factor $\Delta\ln\mathcal{Z}$ of remaining models relative to this model. Adapted from \citet{Stefanov2025a}, Figure 4, under a CC-BY 4.0 licence.
    }
    \label{fig:model_table}
\end{figure}

In answering the question whether GJ~3998~d is a by-product of stellar activity, as thought earlier, or a genuine planet, \citet{Stefanov2025a} employed a model grid as well. Figure~\ref{fig:model_table} displays a slice of this grid, and serves as our example. 
The exploratory analysis of the authors through periodograms was inconclusive regarding the presence of GJ~3998~d. Their periodograms, however, revealed strong powers in RV- and activity indicators in the \SIrange{400}{700}{\day} interval, which could indicate a potential long-term magnetic cycle. It could, in principle, have also been possible that this magnetic-cycle contribution manifested back to \SI{42}{\day} by virtue of a WF peak via \eqref{eq:alias_equation}. To address this, this work restricted itself to cases where eccentricity was monotonically increasing from inner to outer planets. This resulted in nine distinct configurations. At the same time, \citet{Stefanov2025a} recognised that they could independently vary the assumptions on the magnetic cycle (with or without sine contribution, i.e. two configurations) and the complexity of stellar activity (using three approximations of the SEP kernel). This naturally gave rise to a \numproduct{9x2x3} grid.

With such a grid, many tests become available. For example, the question of whether GJ~3998~d is detected no longer depends on computing a Bayes factor between two models alone -- this can now be addressed by comparing $\ln Z$ across family of models. For example, \citet{Stefanov2025a} and \citet{Lamontagne2026} worked with an `average improvement in evidence'
\begin{equation}
    (\ln K)_\text{avg} =
    \operatorname{avg}(\ln\mathcal{Z_\text{B}}) -
    \operatorname{avg}(\ln\mathcal{Z_\text{A}}),
\end{equation}
as well as a `pessimistic improvement in evidence'
\begin{equation}
    (\ln K)_\text{pes} =
    \operatorname{min}(\ln\mathcal{Z_\text{B}}) -
    \operatorname{max}(\ln\mathcal{Z_\text{A}}).
\end{equation}
The operators that are used on such model-family comparisons are reserved to the judgement of their scientific users, and we note that they can reflect different degrees of conservativeness. For example, $(\ln K)_\text{pes}\leq (\ln K)_\text{avg}$ by construction, and the difference between them can be to the order of a few units (e.g. \citealp{Stefanov2025a,Lamontagne2026}).

Bayes-factor comparisons on families of models offer another advantage. Since $\ln\mathcal{Z}$ carries aleatoric uncertainty from numerical integration, most operators applied to the $\ln\mathcal{Z}$ values of a model family should effectively suppress the impact of noise on the Bayes factor. In this sense, such comparisons can be argued to have a higher SNR. From here, it follows that the grid can allow us to estimate the typical error in $\ln\mathcal{Z}$. For instance, in Figure~\ref{fig:model_table}, there does not seem to be a strong preference for any particular GP kernel. If the systematic preference for any given kernel is so weak that it is overwhelmed by noise, the rms of $\ln\mathcal{Z}$ values itself can provide an upper bound on the aleatoric uncertainty in $\ln\mathcal{Z}$ -- even though the grid never repeats the same model configuration.

\subsubsection{Modelling activity signals near planetary periods}\label{sec:pseudo_planets}
Tests of planetary nature typically assume that planets manifest themselves in RV only, and not in activity indicators. However, real cases are seldom straightforward. For example, it is not uncommon for an alias or a harmonic of the stellar rotational period $P_\text{rot}$ to appear near the period of a genuine planetary signal, even though they are not caused by the same physical process (e.g. \citealp{Affer2016,Burt2021,Kane2024,Rescigno2024a,SuarezMascareno2024,Stefanov2025a,Lamontagne2026}). At the same time, it is very realistic that a stellar-activity component induces RV and activity signals, but our GP modelling somehow does not recognise as stellar activity. In this second case, an RV signal will have gone astray in the residual time series, it would stand tall in the residual periodogram, and could be misread as planetary.

The following test tries to address this issue. Suppose there is an excess signal of period $P$ that appears in RV- as well as activity-indicator time series, which our model fails to recognise as stellar activity. In such a case, if our model includes a planetary signal (e.g. in an uninformed RV search), it would erroneously guide its planetary-period parameter to $P$. However, because this excess signal also appears in activity, a similar erroneous result would occur if we instead were to fit an additional sine-like (or Keplerian) signal to the activity data, as if the stellar-activity measurements were being perturbed by a planet. In both scenarios -- the planetary signal in RV time series or a `pseudo-planet' signal in activity time series -- the period $P$ would be incorrectly assigned. The test therefore consists of fitting a planet-like signal in the activity-indicator time series, using the same prior distribution as in the standard RV search. If a well-defined period (or other associated parameters) is recovered in both the RV- and the activity time series, then this likely indicates that the signal originated in stellar activity.

\begin{figure}[t]
    \centering
    \includegraphics[width=4in]{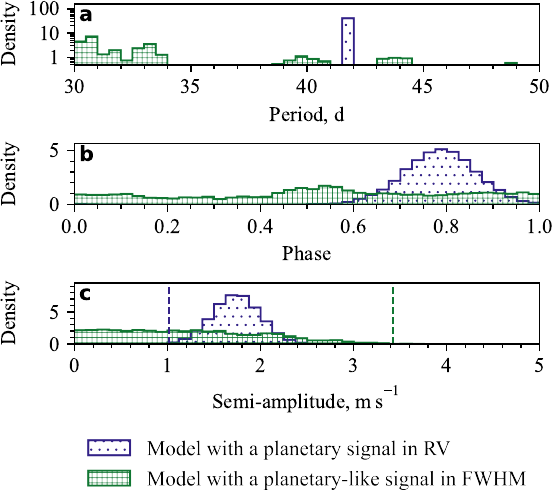}
    \caption{
    Posteriors of GJ~3998~d of the most appropriate model of the GJ~3998 multi-planetary system (blue dot-hatched) against those of an equivalent model that modelled three planet-like sine terms in the FWHM time series (green square-hatched). Adapted from \citet{Stefanov2025a}, Figure 10, under a CC-BY 4.0 licence.
    }
    \label{fig:pseudo_planets}
\end{figure}

Figure~\ref{fig:pseudo_planets} illustrates this test being applied on GJ~3998~d \citep{Stefanov2025a}.
One begins with their most appropriate model with the planetary signal included, and then adjusts the modelling so that the planetary signal in question is modelled in activity instead. Figure~\ref{fig:pseudo_planets} shows a poor, inconsistent agreement between the posteriors of those models. We remark that such an inconsistency implies that the planetary-nature test has passed for that RV signal -- as it means there are no activity signals that agree in period, phase, or strength with the RV signal of GJ~3998~d.

In another variant of this test, instead of modelling a planetary-like signal in activity, a `two-dimensional' signal is fitted in both RV and activity, with a common period $P_\text{common}$ (e.g. \citealp{SuarezMascareno2025a,Stoeva2026,Nari2026,Nari2026a}). In this variant, researchers tend to rely on the strong RV signature to guide inference at $P_\text{common}$, and from there, to examine whether the semi-amplitude in activity is significantly different from zero. While this adjustment may seem minor, it introduces a subtlety. In the original test, we test whether an activity signal can be detected as robustly as an RV signal that we suspect to be a planet. In this variant, we test whether an activity signal -- forced to share a period with an RV analogue -- has a significant semi-amplitude. Experience shows that it is best to perform both tests, as they seem to address the same question in two slightly different approaches.

It bears attention that Bayes factors can also be used for this test and its aforementioned variant. In the original test, three models can be considered:
(i)~a planet-free model, with an associated $\ln\mathcal{Z}_0$;
(ii)~a model with a planetary RV signal, with an associated $\ln\mathcal{Z}_\text{RV}$; and 
(iii)~a model with an planetary-like activity-indicator signal, with an associated $\ln\mathcal{Z}_\text{AI}$.
The comparison between
\mbox{$(\ln\mathcal{Z}_\text{RV}-\ln\mathcal{Z}_0)$} and
\mbox{$(\ln\mathcal{Z}_\text{AI}-\ln\mathcal{Z}_0)$} tells how robust the activity-indicator signal is with respect to the planetary candidate on which we are performing the test. In the variant of the test, the comparison \mbox{$(\ln\mathcal{Z}_\text{RV\&AI}-\ln\mathcal{Z}_\text{RV})$} indicates whether the evidence favours the presence of an activity signal at the exact same period as the RV signal.

A weakness of this family of tests lies in its underlying assumption that planetary signals manifest only in RV data, and not in activity indicators. This may not necessarily be the case. For instance, during star-planet interaction, activity signals can emerge at the orbital (or synodic) period of the planet (e.g. \citealp{ZapateroOsorio2026}).

\subsubsection{False inclusion probability (FIP)}
Some tests are not based on the likelihood integral over parameter space, associated with $\mathcal{Z}$, but on the individual likelihoods of parameter-space vectors that form the final posterior distributions. Such is the case with \citet{Hara2022}, who used the posterior distribution of an inference run to compute the probability of having no planets within a given element in the orbital-frequency space. The authors refer to this quantity as the false-inclusion probability (FIP), and suggest the threshold \mbox{FIP<\num{e-2}} for potential planetary candidates. This test of planetary nature passes if the FIP inequality is satisfied.

In a certain sense, the FIP test is similar to the general Bayes-factor criterion test \mbox{$\ln K>\alpha$} that was discussed in Sect.~\ref{sec:bayes_factor}, but with a different programmatic realisation. A FIP test of $N$ planets over a period domain $\mathbb{P}$ requires the inference of the most appropriate model with $N$ planets, with their orbital-period priors fully encompassing $\mathbb{P}$.\footnote{This does not run into the risk of introducing degeneracies, because the FIP test works at a stage where the parameter space has been already marginalised \citep{Hara2022}.}
The computation proceeds in angular-frequency space with steps of $\Delta\omega=2\pi/nT_\text{baseline}$, with $n$ typically being a natural number between two and five. For each $\omega$, the method evaluates how much posterior probability mass is assigned to signals with periods near that value. This constitutes the true inclusion probability (TIP) -- the frequency that at least one planet resides there in angular-frequency space. FIP is then calculated as the complement of TIP to unity: \mbox{$\text{FIP}=1-\text{TIP}$}.

\begin{figure}
    \centering
    \includegraphics[width=\linewidth]{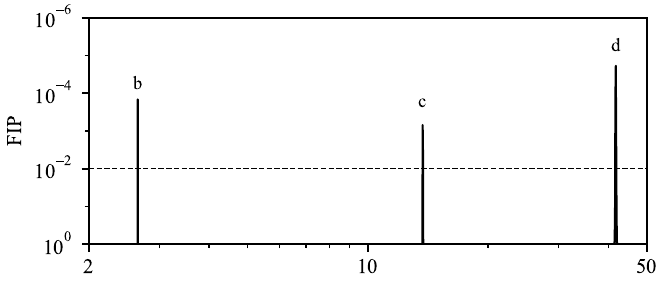}
    \caption{
    FIP-period plot of the selected model in \citet{Stefanov2025a} (solid black line) against the \num{e-2} threshold suggested by \citet{Hara2022} (dashed black line). Adapted from \citet{Stefanov2025a}, Figure 9, under a CC-BY 4.0 licence.
    }
    \label{fig:fip_test}
\end{figure}

Figure~\ref{fig:fip_test} illustrates the FIP test of GJ~3998 as an example.
FIP tests can be viewed as the Bayesian-inference analogues of FAP tests, and are in the process of becoming widely adopted in RV-detection studies (e.g. \citealp{AnnaJohn2023,Standing2023,SuarezMascareno2023,GonzalezHernandez2024,Heidari2024,Baycroft2025}). We would like to reiterate that these tests come directly from likelihood function that is associated with the selected model. As a result, systematics that are inherent to this model will be carried over into the test.

\subsubsection{Apodisation of planetary signals}\label{sec:apodisation}

Planets produce a continuous and uninterrupted signal in stellar RV time series, with a constant semi-amplitude over time. In contrast, stellar activity often comes in bursts or otherwise localised signals. A straightforward test to assess whether an RV signal is localised involves adding a time-variant, Gaussian-like component to the RV semi-amplitude of a planetary candidate:
\begin{equation}
    k_\text{rv}(t) = k\cdot\exp\left[
    \frac{(t-\mu_\text{apo})^2}{2\sigma_\text{apo}^2}
    \right].
    \label{eq:apodisation}
\end{equation}

The transformation described in \eqref{eq:apodisation} is known as the apodisation\footnote{From Greek \emph{apodos}, `footless'.}
of Keplerian signals \citep{Gregory2016,Hara2022a}. It is a remarkably simple test for the localisation of an RV signal over the baseline, and requires only two additional parameters: $\mu_\text{apo}$ and $\sigma_\text{apo}$. In an analogous manner to the modelling of activity signals near suspected planetary periods (Sect.~\ref{sec:pseudo_planets}), the test of planetary nature passes if $\mu_\text{apo}$ and $\sigma_\text{apo}$ are poorly defined. A true planetary signal should have $k_\text{rv}(t)=k$, which occurs if and only if $\sigma_\text{apo}$ is comparably larger than the baseline $T_\text{baseline}$. In this case, $\mu_\text{apo}$ would be poorly defined, and its prior distribution would not be overwhelmed by evidence. On the contrary, a well-defined $\mu_\text{apo}$ and $\sigma_\text{apo}\lesssim T_\text{baseline}$ tell of a localised non-planetary signal.

\begin{figure}
    \centering
    \includegraphics[width=\linewidth]{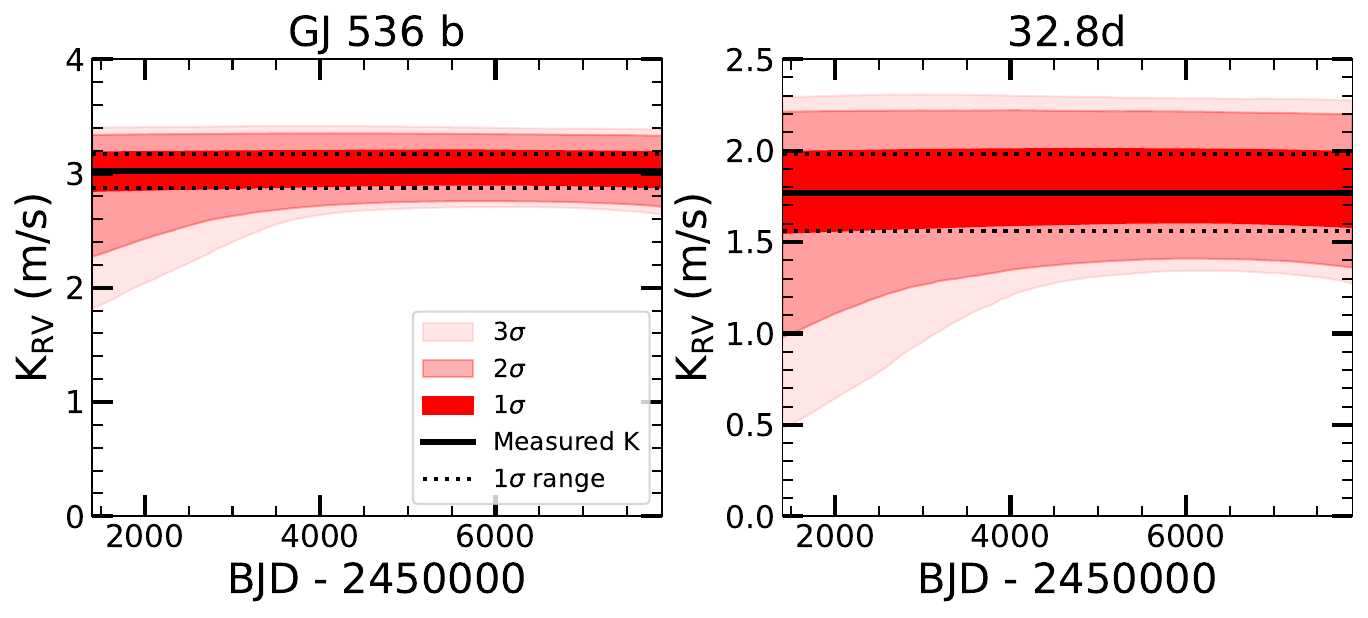}
    \caption{
    The results of an apodised-Keplerian test on the velocimetry of GJ~536. The evolution of $k_\text{rv}(t)$ of the planet GJ~536~b (left) and a \SI{32.8}{\day} signal (right) are given in shaded red bands, which are propagated directly from the posterior distributions of $\mu_\text{apo}$ and $\sigma_\text{apo}$. Both signals pass the apodisation test, and the \SI{32.8}{\day} is eventually announced as a new planet GJ~536~c. Adapted from \citet{SuarezMascareno2025a}, Figure 4, under a CC-BY 4.0 licence.
    }
    \label{fig:apodised_test}
\end{figure}

We illustrate the apodisation test of planetary nature through the case of GJ~536. In a second instance of reduction and analysis, \citet{SuarezMascareno2025a} updated the orbital parameters of the known planet GJ~536~b, and discovered a strong RV signal in \SI{32.8}{\day}. To assess the nature of this signal, the authors performed a series of tests, including the apodisation one. For both GJ~536~b and the new signal, they found that $\mu_\text{apo}$ is undefined, and that \mbox{$\sigma_\text{apo}>18T_\text{baseline}$}. This constitutes a clear pass of the apodisation test, which carried weight in the publication of \SI{32.8}{\day} as a new planet, GJ~536~c \citep{SuarezMascareno2025a}. In addition, the authors use their $\mu_\text{apo},\sigma_\text{apo}$ posteriors to propagate $k(t)$ and its uncertainty over their baseline (Fig.~\ref{fig:apodised_test}).
Such a representation of the evolution of $k(t)$ is more interpretable and convincing than a two-dimensional posterior slice. However, low-likelihood regions that were eventually rejected by sampling may be in the space of low $\sigma_\text{apo}$, especially for time series that were not uniformly sampled. This could introduce a fraction of spurious, localised sinusoidals that that can inflate the confidence intervals of $k(t)$ (e.g. near \SI{2452000}{\julianday}; right panel of Fig.~\ref{fig:apodised_test}). We caution that such features should not be interpreted as strong anti-planetary evidence, and that the posterior distributions of $\mu_\text{apo},\sigma_\text{apo}$ tend to be more robust.

Unsurprisingly, Bayes factors can too be taken into account when doing the apodisation test (e.g. \citealp{Stoeva2026}). The $\ln\mathcal{Z}$ difference from a model with time-invariant semi-amplitude \mbox{$k_\text{rv}(t)=k$} to a model with an apodised $k_\text{rv}(t)$ states directly whether the evidence favours a non-localised RV signal. In this context, the test of planetary nature passes if the resulting $\Delta\ln\mathcal{Z}$ is not overwhelmingly positive; or is, ideally, negative. Other parametrisations of $k_\text{rv}(t)$ also exist, for example by multiplying $k$ by a rectangular function (e.g. \citealp{Nari2026}).

\subsubsection{Detection limits for additional companions}

When we set out to hunt for exoplanets, we often return with one or more planetary discoveries, much like trophies from a distant realm.\footnote{And, occasionally, we return with nothing!-- see Chapter~\ref{chap:gj526}.}
However, it is rarely very certain that we were able to capture every planet in a system. The precision of our observations, our observational-sampling strategy, the influence of stellar activity and its modelling, among other circumstances, all affect how well we can detect planets. For this reason, many studies on exoplanet searches include assessments of their sensitivity to undetected planets. These efforts have produced a family of tests that are not of planetary nature, but instead provide a `detection limit' of a certain study, based on its modelling approach and the available data at hand.

One common detection-limit approach, known as an injection-recovery test, is to inject a synthetic circular-planet signal in the RV data at an arbitrary period $P_\text{syn}$, semi-amplitude $k_\text{syn}$, and phase. Then, the selected model for analysis is re-fitted to these modified time series, not by performing Bayesian inference anew, but through the median posterior distribution from said inference. The model may adopt a portion or the entirety of the synthetic signal as stellar-activity, for example. Alternatively, the model may not recognise the synthetic signal at all, which would lead to a sharp peak in the periodogram of the residual time series. By injecting many of these synthetic signals, one at a time, we can assess how often the signal is recovered in the $P_\text{syn}$--$k_\text{syn}$ space. The definition of recovery can stem from any aggregate statistic of the injection signal after fitting (e.g. \citealp{Konacki2009,Howard2016}). In recent times, the FAP of the strongest peak tends to be popular choice for such a statistic (e.g. \citealp{Cloutier2021,Laliotis2023,GonzalezHernandez2024,Maciejewski2023,Cortes-Zuleta2025,Osborn2026}, among many others).

\begin{figure}
    \centering
    \includegraphics[width=\linewidth]{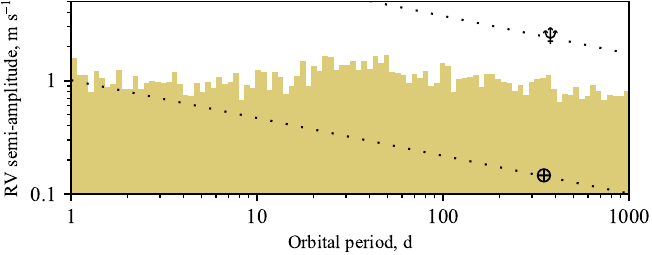}
    \caption{
    The detection limits of \citet{Stefanov2025b} for GJ~526, an early-type M dwarf without confirmed planets at the time of writing. Shaded bands show the $3\sigma$ RV semi-amplitude posterior of a modelled planet in their selected model, in 100 log-spaced orbital-period bins. Two dotted lines show the RV semi-amplitudes that an Earth-mass planet and a Neptune-mass planet would inject in the system. Adapted from \citet{Stefanov2025b}, Figure 5, under a CC-BY 4.0 licence.
    }
    \label{fig:detection_limit}
\end{figure}

Other tests assess detection limits by virtue of Bayesian inference. The simplest check is to take the sampled planetary-parameter posteriors of the most appropriate model, and propagate them directly into a period-mass distribution \citep{Faria2020}. A more sophisticated approach is to take the most appropriate model and condition it to the median of your posteriors, akin to injection-recovery tests, but then to perform a series of many reduced-complexity Bayesian inferences on residual data. One could add a hypothetical circular planet and sample its parameters, while holding fixed all model components except radial-velocity offsets and jitters. This process can be repeated across many orbital-space bins, running a separate inference for each bin.\footnote{It is this part in particular that splits the procedure into many computationally palatable sub-procedures -- in the very spirit of nested sampling.}
The result is a collection of `bin models' that only sampled a few parameters that pertain to RV, most notably the semi-amplitude of the hypothetical planet $k_\text{hyp}$. High percentiles of the $k_\text{hyp}$ posterior from each bin model can tell us the largest semi-amplitude the hypothetical planet could have, if it existed at all \citep{SuarezMascareno2025,SuarezMascareno2025a,Stoeva2026}. For example, \citet{Stefanov2025b} realised this detection-limit test on GJ~526 with the $+3\sigma$ percentile (Fig.~\ref{fig:detection_limit}). In their case, what is presented as the detection limit is the $+3\sigma$ semi-amplitude of a potential undiscovered planet.

We should mind that, similar to the technicalities around modelling activity signals near planets in Sect.~\ref{sec:pseudo_planets}, the different design of these detection-limit tests changes the very meaning of what a detection limit is. For example, injection-recovery tests are fundamentally based on the inability of the selected model to recognise an RV signal. Bayesian-inference variants try to mitigate for this by not injecting a synthetic signal that they aim to track, but cannot go around the modelling sensitivity.

Detection-limit tests are valuable for assessing the selection bias that is inherent in current exoplanetary-search efforts. This is particularly important to a related field of ours -- that of exoplanetary demographics, where selection biases influence observed planetary occurrences. However, we stress that detection-limit tests can still serve us as weak indicators of planetary nature, even though they do not strictly pertain to it. Planetary candidates are typically detected above the detection-limit threshold -- and that is expected, since the signal they produced was strong enough to be identified  \citep{Tuomi2014}. In fact, it would be concerning if a detected signal fell below said threshold, as such case would suggest that observations did not favour the neighbourhood in $P$--$k_\text{rv}$ space around our solution. In such a scenario, the planetary candidate would appear highly unusual -- and more often than not, such anomalies are non-planetary. We can extend this reasoning to all comparisons between a potential planetary candidate and the distributions of known planets in the literature. If a candidate lies far from the bulk of any observed distributions, this begs the question of whether this is a sign of the most marvellous planet, or of an honest mistake of ours.

\section{Aims and objectives of this doctoral programme}
The main objective of this doctoral programme is to search for low-mass planetary companions within the habitable zones of late-type stars, with a particular emphasis on M dwarfs. This implies a detailed study and understanding of the activity of these stars. We shall utilise both archival and contemporaneous spectroscopic data of nearby M dwarfs in order to:
(i)~assess their stellar activity using spectroscopic indicators,
(ii)~gain an understanding of how stellar activity manifests in RV and activity indicators,
(iii)~design a methodology that interprets or separates stellar-activity signals from those potentially attributable to planetary companions, including modelling of stellar-surface inhomogeneities.
We shall take advantage of optical and near-infrared spectroscopy (Table~\ref{tab:spectrographs}), as well as multi-band photometry (Table~\ref{tab:photographs}).

To meet these objectives, the following aims will be pursued. We will collect \num{e2} optical and near-infrared spectroscopic measurements per star for a selected sample of nearby, low-activity M-dwarfs. These observations will enable a detailed analysis of stellar activity. We will propose a set of targets for robotic-telescope monitoring. This monitoring will produce photometric time series that will support a more comprehensive model of their stellar activity. We will employ existing exploratory and periodogram-based routines (Sect.~\ref{sec:pgram_tests}) to detect periodic signals, GP frameworks (Sect.~\ref{sec:gp_frameworks}) to model the underlying stellar activity, and Bayesian-inference methods (Sect.~\ref{sec:bayesian_tests}) to propose or confirm planetary candidates. We will derive velocimetry from available spectra through established RV-determination techniques (CCF, template matching, line-by-line routines; Sect.~\ref{sec:rv_measurement}) to improve the accuracy, the identification and subsequent subtraction of stellar-activity signals from RV data. Finally, we will search for planetary companions around our heterogeneous sample of M dwarfs, and use all available knowledge to isolate planetary signals from stellar activity, which typically varies at the order of \SI{e1}{\metre\per\second}.

In the chapters to follow, I outline my main contributions as a leading author in studies that aimed at achieving these objectives. In Chapter~\ref{chap:gj3998d}, we report the discovery of GJ~3998~d, a super-Earth within the habitable zone of the nearby M dwarf GJ~3998. In Chapter~\ref{chap:gj526}, we present a comprehensive analysis of the stellar activity of the nearby M dwarf GJ~526 in RV, FWHM and Mount Wilson S-index time series. In Chapter~\ref{chap:barnard_letter}, we search for transits of planets in the multi-planetary system of Barnard's star during its first and to date only TESS sector. Chapter~\ref{chap:conclusions} summarises the main findings of these studies and their logical lines of continuation. Appendix~\ref{chap:appendix} discusses additional related works that were carried out alongside this doctoral programme.

\begin{table}[t!]
\centering
\caption{Summary of the spectrographs that were used in the works that comprise or are associated with this dissertation.}
\vspace{4pt}
\label{tab:spectrographs}
\begin{tabular}{lcccc}
\hline\hline
\multicolumn{1}{c}{Instrument} &
\multicolumn{1}{c}{Telescope} &
\multicolumn{1}{c}{$\lambda$, nm} &
\multicolumn{1}{c}{$R$, \num{e3}} &
\multicolumn{1}{c}{Ref.} \\ \hline
CARMENES & Calar Alto \SI{3.5}{\metre} & \numrange{520}{960} & 93 & 1 \\
 &  & \numrange{960}{1710} & 82 & 1 \\
ESPRESSO & VLT \numproduct{4x8.2}$\,\unit\metre$ & \numrange{378}{789} & \numrange{140}{190} & 2 \\
HARPS & ESO \SI{3.6}{\metre} & \numrange{378}{691} & 115 & 3 \\
HARPS-N & TNG \SI{3.6}{\metre} & \numrange{383}{693} & 115 & 4 \\
HIRES & Keck~I \SI{10}{\metre} & \numrange{300}{1100} & 67$^\dagger$ & 5 \\
NIRPS & ESO \SI{3.6}{\metre} & \numrange{970}{1810} & 100 & 6 \\
\hline
\multicolumn{5}{p{0.95\linewidth}}{
\footnotesize
$^\dagger$depends on slit width. References:
(1)~\citet{carmenes_instrument},
(2)~\citet{espresso},
(3)~\citet{harps},
(4)~\citet{harpsn},
(5)~\citet{hires},
(6)~\citet{Bouchy2025}.
}
\end{tabular}
\end{table}

\begin{table}[t!]
\centering
\caption{Summary of photometric instruments that were used in the works that comprise or are associated with this dissertation.}
\vspace{4pt}
\label{tab:photographs}
\begin{tabular}{lcccc}
\hline\hline
\multicolumn{1}{c}{Instrument} &
\multicolumn{1}{c}{Telescope} &
\multicolumn{1}{c}{Band} &
\multicolumn{1}{c}{Pixel scale, $''$} &
\multicolumn{1}{c}{Ref.} \\ \hline
TESS & Satellite, \numproduct{4x10.5}$\,\unit{\centi\metre}$ & TESS G & 21 & 1 \\
ASAS & Ground net'k, \SIrange{7}{25}{\centi\metre} & ASAS V & \numrange{14}{15} & 2 \\
 & & ASAS I & \numrange{14}{15} & 2 \\
ASAS-SN & Ground net'k, \SI{14}{\centi\metre} & Johnson V & 8 & 3,4 \\
 & & Sloan g & 8 & 3,4 \\
\hline
\multicolumn{5}{p{0.95\linewidth}}{
\footnotesize
References:
(1)~\citet{tess},
(2)~\citet{asas},
(3)~\citet{asassn1},
(4)~\citet{asassn2}.
}
\end{tabular}
\end{table}

%% file: src/chapters/gj3998d.tex
\chapter{A super-Earth in the habitable zone of GJ~3998}
\chaptermark{A super-Earth around GJ~3998}
\label{chap:gj3998d}
\begin{verse}[60mm]
\noindent You must persevere, must develop new strength by continuous study, until that which is only a good inclination becomes a good settled purpose.
\end{verse}
\attrib{Seneca, \textit{Moral Letters to Lucilius}, XVI.1 (R. M. Gummere, trans.), circa AD 65}
\vspace{7mm}
\lettrine[nindent=4pt,findent=1pt,lraise=-0.0,loversize=-0.21]{\starburst{S}}{cientific} insight is not exhausted by a single analysis of a dataset. Although it may initially appear that a scientific 
investigation concludes with the acquisition and processing of data, this is often just the beginning. 
Indeed, in certain cases, a small number of additional measurements can prompt hypotheses that are later substantiated by the broader body of available data. At times, however, it happens that observational data do not come out `at the right time', and have to await several years in their archives for their relevant methods and mathematical machinery to be constructed, so that they serve to be useful.

Such was the case with GJ~3998, a quiet early M-dwarf in our cosmic neighbourhood (\SI{18.3}{\parsec}; \citealp{gaiaDR3}). Since 2012, GJ~3998 had been a target of the HADES programme \citep{Affer2016}, an RV uninformed search of planets around nearby M dwarfs for the purposes of demographics in the near vicinity. The HADES programme was realised with the high-resolution ultra-stable HARPS-N spectrograph at the \SI{3.58}{\metre} \textit{Telescopio Nazionale Galileo} (TNG; \citealp{harpsn}) at the Roque de Los Muchachos Observatory, La Palma, Spain. \citet{Affer2016} utilised 138 HARPS-N spectra to discover two planets around GJ~3998:
GJ~3998~b, with a period near
\SI{2.65}{\day}, and eccentricity of
zero, and a minimum mass of
\mbox{$2.47\pm 0.27\,\unit\earthmass$};
and GJ~3998~c, with a period near
\SI{13.74}{\day}, an eccentricity of
\mbox{$0.049^{+0.052}_{-0.034}$},
and a minimum mass of
\mbox{$6.26^{+0.79}_{-0.76}\,\unit\earthmass$}. The authors noted the presence of two other significant peaks in the RV periodogram.
The first was at \SI{30.7}{\day}, which the authors attributed to the stellar rotational period $P_\text{rot}$ of GJ~3998, and successfully constrained to 
\mbox{$P_\text{rot}=31.8^{+0.6}_{-0.5}\,\unit\day$} through a one-dimensional GP.
The second peak was at \SI{42.5}{\day}, which at the time, was understood to be `indicative of a modulation in stellar variability due to differential rotation', i.e. a by-product of stellar activity.

Once initiated, the HADES programme would in fact produce a series of 15 subsequent articles to date, with the last being our own re-analysis on GJ~3998. In 2024, armed with more spectra (204 in total), more sophisticated mathematical machinery (e.g. $FF'$ GP regression; \citealp{Rajpaul2015,spleaf1,spleaf2}), improved RV determination techniques (e.g. the LBL implementation by \citealp{Artigau2022}), and the natural arrogance of a first-year doctoral student, we set out to re-analyse this spectroscopy and look for more.

I identified GJ~3998 as a candidate for re-analysis. I designed the analysis methodology, with guidance from my supervisors and my research group on the selection of research questions to address. This methodology led me to develop several analysis tools, such as \textsc{nieva} (Stefanov et al. in prep; see Sect.~\ref{sec:future_work}). I performed the entirety of the analysis, which led to the confirmation of GJ~3998~d, with the exception of the RV reduction using SERVAL \citep{serval}, a task carried out by Nicola Nari. I authored all components of this work, including the text and figures.

\cleardoublepage
\includepdf[
    pages=-,
    offset=6mm 0,
]{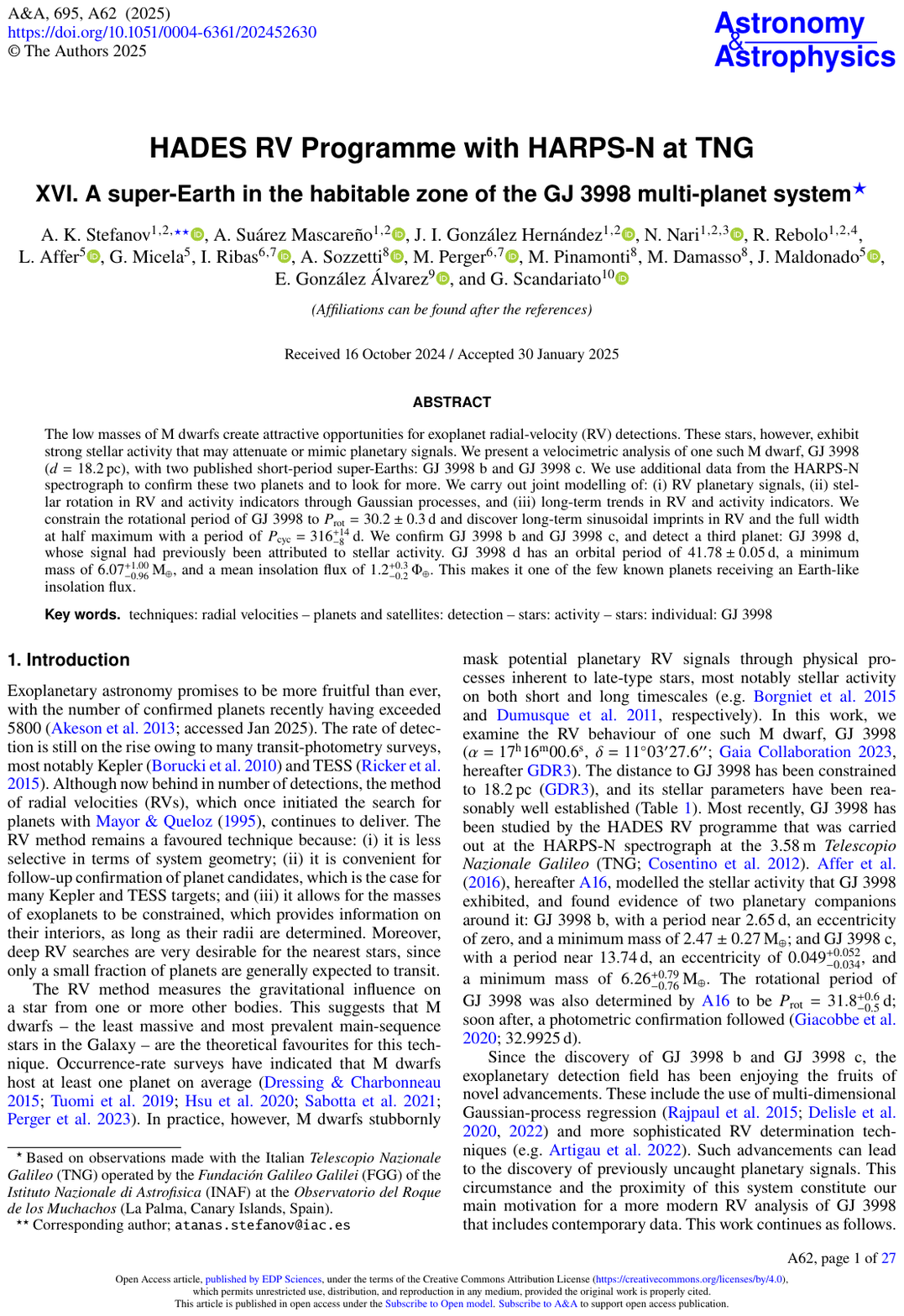}

%% file: src/chapters/gj526.tex
\chapter{Stellar-activity analysis of the nearby M dwarf GJ~526}
\chaptermark{Stellar-activity analysis of GJ~526}
\label{chap:gj526}
\begin{verse}[60mm]
\noindent Materials are indifferent, but the use that one makes of them is by no means indifferent. How, then, can one preserve firmness and calmness of mind, and at the same time the attentiveness that saves us from careless and thoughtless action? By following the example of those who play at dice. The counters are indifferent, the dice are indifferent. How can I know in what way the throw will fall? But to be attentive and skilful in making use of whatever does fall, that is now my task.
\end{verse}
\attrib{Epictetus, \textit{Discourses}, II.5 (R. Hard, trans.), circa AD 108}
\vspace{7mm}
\lettrine[nindent=4pt,findent=1pt,lraise=-0.0,loversize=-0.21]{\starburst{A}}{lthough} the discovery of exoplanets remains a major driver of observational efforts in the field, cool dwarfs with no detectable planetary companions deserve comparable attention. From a present-day perspective, each cool dwarf appears to be a scientifically rich laboratory in its own right and with its own unique activity signature. Nature provides us with stars whose potential orbital companions are so elusive that the stars themselves become excellent subjects for the study of stellar activity -- free from the complications that arise in planet-star interaction scenarios. Therefore, it would be improvident not to analyse such stars with the same rigour as those showing potential planetary signals. One day, the cumulative knowledge we gain from such studies will be sufficient to enable the classification of stars into well-defined groups based on their activity characteristics, rather than continuing the current artisanal, star-by-star approach.

The article that follows, \citet{Stefanov2025b}, tries to conduct such analysis on the nearby M dwarf GJ~526 ($d=\SI{5.4}{\parsec}$). GJ~526 has been known for its clear rotational activity ($P_\text{rot}=\SI{52.1(12.0)}{\day}$; \citealp{SuarezMascareno2017}), and its stable, slowly evolving activity has made it a valuable spectral standard (e.g. \citealp{Binks2016,Pancino2017}). Over the past two decades, GJ~526 has been observed by several high-precision RV programmes: Lick, HIRES, HARPS, CARMENES, and ESPRESSO -- yielding a continuous \SI{26}{\year} spectroscopic record. Its very high-resolution spectroscopic observations with ESPRESSO, in particular, yielded exceptionally precise RV measurements, with uncertainties as low as \SI{10}{\centi\metre\per\second}. Despite the depth and consistency of these data, no comprehensive analysis had previously been undertaken for this star. We recognised that the long-term high-fidelity monitoring of this object could make it a key benchmark for understanding stellar activity, and we therefore initiated a detailed analysis.

I designed the analysis methodology of this work, with guidance from my supervisors and my research group on the selection of research questions to address. The only components of the analysis that I did not perform myself were: the derivation of stellar parameters (Sect.~3), carried out by Sérgio G. Sousa, Hugo M. Tabernero, and María-Rosa Zapatero Osorio; and the proper-motion anomaly technique to refine the detection limit of planets around GJ~526 (parts of Sect.~6.3), carried out by Alessandro Sozzetti. I authored all components of this work, including the text and figures, except Fig.~B.1 (auth. María-Rosa Zapatero Osorio).
 
\cleardoublepage
\includepdf[
    pages=-,
    offset=6mm 0,
]{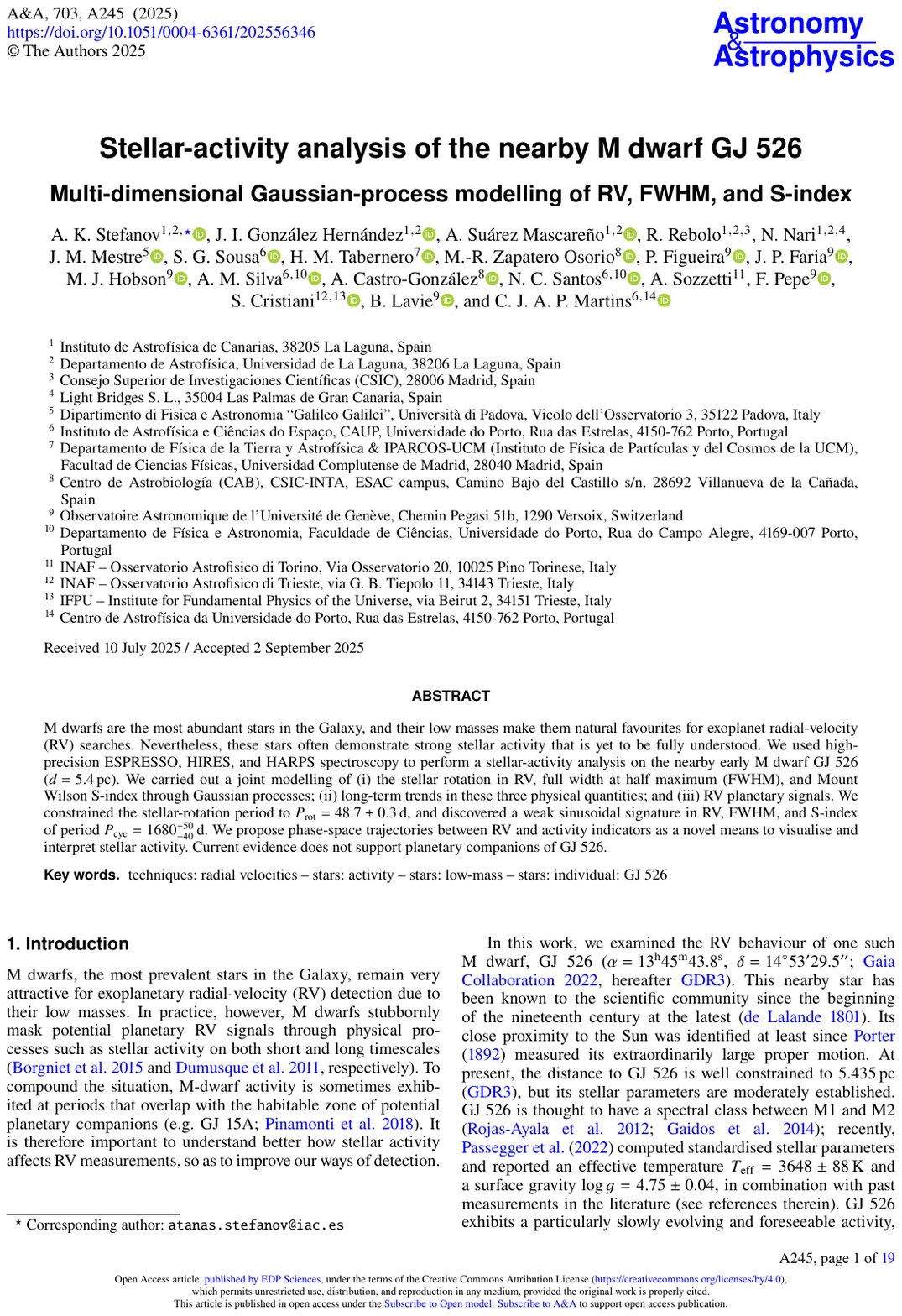}

%% file: src/chapters/barnard_letter.tex
\chapter{No transits of Barnard's star in TESS photometry}
\chaptermark{No transits of Barnard's star}
\label{chap:barnard_letter}
\begin{verse}[60mm]
\noindent In joining battle, seek the quick victory. If battle is protracted, your weapons will be blunted and your troops demoralized \textellipsis\ Thus in war, I have heard tell of a foolish haste, but I have yet to see a case of cleverly dragging on the hostilities.
\end{verse}
\attrib{Sun Tzu, \textit{The Art of Warfare}, II.3--6 (R. T. Ames, trans.), fifth century BCE}
\vspace{7mm}
\lettrine[nindent=4pt,findent=1pt,lraise=-0.0,loversize=-0.21]{\starburst{O}}{ften,} the paths we lay out for ourselves are fraught with difficulties. At times, however, unexpected opportunities come along the way, and wait for those fortunate enough to seize them. Such was the case with the following letter to \textit{Astronomy \& Astrophysics} \citep{Stefanov2025}, which came about quite unexpectedly. It was 17 September 2024, a lazy Tuesday morning, and our research group was holding a regular weekly meeting. Only a few weeks had passed since the acceptance of \citet{GonzalezHernandez2024}, the unequivocal discovery of the sub-Earth Barnard~b, that was led by my supervisor Jonay. 
The detection of a planet orbiting Barnard's star was a feat the astronomical community had been trying to achieve since the 1960s -- and though I did not at the time fully grasp the importance of this discovery, I could still sense the tension in the room that day.

The press release had been scheduled for 1 October, exactly two weeks later. Most of the work on the article had already been completed, but we kept discussing it -- perhaps because its weight on our minds felt too great to ignore. That Tuesday, we were discussing new lines of work on Barnard's star, with the idea of investigating its photometric behaviour more thoroughly. In that meeting, I checked the TESS coverage of Barnard's star out of habit, and I happened to come across its first TESS sector that had been published just a few days after the acceptance of \citet{GonzalezHernandez2024}. This high-quality photometry dataset appeared to contain nothing conspicuous. Yet, we were faced with a decision. It was too late to revise the already approved article, so we had to take the initiative -- otherwise, the opportunity would be missed. We decided to prepare a self-contained letter on this photometric data, which would be released alongside the press release of Barnard~b on 1 October. With only two weeks remaining, it was necessary to work with expedition. Jonay asked me, `Can you do it?' and I responded automatically, `Yes,' without being sure myself if I was telling the truth. I went back home, pulled the shutters down, closed the curtains... and 14 days later, \citet{Stefanov2025} appeared on arXiv alongside \citet{GonzalezHernandez2024}. This letter was submitted to \textit{Astronomy \& Astrophysics} on the same day, and it was accepted on 28 November 2024, with no revisions requested by the referee.

Barnard's star is undoubtedly an interesting object by itself, but its planetary system would have likely become one of the most significant in the night sky, had even a single planet been observed to transit. Transits would have opened the door to full physical characterisation, atmospheric, and dynamical studies in a system so close to our own. \citet{GonzalezHernandez2024} estimated the transit probability of Barnard~b to be 3.7 per cent. This may seem modest, but given the attractive scientific opportunity, the risk-reward ratio of the letter was nevertheless very favourable. After all, rolling a pair of sixes is about as probable -- and we are all too familiar with how often this happens in games of chance.

Barnard's star is by no means faint -- for it has been monitored photometrically for decades, primarily for the purpose of flare- and activity studies (e.g. \citealp{Toledo-Padron2019} and the data presented therein). We estimated the maximum potential transit depths of all four planets to fall within \SIrange{1.0}{1.4}{\ppt}. Recurring transits of such depth once every few days would have perhaps produced detectable signals that would have eventually prompted dedicated follow-up observations. Most likely, the challenge was in the coverage of those historical datasets, for even such `few-day' transits would require repeated observations at precise phases. In this regard, TESS photometry stands out as perhaps the first homogeneous and nearly-uninterrupted dataset that enabled clean and robust tests for transits around Barnard's star. We took advantage of this circumstance and carried out a timely and methodologically robust check of this nearby planetary system. 

In any case, the preparation of \citet{Stefanov2025} proved fundamental to my academic development, for it was shown to me that I could take on a scientific problem from conceptualisation to publication in a short amount of time. And so, this whole effort turned out to be a very good training in the life cycle of a scientific article, in a controlled and highly reputable environment, and in the company of sterling co-authors, who helped me greatly and whom I thank for their patience and prompt responses.

I identified the novel photometry of Barnard's star for analysis, which led to the realisation of this work. I led all aspects of the analysis, with guidance from my supervisors, my research group and the broad circle of authors. The only components of the analysis that I did not perform myself were: the use of Wōtan \citep{wotan} and transit least-squares periodograms (TLSPs; \citealp{tls}) for the search of transits in SAP photometry (parts of Sect.~3.1), carried out by Mario Damasso; as well as the transit-curve fitting (Sect.~3.2), carried out by Alejandro Suárez Mascareño. I authored all components of this work, including the text and figures.

\cleardoublepage
\includepdf[
    pages=-,
    offset=6mm 0,
]{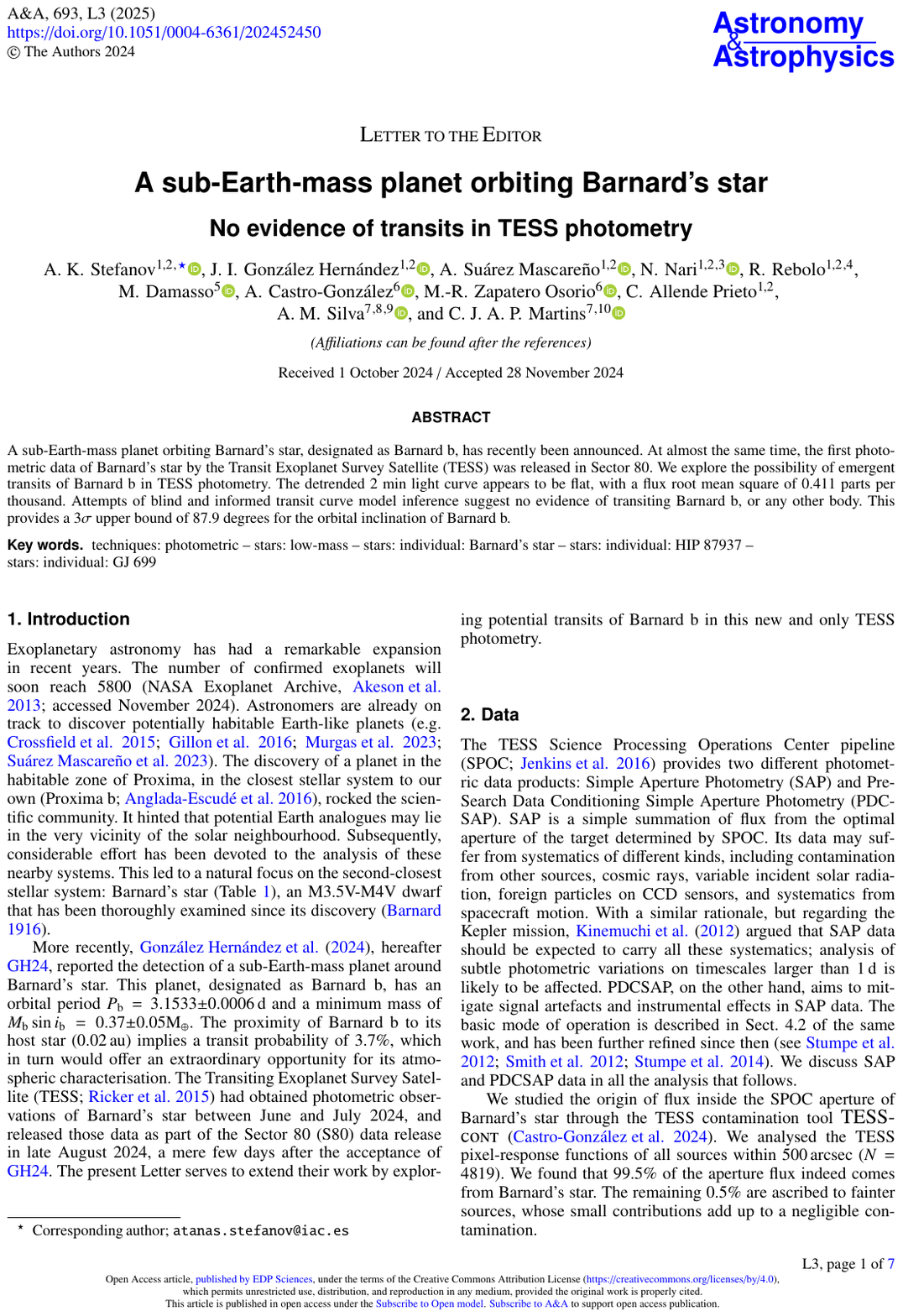}

%% file: src/chapters/conclusions.tex
\chapter{Conclusions}\label{chap:conclusions}
\begin{verse}[90mm]
\raggedleft
Come ease, or come travail; come pleasure or pain,\\
My warst word is---`Welcome, and welcome again!'
\end{verse}
\attrib{Robert Burns, \textit{Contented wi' little} (Alexander Smith, edit.), AD 1794}
\vspace{7mm}
\lettrine[nindent=4pt,findent=1pt,lraise=-0.0,loversize=-0.21]{\starburst{I}}{n} this dissertation, I employed high-resolution ultra-stable spectroscopy and high-precision photometry to understand stellar activity through GP modelling, as well as to search for low-mass rocky planets in the habitable zones of M dwarfs. The main conclusions of this work now follow.

I used 204 high-precision HARPS-N spectra of the nearby early M dwarf GJ~3998. I re-reduced these spectra through the line-by-line framework by \citet{Artigau2022} that uses the full spectral information to measure RVs. My primary goal was to confirm two published short-period super-Earths: GJ~3998~b and GJ~3998~c. This endeavour required the joint modelling of: (i)~RV planetary signals, (ii)~stellar rotation in RV and activity indicators through GPs, and (iii)~long-term trends in RV and activity indicators. Through Bayesian inference, I formally confirmed GJ~3998~b and GJ~3998~c, and detected a third planet: GJ~3398~d, which had previously been interpreted as a stellar-activity by-product \citep{Affer2016}. I measured GJ~3998~b, GJ~3998~c, and GJ~3998~d to have orbital periods of
\mbox{$2.65033^{+(22)}_{-(19)}$\,\unit\day},
\mbox{$13.727^{+0.003}_{-0.004}$\,\unit\day}, and
\mbox{$41.78\pm 0.05\,\unit\day$}, respectively, and
minimum masses of
\mbox{$2.50^{+0.30}_{-0.29}$\,\unit\earthmass},
\mbox{$6.82^{+0.78}_{-0.75}$\,\unit\earthmass}, and
\mbox{$6.07^{+1.00}_{-0.96}$\,\unit\earthmass}, respectively. This new discovery, GJ~3998~d, receives incident flux comparable to that of Earth (\mbox{$1.2^{+0.3}_{-0.2}\,\unit\earthflux$}), and is one of the few known exoplanets to do so. Moreover, GJ~3998~d is a rare case of a false-negative exoplanetary detection, and required a multitude of planetary-nature tests, so as to address concerns raised by \citet{Affer2016}. Most tests have been reviewed in Sect.~\ref{sec:hunter_guide}, and have significantly inspired the development of this section.
My detection-limit test indicated that if a fourth planet existed, it would have induced an RV semi-amplitude smaller than \SI{1}{\metre\per\second}. Contemporaneous velocimetry suggested that GJ~3998 exhibited magnetic activity with a period of
\mbox{$316^{+14}_{-8}$\,\unit\day}, which would place it among the stars with rather short magnetic-cycle periods. I observed this longer-term stellar activity in three activity indicators: FWHM, H$\alpha$, and Na~I. Finally, I updated the constraint on the rotational period of GJ~3998 to \mbox{$P_\text{rot}=30.2\pm 0.3\,\unit\day$}.

I utilised the spectral information from 75 ESPRESSO, 56 HIRES, and 32 HARPS high-precision spectra of the nearby early M dwarf GJ~526 over a baseline of \SI{26}{\year}. Using the aforementioned joint-modelling framework, I conducted a stellar-activity analysis of GJ~526 in three physical quantities: RV, FWHM, and Mount Wilson S-index. I determined the stellar-rotation period of GJ~526 to be \mbox{$P_\text{rot} = 48.7 \pm 0.3\,\unit\day$}, and deduced that its active regions are dark spots on the stellar surface. I found that GJ~526 exhibited weak magnetic activity with a period of \mbox{$P_\text{cyc} = 1680^{+50}_{-40}\,\unit\day$}. I showed that there was no current evidence in favour of a planetary companion around GJ~526. If such a companion were present, its RV semi-amplitude would be expected to be less than 
\SI{1.5}{\metre\per\second}. I introduced phase-space trajectories as a tool for visualising and interpreting stellar activity. These trajectories offer advantages over traditional correlations between RVs and activity indicators by virtue of their encoded temporal information. I qualitatively compared phase-space trajectories of GJ~526 in RV, FWHM, 
and S-index over two time intervals separated by nearly $P_\text{cyc}/2$. I demonstrated that these trajectories could provide meaningful insights in temporal intervals that contained sufficiently many data points.

Through the course of this doctoral programme, \citet{GonzalezHernandez2024} reported the detection of a sub-Earth-mass planet orbiting Barnard's star (Barnard~b; \mbox{$P_\text{b}=\SI{3.1533(0.0006)}{\day}$}), as well as three planetary candidates (Barnard~c,d,e; \SIrange{2.3}{6.7}{\day}). Days after this work was accepted, TESS released its first photometric observations of Barnard's star. In response, I undertook an analysis of this newly available photometric data, with the objective of extending the findings of \citet{GonzalezHernandez2024} under a 14-day time constraint. Neither PDCSAP nor SAP photometry exhibited any apparent transit features within the measurement precision, and the flux distribution of detrended SAP photometry was remarkably Gaussian-like. I applied a similar methodology to the detrended PDCSAP photometry of \SI{20}{\second} and \SI{2}{\minute} cadence. In both cases, I could detect no significant transit signals. On that basis, I concluded that the current TESS photometry did not contain transits of Barnard~b, or any other potential planetary candidate. From this result, I derived a 3$\sigma$ upper bound of \SI{87.9}{\degree} for the orbital inclination of Barnard~b.

These three works (Chapters~\ref{chap:gj3998d}, \ref{chap:gj526}, \ref{chap:barnard_letter}) and all other associated works to this doctoral programme (Chapter~\ref{chap:appendix}) reinforce that stellar activity is not simply another form of white noise. Instead, it is a dynamic, often dominant source of variability for which we cannot prepare much in advance. When searching for low-amplitude planets, conventional methods such as periodogram-based or simple-correlation techniques often fail to capture the full complexity of stellar activity. The prudent application of GPs appears to offer a more effective approach, at least in the current phase of detection of sub-$\unit{\metre\per\second}$ semi-amplitudes. Even if this does not remain to be the case in the future, we have already got to an important insight. Stellar activity is not another nuisance we should mind -- it has already redefined how we interpret RV data in cool-dwarf systems. This recognition must be accepted, as it is essential for the successful detection of low-mass exoplanets.

\section{Future work}\label{sec:future_work}
The work that led to the results in this dissertation was very multi-faceted, and required extensive numerical modelling and working with large amounts of data. This paved the way of future perspectives that pertain both to the discovery of new systems as well as general problems in the field.

\subsubsection{Immediate priorities}

Stellar activity influences spectral lines in an uneven way, as touched upon in Sect.~\ref{sec:activity_handbook}. On the other hand, planetary motion causes uniform Doppler shifts in all lines. During my research stay in \emph{Institut Trottier de recherche sur les exoplanètes}, Montréal, Canada, I examined the LBL correlation in pixel space between the template flux and its temperature derivative. The slope of this correlation, combined with the normalised fractional change in black-body radiance, defines a new metric $\rho$, which I treated as an intrinsic property of a spectral line. My team and I found that the spectral lines of two well-known and active late-type stars -- AU~Mic and V1298~Tau -- produce systematically different RVs depending on their $\rho$. For small temperature variations, this dependence can be assumed linear, and I showed that its slope $\zeta$ can function as a stand-alone activity indicator. I provided a physical justification for $\zeta$ and demonstrated using SPIRou data of the aforementioned stars that it performs effectively within a multi-dimensional $FF'$ GP model (Sect.~\ref{sec:ff_gps}). $\zeta$ is directly linked to RV, unaffected by planetary motion, and inexpensive to calculate within the LBL framework implemented by \citet{Artigau2022}. This work is already submitted to \emph{Astronomy \& Astrophysics} and is expected to be in its second round of peer review by the time of my defence.

At the same time, I advanced individual analyses of three particular early-type M dwarfs: StKM~1-1877, GJ~21, and GJ~1. StKM~1-1877 has been known to the community for its young age and strong stellar activity, and had been continuously monitored through high-resolution spectroscopy. I succeeded in modelling its strong \mbox{$10^2\,\unit{\metre\per\second}$} RV signal in activity, and identified a subtle \mbox{$10^0\,\unit{\metre\per\second}$} RV signal, StKM~1-1877.I, in the time series. StKM~1-1877.I passed all planetary-nature tests described in Sect.~\ref{sec:hunter_guide} except Sect.~\ref{sec:bayes_factor}. GJ~21 has got a previously proposed \mbox{$10^0\,\unit{\metre\per\second}$} planetary candidate \citep{Pinamonti2022}. Current observational evidence supports the planetary nature of this candidate, but the test described in Sect.~\ref{sec:apodisation} returned inconclusive results. I have been awarded \SI{33}{\hour} of observational time on the HARPS-N spectrograph to confirm the planetary candidates around StKM~1-1877 and GJ~21. This observational programme is due to end in September 2026, and the analysis of its data shall be my primary pastime during my arranged post-doctoral appointment. GJ~1 shows no evidence of planetary companions, but has been studied by several RV programmes, including: ESPRESSO, CARMENES, HARPS, and most notably, NIRPS \citep{Bouchy2025}. I have reached two well-described, self-contained models of the stellar activity of GJ~1 in the infrared and visible spectral band. I am currently trying to combine these two models for the purposes of comparison and commentary on the stellar activity in both spectral ranges. I have committed to completing this work as a NIRPS-consortium manuscript in due course, and it shall form a key component of the work package on uninformed planetary searches (NIRPS WP1).

\subsubsection{Long-term prospects}

While the exploration of phase-space trajectories in Chapter~\ref{chap:gj526} revealed significant potential to interpret stellar activity, further work will be needed to evaluate their practical utility. I shall examine whether specific trajectory features such as the direction of motion or geometric properties (e.g. angular speed or enclosed area) carry physical meaning. I shall examine these possibilities empirically, or potentially by comparing observed phase-space trajectories of other cool dwarfs with those produced by synthetic stellar-activity simulations.

In exoplanetary searches, many stars are found without planetary companions. I have a strong conviction that these opportunities must not be wasted, for they offer valuable insight into stellar activity -- much like the example we have set with GJ~526 (Chapter~\ref{chap:gj526}). I have modelled and characterised the stellar activity of 11 other late-type stars using archival high-precision spectroscopy from CARMENES, HARPS, HARPS-N, and HIRES (Table~\ref{tab:ongoing_dwarfs}).
\begin{table}[t!]
\centering
\caption{
A list of cool dwarfs, for which my stellar-activity models are sufficiently well-behaved to be published independently, or to form a composite paper. For each star, I mention my current best model, as well as its preliminary measure of the stellar-rotation period.
}
\label{tab:ongoing_dwarfs}
\begin{tabular}{lllc}
\hline\hline
\multicolumn{1}{c}{Star} &
\multicolumn{1}{c}{Used spectroscopy} &
\multicolumn{1}{c}{Current best model} &
\multicolumn{1}{c}{$P_\text{rot}$, d}
\\ \hline
GJ~2 & CARMENES & $FF'$ GP; RV\&$\Delta T$ & $21.4\pm 0.2$ \\
 & HARPS-N & & \\
 & HIRES & & \\[0.8ex]
GJ~26 & CARMENES & $FF'$ GP; RV\&$\Delta T$ & $54.8^{+4.1}_{-5.5}$ \\
 & HARPS-N & & \\
 & HIRES & & \\[0.8ex]
GJ~162 & HARPS-N & $FF'$ GP; RV\&$\Delta T$ & $32.6^{+2.1}_{-1.3}$ \\[0.8ex]
GJ~169 & CARMENES & $FF'$ GP; RV\&H$\alpha$ & $25.6^{+0.4}_{-0.3}$ \\[0.8ex]
GJ~208 & CARMENES & $FF'$ GP; RV\&H$\alpha$ & $12.1\pm 0.1$ \\
 & HARPS & & \\
 & HIRES & & \\[0.8ex]
GJ~212 & CARMENES & $FF'$ GP; RV\&H$\alpha$ & $17.9\pm 0.1$ \\[0.8ex]
GJ~450 & CARMENES & $FF'$ GP; RV\&H$\alpha$ & $23.1^{+0.4}_{-0.3}$ \\
 & HARPS & & \\
 & HIRES & & \\[0.8ex]
GJ~606 & CARMENES & $FF'$ GP; RV\&$\Delta T$ & $21.2\pm 0.1$ \\
 & HARPS & & \\
 & HARPS-N & & \\[0.8ex]
GJ~1074 & HARPS-N & $FF'$ GP; RV\&Phot & $18.8\pm 1.0$ \\[0.8ex]
GJ~4063 & CARMENES & $FF'$ GP; RV\&H$\alpha$ & $39.2^{+15.4}_{-4.8}$ \\
 & HIRES & & \\[0.8ex]
GJ~4306 & HARPS-N & $FF'$ GP; RV\&$\Delta T$ & $30.2\pm 1.1$ \\ \hline
\end{tabular}
\end{table}
These objects have reached their final stage of analysis. I have the liberty of publishing them separately, with individual focus on their behaviour; or in groups, as representative examples of broader advances in the stellar-activity interpretation through GP frameworks (e.g. phase-space trajectories). Their manner of publication shall be shaped by their opportunities for broader impact, and by my own future priorities.

\subsubsection{Emergent software infrastructure}

The work that led to this dissertation involved a sequence of seemingly disparate steps that ranged from low-level tasks (e.g. programmatically reading, exploring, and manipulating tabular data) to high-level ones (e.g. modelling complex stellar activity). During a past formal programming training of mine, I had been consistently exposed to the principle of avoiding code duplication. This has influenced my approach to software design, and the present academic efforts naturally resulted in a self-contained Python codebase that benefits from its underlying C-based dependencies (\textsc{nieva};\footnote{A conjugated form of the Spanish \emph{nevar}, `the act of snowing'.} Stefanov et al., in prep). I have prioritised computational efficiency, numerical accuracy and broad usability in the development of \textsc{nieva}.

To illustrate that, Listing~\ref{lst:nieva} 
\begin{listing}[t!]
\caption{The fundamental Python code that generates the best model in \citet{Stefanov2025a}.}
\begin{minted}{python}
# Load time series (automatic parsing beyond CSV files)
harpsn = nieva.dataclasses.Timeseries("data/GJ3998/harpsn_cds.csv")
# Define model
model = nieva.models.MEPModel(
    # Include time series
    source_bundle=nieva.models.SourceBundle(
        sources={
            "harpsn":{
                "timeseries":harpsn,
                "dimensions":["rv_lbl", "fwhm_ccf"],
                "dimensions_err":["e_rv_lbl", "e_fwhm_ccf"],
            },
        },
    ),
    # Include N sine-like/Keplerian signals
    plasins=3,
    # Define FF' GP
    mode="multidimensional",
    include_grad=[True, False],
    # Define any long-term function in a modular manner
    corrector_bundle=nieva.correctors.CorrectorBundleSum([
        nieva.correctors.CSineCorrectorBundle(),
        nieva.correctors.NDimCorrectorBundle([
            nieva.correctors.PolyCorrector(2),
            nieva.correctors.PolyCorrector(2)
        ])
    ])
)
# Define priors
model.priors["corr_period"] = nieva.distributions.LUniformDistribution(200,2000)
model.priors["corr_dim1_ksin"] = nieva.distributions.UniformDistribution(0,40)
model.priors["corr_dim2_ksin"] = nieva.distributions.UniformDistribution(0,40)
model.priors["corr_dim1_phase"] = nieva.distributions.UniformDistribution(0.25,1.25)
model.priors["corr_dim2_phase"] = nieva.distributions.UniformDistribution(0.5,1.5)
model.priors["plasin1_period"] = nieva.distributions.UniformDistribution(2,3)
model.priors["plasin1_krv"] = nieva.distributions.UniformDistribution(0,5)
model.priors["plasin1_phase"] = nieva.distributions.UniformDistribution(-0.4,0.6)
# ... priors for other planets omitted for the sake of brevity
model.priors["mep_period"] = nieva.distributions.LUniformDistribution(2,200)
model.priors["mep_timescale"] = nieva.distributions.LUniformDistribution(40,1e4)
# The model object generates functions that can be used with any sampler
sampler = ultranest.ReactiveNestedSampler(
    param_names=cm.get_parameter_names(),
    wrapped_params=cm.get_parameter_wraps(),
    loglike=cm.loglike,
    transform=cm.transform
)
\end{minted}
\label{lst:nieva}
\end{listing}
contains the fundamental Python code that generates the best model of my analysis of GJ~3998 \citep{Stefanov2025a}: a long-term function with independent quadratic functions in RV and FWHM and a common-sine component, three circular-orbit planets, and a MEP QP kernel in multi-dimensional $FF'$ form. The entirety of the code is contained in just a few dozen lines of code, and remains fully in memory, meaning it can be `bent' and played with by the user. \textsc{nieva} is built in an abstract, modular manner, and can serve many nested-sampling implementations. I shall see to it that \textsc{nieva} is published in due time, following the submission of this dissertation.

\section{Personal outlook for the future of the field}
The achievement of reaching RV precision better than \SI{1}{\metre\per\second} (the metre-per-second barrier) was convincingly surpassed with the commissioning of HARPS \citep{harps}. This took place more than two decades ago. Since then, there have been only 14 non-controversial RV detections with semi-amplitudes below \SI{1}{\metre\per\second} and with a precision exceeding $3\sigma$.
If we disregard the RV-detection criterion, the number of such detections increases to 19. For a comparison, the total number of non-controversial exoplanet discoveries for the same time period is over 6000 \citep{nasaexoplanettable2}. Several years ago, the decimetre-per-second barrier was also overcome with the introduction of ESPRESSO \citep{espresso}. However, even with ESPRESSO, the planet with the smallest measured RV semi-amplitude, Barnard~e, stands just above \SI{20}{\centi\metre\per\second} \citep{GonzalezHernandez2024,Basant2025}, and required considerable observational effort.

A consistent RV precision of \SI{10}{\centi\metre\per\second} is a remarkable engineering feat, and will undoubtedly reduce observational effort by minimising the amount of exposure time per detection. However, I believe this is not the way out of the problem, and that we should not be relying on precision alone. It has become evident that the main challenge in the detection of sub-\unit{\metre\per\second} planets lies in stellar activity. This is not a novel observation -- late dwarfs have been documented on many occasions to exhibit highly individual, time-variable, and unpredictable behaviour. No two late dwarfs seem to behave in the same way, and we sometimes jokingly refer to them as having their own personalities. This is why current stellar-activity modelling is largely carried out on a star-by-star basis, akin to artisanry.

Nature cannot have been so wasteful in providing a unique activity behaviour for every star. There must be a set of fundamental, simple physical laws that give rise to such diverse behaviour. These laws should be understood, much like how chaos theory helped understand phenomena in heat transfer, ecological modelling, and other dynamical systems. Therefore, to me, it is imperative that we seek ways to scale these star-by-star analyses, and transform them into an industrialised, systematic process. One potential path is to apply a homogeneous framework to a sample of \num{e2} cool dwarfs (e.g. $FF'$ GPs with the same activity indicator).
This would allow the community to identify common patterns, understand the nature of stellar activity, and most importantly, foretell its strong and weak `seasons' for the purposes of sub-\unit{\metre\per\second} planet hunting. I maintain that addressing this last issue is critical. We have yet to fully explore the exoplanetary population with sub-\unit{\metre\per\second} RV semi-amplitudes, and our instrumental-precision capabilities are already at \SI{10}{\centi\metre\per\second}. Without a more clear understanding of the ways of stellar activity, I believe that such discoveries will remain out of reach.

There is another issue in the field that I am afraid is not discussed often enough. We have been predominantly prioritising precision over recall of RV detections. Stellar activity gives a very good reason to do so, and much of this dissertation aimed to facilitate this choice. However, despite our efforts and the many documented false positives in our field, I suspect the number to be even greater among the current sample of confirmed planets. The systematic work on false positives remains to be a demanding and underappreciated task. Identifying false positives comes with the usual drudgery, but receives little media attention, can strain professional relationships, and is less likely to be published on account of being a negative result.

This should not dishearten us, but only inspire a greater commitment to consistency and rigour in the application of detection tools and evaluation criteria. Our ultimate goal should be to establish a catalogue of robustly confirmed exoplanets. This can include both the homogeneous re-analysis of archival data (e.g. \citealp{Osborne2025}), or a study of new systems in the same manner. Observational data originates from diverse instruments, spans different epochs, and often comes in many forms (e.g. iodine-cell RVs, CCF RVs, raw spectra). Yet, recent literature demonstrates that Bayesian inference and modern mathematical machinery have proven capable of addressing these inconsistencies. Among our caught planetary trophies, there may still be hundreds of tins, bottles, and other rubbish fished from deep stellar waters -- and those await their proper identification and correction.

%% file: src/chapters/other_contribs.tex
\chapter{Other contributions}\label{chap:appendix}
\lettrine[nindent=4pt,findent=1pt,lraise=-0.0,loversize=-0.21]{\starburst{D}}{uring} my doctoral programme, I was able to contribute to 21 peer-reviewed scientific publications beyond the three already included in this manuscript. What follows is a list of this work in chronological order. Appendix~\ref{sec:contributions_major} provides a list of my more substantial contributions, including the published article abstract and a brief description of my involvement. Appendix~\ref{sec:contributions_minor} lists other publications where my contributions were limited to proofreading, data supply, results interpretation, and general advice.

\section{Major contributions}\label{sec:contributions_major}
\textbf{\fullcite{GonzalezHernandez2024}.}
Barnard's star is a primary target within the ESPRESSO guaranteed time observations (GTO) as it is the second closest neighbour to our Sun after the $\alpha$ Centauri stellar system. We present here a large set of 156 ESPRESSO observations of Barnard's star carried out over four years with the goal of exploring periods of shorter than \SI{50}{\day}, thus including the habitable zone (HZ). Our analysis of ESPRESSO data using Gaussian process (GP) to model stellar activity suggests a long-term activity cycle at \SI{3200}{\day} and confirms stellar activity due to rotation at \SI{140}{\day} as the dominant source of radial velocity (RV) variations. These results are in agreement with findings based on publicly available HARPS, HARPS-N, and CARMENES data. ESPRESSO RVs do not support the existence of the previously reported candidate planet at \SI{233}{\day}. After subtracting the GP model, ESPRESSO RVs reveal several short-period candidate planet signals at periods of \SI{3.15}{\day}, \SI{4.12}{\day}, \SI{2.34}{\day}, and \SI{6.74}{\day}. We confirm the \SI{3.15}{\day} signal as a sub-Earth mass planet, with a semi-amplitude of \SI{55(7)}{\centi\metre\per\second}, leading to a planet minimum mass $m_p\sin i$ of \SI{0.37(5)}{\earthmass}, which is about three times the mass of Mars. ESPRESSO RVs suggest the possible existence of a candidate system with four sub-Earth mass planets in circular orbits with semi-amplitudes from \SI{20}{\centi\metre\per\second} to \SI{47}{\centi\metre\per\second}, thus corresponding to minimum masses in the range of \SIrange{0.17}{0.32}{\earthmass}. The sub-Earth mass planet at \SI{3.1533(6)}{\day} is in a close-to circular orbit with a semi-major axis of \SI{0.0229(3)}{\au}, thus located inwards from the HZ of Barnard's star, with an equilibrium temperature of \SI{400}{\kelvin}. Additional ESPRESSO observations would be required to confirm that the other three candidate signals originate from a compact short-period planet system orbiting Barnard's star inwards from its HZ.
\textbf{Personal contribution:} I worked on a BERV undercorrection on Barnard's star from the ESPRESSO DRS, which was expected to leave a \SI{1}{\year}, \SI{20}{\centi\metre\per\second} peak-to-peak signal (J.~P.~Faria, priv.comm.). I found a proper-motion truncation issue in spectral headers that in fact left a \SI{35}{\centi\metre\per\second} peak-to-peak signal.\\\\
\textbf{\fullcite{SuarezMascareno2025}.}
We obtained 420 high-resolution spectra of Proxima, over 159 nights, using the Near Infra Red Planet Searcher (NIRPS). We derived 149 nightly binned radial velocity measurements with a standard deviation of \SI{1.69}{\metre\per\second} and a median uncertainty of \SI{55}{\centi\metre\per\second}, and performed a joint analysis combining radial velocities, spectroscopic activity indicators, and ground-based photometry, to model the planetary and stellar signals present in the data, applying multi-dimensional Gaussian process regression to model the activity signals. We detect the radial velocity signal of Proxima~b in the NIRPS data. All planetary characteristics are consistent with those previously derived using visible light spectrographs. In addition, we find evidence of the presence of the sub-Earth Proxima~d in the NIRPS data. When combining the data with the HARPS observations taken simultaneous to NIRPS, we obtain a tentative detection of Proxima d and parameters consistent with those measured with ESPRESSO. By combining the NIRPS data with simultaneously obtained HARPS observations and archival data, we confirm the existence of Proxima~d, and demonstrate that its parameters are stable over time and against change of instrument. We refine the planetary parameters of Proxima~b and d, and find inconclusive evidence of the signal attributed to Proxima~c \mbox{($P =\SI{1900}{\day}$)} being present in the data. We measure Proxima~b and d to have minimum masses of
\SI{1.055(55)}{\earthmass}, and
\SI{0.260(38)}{\earthmass}, respectively. Our results show that, in the case of Proxima, NIRPS provides more precise radial velocity data than HARPS, and a more significant detection of the planetary signals. The standard deviation of the residuals of NIRPS after the fit is $~\SI{80}{\centi\metre\per\second}$, showcasing the potential of NIRPS to measure precise radial velocities in the near-infrared.
\textbf{Personal contribution:} I led the transit-curve analysis that is contained in Appendix~H. It is conceptually analogous to \citet{Stefanov2025}, Sect.~3.1.\\\\
\textbf{\fullcite{Lamontagne2026}.}
We present an updated characterization of the planetary system orbiting the nearby M2 dwarf GJ~3090 (TOI-177; $d=\SI{22}{\parsec}$), based on new high-precision radial velocity (RV) observations from NIRPS and HARPS. With an orbital period of \SI{2.85}{\day}, the transiting sub-Neptune GJ~3090~b has a mass we refine to \SI{4.52(0.47)}{\earthmass}, which, combined with our derived radius of \SI{2.18(0.06)}{\earthradius}, yields a density of \mbox{$2.40^{+0.33}_{-0.30}\unit{\gram\per\centi\metre\cubed}$}. The combined interior structure and atmospheric constraints indicate that GJ~3090~b is a compelling water-world candidate, with a volatile-rich envelope in which water likely represents a significant fraction. We also confirm the presence of a second planet, GJ~3090~c, a sub-Neptune with a \SI{15.9}{\day} orbit and a minimum mass of \SI{10.0(1.3)}{\earthmass}, which does not transit. Despite its proximity to the star's \SI{18}{\day} rotation period, our joint analysis using a multidimensional Gaussian process (GP) model that incorporates TESS photometry and differential stellar temperature measurements distinguishes this planetary signal from activity-induced variability. In addition, we place new constraints on a non-transiting planet candidate with a period of \SI{12.7}{\day}, suggested in earlier RV analyses. This candidate remains a compelling target for future monitoring. These results highlight the crucial role of multidimensional GP modelling in disentangling planetary signals from stellar activity, enabling the detection of a planet near the stellar rotation period that could have remained undetected with traditional approaches.
\textbf{Personal contribution:} My personal workflow naturally grew into a self-contained codebase, with which I explore RV data efficiently, and with which I carry out joint modelling of: (i) RV planetary signals, (ii) stellar rotation in RV and activity indicators through Gaussian processes, (iii) long-term trends in RV and activity indicators. Mr Lamontagne used this codebase to validate their overall results and to generate Figure~8 in their work.\\\\
\textbf{\fullcite{Stoeva2026}.}
Detection and characterisation of Jovian analogues in precision radial velocity (RV) measurements is gaining momentum due to the constantly increasing observational baseline of Doppler surveys. The occurrence rate of Jovian-mass exoplanets is crucial for understanding the architecture of planetary systems. However, long-period RV signals in Doppler surveys could also be induced by stellar magnetic cycles, leading to misinterpretations of planetary candidates. We investigate the long-term RV variability in the K-dwarf star GJ~1137 (HD~93083, HIP~52521), a known Saturn-mass exoplanet host, and assess the role of stellar activity in shaping the observed signals. We analyse 13 years of archival high-precision spectroscopic observations obtained with the High Accuracy Radial velocity Planet Searcher spectrograph (HARPS). We performed an extensive spectroscopic analysis of the stellar activity indicators and applied an RV modelling approach, incorporating Keplerian fits, Gaussian process regression as a proxy for stellar activity, and other stellar activity diagnostics. Furthermore, we refined the orbital parameters and the minimum mass of the known exoplanet GJ~1137~b and searched for additional planetary candidates in the system. We detect a long-period RV signal that, if interpreted as planetary, would suggest the presence of a Jovian-analogue companion. However, our spectroscopic activity analysis provides strong evidence that this variability is induced by the star’s long-term magnetic cycle \mbox{( $P_\text{cyc} = 5870^{+480}_{-350}$ days)} rather than by an orbiting planet. The signal is detected in both the full width at half maximum (FWHM) of the cross-correlation function and the chromospheric activity index \mbox{$\log R'_{\rm HK}$}. We measure the stellar rotation period to $P_\text{rot}=32.3^{+1.2}_{-1.3}\,\unit\day$ and identify a significant short-period RV signal, which we attribute to a Super Earth with a period of $9.6412^{+(12)}_{-(11)}\,\unit\day$ and a minimum mass of \mbox{$5.12^{+0.70}_{-0.69}\,\unit\earthmass$}, making GJ~1137 a multiple-planet system.
\textbf{Personal contribution:} I led the efforts towards the stellar-activity modelling with multi-dimensional GPs of GJ~1137. This was ultimately instrumental in detecting GJ~1137~c (Sects.~4.1--4.4).

\section{Minor contributions}\label{sec:contributions_minor}
~\vspace{-\baselineskip}

\fullcite{SuarezMascareno2024}

\fullcite{Nari2025}

\fullcite{Allart2025}

\fullcite{Vaulato2025}

\fullcite{Bouchy2025}

\fullcite{GomesDaSilva2025}

\fullcite{Bazinet2025}

\fullcite{SuarezMascareno2025a}

\fullcite{Parc2025}

\fullcite{Vaulato2025a}

\fullcite{Nari2026}

\fullcite{Frensch2026}

\fullcite{Osborn2026}

\fullcite{Srivastava2026}

\fullcite{Nari2026a}

\fullcite{Weisserman2026}

\fullcite{Pelletier2026}